\documentclass[prd,aps,a4paper,showkeys,nofootinbib]{revtex4-1}

\usepackage{graphicx,psfrag}
\usepackage{mathrsfs}
\usepackage{amsmath,amsfonts,amssymb}
\usepackage{multirow}
\usepackage{comment}
\usepackage{ulem} \usepackage{hyperref}

\usepackage{cleveref}  \Crefname{equation}{Eq.}{Eqs.}
\Crefname{figure}{Fig.}{Figs.}
\Crefname{tabular}{Tab.}{Tabs.}

\usepackage{enumitem}

\usepackage{xspace}

\usepackage[caption=false]{subfig}

\newcommand{\be}{\begin{equation}}
\newcommand{\ee}{\end{equation}}
\newcommand{\bea}{\begin{eqnarray}}
\newcommand{\eea}{\end{eqnarray}}
\newcommand{\bel}{\begin{align}}
\newcommand{\eel}{\end{align}}

\def\d{{\rm d}}

\def\GMc2{{\rm G M_{\odot} c^{-2}}}

\def\kt2{\kappa^\text{T}_2}

\def\D{\mathrm{D}}
\def\pd{\partial{}}

\def\2nd{2^\mathrm{nd}}
\def\4th{4^\mathrm{th}}
\def\6th{6^\mathrm{th}}
\def\8th{8^\mathrm{th}}

\def\defG{\widehat{\Gamma}}

\def\z4c{$\mathrm{Z}4\mathrm{c}$}
\def\z4oc{$\mathrm{Z}4(\mathrm{c})$}
\def\z4{$\mathrm{Z}4$}
\def\ccz4{$\mathrm{CCZ}4$}

\def\hpb{\hphantom{(}}

\newcommand{\tGRAthena}{\texttt{GR-Athena++}}
\newcommand{\GRAthena}{\tGRAthena\xspace}

\newcommand{\Cpp}{\texttt{C++}\xspace}
\newcommand{\AwA}{\texttt{AwA}\xspace}

\usepackage{pifont} %

\newcommand{\GRA}{\texttt{GR-Athena++}}

\def\z4c{$\mathrm{Z}4\mathrm{c}$}
\def\z4oc{$\mathrm{Z}4(\mathrm{c})$}
\def\z4{$\mathrm{Z}4$}
\def\ccz4{$\mathrm{CCZ}4$}

\usepackage{color}
\definecolor{cyan}{rgb}{0,0.9,0.9}
\definecolor{orange}{rgb}{0.9,0.5,0}
\definecolor{magenta}{rgb}{1,0,1}
\definecolor{purple}{rgb}{0.8,0.4,0.8}
\definecolor{gray}{rgb}{0.8242,0.8242,0.8242}

\begin{document}

\title{Spectrally-tuned compact finite-difference schemes with domain decomposition and applications to numerical relativity}

\author{Boris \surname{Daszuta}$^{1}$}
\affiliation{${}^1$Theoretisch-Physikalisches Institut, Friedrich-Schiller-Universit{\"a}t Jena, 07743, Jena, Germany}

\date{\today}

\begin{abstract}
    Compact finite-difference (FD) schemes specify derivative approximations implicitly, thus to achieve parallelism with domain-decomposition suitable partitioning of linear systems is required.
      Consistent order of accuracy, dispersion, and dissipation is crucial to maintain in wave propagation problems such that deformation of the associated spectra of the discretized problems is not too severe. In this work we consider numerically tuning spectral error, at fixed formal order of accuracy to automatically devise new compact FD schemes. Grid convergence tests indicate error reduction of at least an order of magnitude over standard FD. A proposed hybrid matching-communication strategy maintains the aforementioned properties under domain-decomposition. Under evolution of linear wave-propagation problems utilizing exponential integration or explicit Runge-Kutta methods improvement is found to remain robust. A first demonstration that compact FD methods may be applied to the \z4c{} formulation of numerical relativity is provided where we couple our header-only, templated \Cpp{} implementation to the highly performant \GRA{} code. Evolving \z4c{} on test-bed problems shows at least an order in magnitude reduction in phase error compared to FD for propagated metric components. Stable binary-black-hole evolution utilizing compact FD together with improved convergence is also demonstrated.
\end{abstract}


\maketitle

\section{Introduction}
\label{sec:intro}
Finite difference (FD) methods provide a well-known and flexible approach on structured grids furnishing derivative approximants at target grid-points. Allowing for an implicit, linear relation between known function samples and sought derivative values leads to compact finite difference (CFD) schemes \cite{Lele:1992cd}. At fixed formal order of accuracy and total number of coupled points, CFD allows for construction of narrower stencils \cite{hirsh1975higherorderaccurate,carpenter1994timestableboundaryconditionsa} when compared with explicit FD together with improved resolution characteristics over a wider range of (spatial) scales \cite{Lele:1992cd,mehra2017algorithm986suite}. Numerical solution of the associated linear systems (typically banded tri- or penta-diagonal \cite{hirsh1975higherorderaccurate,Lele:1992cd,fu1997highorderaccurate,wang2013highordercfdmethods,qin2014highaccuracynumerical,chen2021novelparallelcomputing}) is more expensive than the direct evaluation of explicit FD. Solution of such banded systems is of linear algorithmic complexity \cite{higham2002accuracy,golub2013matrix,askar2015solvingpentadiagonallinear} and this computational overheard is mitigated through improved resolving efficiency. Indeed this efficiency at resolving widely disparate length-scales has led CFD-based techniques to be extensively widely in areas such as e.g.~computational aeroacoustics \cite{kim2007optimisedboundarycompact}, direct numerical simulations of Navier-stokes without turbulence modeling \cite{jagannathan2016reynoldsmachnumber}, and large eddy simulations \cite{bodony2005usinglargeeddysimulation}; see also \cite{song2022robusthighresolutionsimulations}.

Given problem-specific requirements it may be preferable to quantify and minimize error at specific spatial scales or alternatively over a desired wavenumber range when viewed in the frequency domain. Indeed increasing resolution capability of CFD further rather than trunction order can enchance overall accuracy more effectively \cite{kim1996optimizedcompactfinite,kim2007optimisedboundarycompact}. One approach in derivation of CFD schemes is to impose a specific (implicit) stencil arrangement and by Taylor matching derive coefficients that yield a scheme of specific formal order of accuracy \cite{hirsh1975higherorderaccurate,Lele:1992cd}. Unfortunately this does not give direct access to control over desired spectral behaviour at particular wavenumbers. Instead rather than matching all possible coefficients a degree of freedom may be left underdetermined. This can be exploited through minimization against a suitably chosen cost function characterizing an error profile which allows for spectral-tuning \cite{kim1996optimizedcompactfinite,liu2008optimizedcompactfinite,zhang2013optimizedexplicitfinitedifference}. A unified approach to this for construction of CFD schemes with arbitrary derivative degree, order, and stencil size (and bias) was recently introduced in \cite{Deshpande:2019uf}. The advantage of this latter approach in addition to the aforementioned generalizations is that various properties such as coefficient (skew-)symmetries do not need to be imposed a priori but are a consequence of the optimization procedure. While quite general, the framework of \cite{Deshpande:2019uf} does not immediately embed derivation of multi-derivative such as those of \cite{fu1997highorderaccurate,qin2014highaccuracynumerical} or Hermite-FD methods \cite{Fornberg:2020ch}.

In realistic problems that significantly deviate from non-ideal conditions parallelizing simulations can be extremely demanding both from the point of view of problem reformulation aspects together with implementation such that high performance computing (HPC) infrastructure may be efficiently utilized \cite{huerta2019supportinghighperformancehighthroughput}. Implicit CFD schemes introduce data-dependency which must be treated when one seeks to introduce parallelism. Two broad categories of approaches can be envisaged \cite{kim2013quasidisjointpentadiagonalmatrix,chen2021novelparallelcomputing}: the algorithmic approach and the boundary approximation approach. In the algorithmic approach the implicit system is solved in parallel utilizing techniques such as the pipeline Thomas algorithm \cite{povitsky2000higherordercompactmethod} or parallel diagonal dominance algorithm \cite{sun1995applicationaccuracyparallel,terekhov2016highlyscalableparallel}. For the boundary approximation approach (BAA) a computational domain is partitioned and closures are imposed on the CFD scheme such that inversion of linear systems may be performed in a decoupled fashion on each sub-domain. Unfortunately this can lead to changes in the resolving efficiency in the vicinity of boundaries and potentially introduce artifacts. Examples of treating partitioned sub-domains include partial overlapping with implicit closures supplemented by one-sided filtering as in \cite{sengupta2007newcompactscheme} or sub-domain ghost layer extension with closures prescribed based on a combination of optimization \cite{kim2013quasidisjointpentadiagonalmatrix}. It is known that regardless of the style of approach when fluid flow or wave-propagation problems are treated numerical dispersion relation preservation (DRP) is crucial \cite{tam1993dispersionrelationpreservingfinitedifference}. It was recently demonstrated in \cite{chen2021novelparallelcomputing} that for first degree derivative, upwind CFD schemes DRP can be achieved exactly. This was illustrated in a $\4th$ order scheme that showed excellent consistency during numerically evolved flow problems under domain partitioning preserving accuracy particularly as sub-domain sampling was increased.

As in the hydrodynamical context an important concern for numerical relativity (NR) investigations of the binary black hole (BBH) merger problem is careful treatment of the widely-varying range of length and time-scales involved \cite{ashtekar2015general}. The numerical solution of the BBH merger problem crucially complements observational gravitational wave (GW) based detection efforts \cite{Abbott:2016blz,TheLIGOScientific:2016wfe,TheLIGOScientific:2017qsa}. A variety of successful NR based approaches for the BBH merger problem are known. Time discretization is generally achieved through a method of lines prescription whereas the most common spatial treatment has utilized Cartesian grid coordinatization and been based on a combination of FD and adaptive mesh refinement (AMR) in \cite{Pollney:2009yz,Reisswig:2012nc,brown2009turduckeningblackholes,sperhake2007binaryblackholeevolutions,zlochower2005accurateblackhole,herrmann2007unequalmassbinary} built upon \cite{Goodale:2002a} -- see also the independent codes of \cite{Brugmann:2008zz,cao2008reinvestigationmovingpunctured,Clough:2015sqa}. Pseudo-spectral methods involving multi-patch decomposition of the computational domain through a combination of topological spheres and cylinders have also found excellent success \cite{Szilagyi:2009qz} and the related discontinuous Galerkin based approaches also show future promise \cite{Hilditch:2015aba,Bugner:2015gqa,Kidder:2016hev}. In light of recent HPC trends towards massively parallel systems a particular emphasis has been placed within new codes on domain-decomposition strategies compatible with excellent scaling properties. In particular domain-decomposition described through oct-tree based grids has recently been explored in the FD-based codes of \cite{Fernando:2018mov,Daszuta:2021ecf}. Curiously for NR while AMR based FD approaches, together with (pseudo)-spectral and discontinuous Galerkin methods have been pursued it does not appear that CFD methods have been previously utilized for the evolution problem. On the GW observational front the continuing effort towards improving operating sensitivity of current detectors \cite{Abbott_2020} and developing new detectors \cite{akutsu2020overviewkagracalibration,amaroseoane2017laserinterferometerspace,Punturo:2010zz,Evans:2016mbw} allows probing of ever more extreme regions of the underlying parameter space and consequently motivates a search for alternative numerical methods and hence we seek to take a first step towards bridging this gap here.

The first goal of this work is to obviate the need for repeated, hand construction of CFD schemes by automatizing the procedure of \cite{Deshpande:2019uf} numerically and extending it to multi-derivative schemes thereby allowing for rapid experimentation on model problems prior to general application. A second goal is to combine this framework with the BAA technique proposed in \cite{chen2021novelparallelcomputing}, extend it to centered schemes and consider how it may be further refined through an iterative procedure. A third goal is to demonstrate that the CFD methods ensuing from this lead to improved accuracy in solution of wave-propagation problems while remaining robust in the parallel context. Finally we seek to demonstrate that CFD shows future promise in application to the evolution problem of numerical relativity in simulation of binary mergers. The rest of this paper is organized as follows: In \S\ref{sec:method} we describe the overall method for construction of general FD schemes, providing a characterization of spectral error and description of how it may be numerically tuned subject to problem requirements. We describe our hybrid-communication strategy that builds upon DRP and can be leveraged during problems involving domain-decomposition. In \S\ref{sec:res} we consider application to a variety of wave-propagation problems whereupon the \z4c{} system as implemented in \cite{Daszuta:2021ecf} is investigated in \S\ref{ssec:z4c}. Section \ref{sec:conc} concludes. For convenience we have also collected a selection of stencils investigated in the text in \S\ref{app:sch_tuned_coeff}.

\section{Method}
\label{sec:method}

The aim of this section is to provide a method to optimize finite-difference based numerical evaluation of an unknown derivative of a function based on known (derivative) data. Our approach builds upon the framework of \cite{Deshpande:2019uf} and extends it to allow incorporation of data from multiple derivatives in addition to function data in specification of stencils.

\begin{figure}[htbp]
  \includegraphics[width=0.38\textwidth]{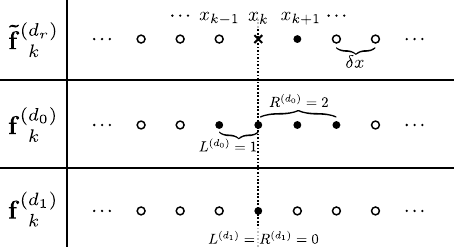}%
  \hfill
  \caption{Coupling of (derivative)-function data at salient nodes (filled circles) in specification of the $d_r$-derivative at the target base-point $x_k$ (cross position) for an example multi-derivative stencil. Here the scheme is implicit as $L^{(d_r)}=0$ but $R^{(d_r)}=1$. The method is also biased as $L{}^{(d_K)}\neq R{}^{(d_K)}$. One restriction we impose for convenience is that the central nodes are included for all classes of function derivative involved in specification of a stencil.
      }
  \label{fig:stencil_sketch}
\end{figure}

Consider a discretization of the interval $\Omega$ of uniform spacing $\delta x$ where sampled points are denoted $x_k$.
Suppose $f\in C{}^{\infty}(\Omega)$ then function (derivative) samples are denoted $f{}_{\hpb k}^{(d)}:=\left.\partial{}^d_x[f(x)]\right|_{x_k}$. A generalized finite-difference stencil may be viewed as a relation between linear combinations of unknown and specified samples. We therefore introduce the stencil coefficient weights\footnote{Typically we suppress indicial ranges when evident from context.}\footnote{In this work we assume that coefficients take values in $\mathbb{F}:=\mathbb{R}$.}:
\begin{alignat}{2}
  \label{eq:coeffTupDefn}
  \boldsymbol{\alpha}^{(d_K)}
  &:=
  \left(
    \alpha_{\hpb n}^{(d_K)}
  \right)_{n=-L^{(d_K)}}^{R^{(d_K)}}
  \in \mathbb{F}^{L^{(d_K)} + R^{(d_K)} + 1}
  &
  \quad
  (0 \leq L^{(d_K)}) \wedge (0\leq R^{(d_K)});
\end{alignat}
together with function samples:
\begin{equation}
  \label{eq:fcnTupDefn}
  \mathbf{f}^{(d_K)}_{\hpb k}
  :=
  \left(
    f{}_{\hpb k+m}^{(d_K)}
  \right)_{m=-L^{(d_K)}}^{R^{(d_K)}}.
\end{equation}
The product is interpreted through summation $\boldsymbol{\alpha}^{(d_K)}\cdot\mathbf{f}^{(d_K)}_{\hpb k} = \sum_{m=-L^{(d_K)}}^{R^{(d_K)}} \alpha_{\hpb m}^{(d_K)} f{}_{\hpb k+m}^{(d_K)}$. As ansatz for the $d_r$-derivative of $f$ at the base-point $x_k$ denoted $\tilde{f}{}_{\hpb k}^{(d_r)}$ we form:
\begin{equation}
  \label{eq:mdAnsatz}
  \delta x^{d_r} \boldsymbol{\alpha}^{(d_r)} \cdot \tilde{\mathbf{f}}{}_{\hpb k}^{(d_r)}
  =
  \sum_{I\in\mathcal{I}}
  \delta x^{d_I}
  \boldsymbol{\alpha}^{(d_I)} \cdot \mathbf{f}{}_{\hpb k}^{(d_I)},
\end{equation}
where $\mathcal{I}$ is an indexing set with number of elements denoted $|\mathcal{I}|$ and $\mathcal{D}_\mathcal{I}:=(d_I)_{I\in\mathcal{I}}$ is ordered such that $0\leq d_0<d_1<\cdots<d_{|\mathcal{I}|-1}$. We assume that $d_r\notin \mathcal{D}_\mathcal{I}$. Notice that we have not restricted the number of elements in individual coefficient weights which allows for consideration of biased schemes. Our setup will allow us to form explicit or implicit stencils. Prescriptions that involve function data and also derivatives as known samples such as the Hermite methods of \cite{Fornberg:2020ch} together with their implicit extensions may also be constructed. As a schematic of multi-derivative, implicit stencil node coupling see Fig.\ref{fig:stencil_sketch}.

In order to determine the $\boldsymbol{\alpha}$-coefficients of a particular scheme we need to impose constraints. Suppose that $\tilde{\mathbf{f}}{}_{\hpb k}^{(d_r)}$ and $\mathbf{f}{}_{\hpb k}^{(d_I)}$ are known then Eq.~\eqref{eq:mdAnsatz} may be viewed as a linear system of equations for %
$\mathcal{C}_S:=L^{(d_r)}+R^{(d_r)}+\sum_{I\in\mathcal{I}}(L^{(d_I)}+R^{(d_I)}) + |\mathcal{I}| + 1$ %
unknowns specifying the stencil.
To achieve this notice that in Eq.\eqref{eq:mdAnsatz} we also have a freedom to fix an overall scaling and so we impose $\alpha{}^{(d_r)}_{\hpb 0} = 1$. Next note that for a stencil of formal order $p$ we require the derivative approximant $\tilde{f}{}_{\hpb k}^{(d_r)}$ to have truncation error $\varepsilon_T:=\tilde{f}{}_{\hpb k}^{(d_r)}-f{}_{\hpb k}^{(d_r)}=\mathcal{O}(\delta x^p)$. Therefore we consider Taylor series expansion of $f{}^{(d_K)}$ about $x_k\in \Omega$ to order $r$:
\begin{equation}
  \label{eq:Tcons}
  f_{\hpb k+m}^{(d_K)}
  =
  \sum_{l=0}^{r}\frac{(m\delta x)^l}{l!}
  \left.
    \partial_x^{d_K+l}[f(x)]
  \right|_{x=x_k}
  + \mathcal{O}(\delta x^{r+1}).
\end{equation}
Substitution of Eq.\eqref{eq:Tcons} into Eq.\eqref{eq:mdAnsatz} and matching coefficients order by order leads to $\mathcal{C}_T$ additional linear constraints on the $\boldsymbol{\alpha}$-coefficients. If it is the case that $r$ is selected such that $\mathcal{C}_S=\mathcal{C}_T+1$ then we have a fully determined linear system that may be solved uniquely resulting in a scheme for numerical calculation of a derivative approximant of fixed formal order. For a general, implicit, multi-derivative scheme this can be a potentially tedious procedure and consequently we make use of a computer algebra system (CAS) to automate construction of these constraints \cite{notebook_repo}.
As a example consider Fig.\ref{fig:stencil_sketch} for $d_r=1$, $d_0=0$, and $d_1=2$. In this case we have $C_S=7$, $W=2$, and can find $C_T=6$ linear conditions:
\begin{equation}
\begin{aligned}
  0 &= \sum_{m=-1}^2 \alpha{}^{(0)}_{\hpb m}, \\
  0 & = \alpha{}^{(1)}_{\hpb 0} + \alpha{}^{(1)}_{\hpb 1}
  + \alpha{}^{(0)}_{\hpb -1}
  - \alpha{}^{(0)}_{\hpb 1}
  - 2\alpha{}^{(0)}_{\hpb 2}, \\
  0 & = 2\alpha{}^{(1)}_{\hpb 1}
  - \alpha{}^{(0)}_{\hpb -1}
  - \alpha{}^{(0)}_{\hpb 1}
  - 4 \alpha{}^{(0)}_{\hpb 2}
  - 2 \alpha{}^{(2)}_{\hpb 0}, \\
  0 & = 3 \alpha{}^{(1)}_{\hpb 1}
  + \alpha{}^{(0)}_{\hpb -1}
  - \alpha{}^{(0)}_{\hpb 1}
  - 8 \alpha{}^{(0)}_{\hpb 2}, \\
  0 & = 4 \alpha{}^{(1)}_{\hpb 1}
  - \alpha{}^{(0)}_{\hpb -1}
  - \alpha{}^{(0)}_{\hpb 1}
  -16 \alpha{}^{(0)}_{\hpb 2}, \\
  0 & =
  5\alpha{}^{(1)}_{\hpb 1}
  + \alpha{}^{(0)}_{\hpb -1}
  - \alpha{}^{(0)}_{\hpb 1}
  - 32 \alpha{}^{(0)}_{\hpb 2}.
\end{aligned}
\end{equation}
These together with the normalization condition immediately yields:
\begin{align}
  \boldsymbol{\alpha}^{(1)} &= \left(
    1,\, 2/3
  \right), &
  \boldsymbol{\alpha}^{(0)} & = \frac{1}{3}\left(
    1/3,\, -11/2,\, 5,\, 1/6
  \right), &
  \alpha^{(2)}_{\hpb 0} &= -1/3;
\end{align}
which can be verified to be of formal order $\mathcal{O}(\delta x^5)$ for computing $\tilde{f}^{(1)}_{\hpb k}$ implicitly. Such order matching strategies are often employed in the construction of the classical Pad\'{e}
schemes \cite{Lele:1992cd}.

Of more interest is the case of $\mathcal{C}_S>\mathcal{C}_T+1$ which yields an \textit{underdetermined system}. This property can be exploited. We can consider further introduction and constrained minimization against a well-defined objective function $\varepsilon$ (e.g. characterizing some error metric of interest) so as to yield $\boldsymbol{\alpha}^{(d_K)}$ at some desired fixed formal order tuned against $\varepsilon$ \cite{Deshpande:2019uf}. One can also impose constraints that fix the values of constants that appear in the truncation error $\varepsilon_T$ directly which is of use during domain-decomposition (\S\ref{ssec:domdec_drp} and \S\ref{ssec:domdec_iter}).

\subsection{Characterizing and tuning derivative spectral error}
\label{ssec:sperr}

In addition to imposing control on formal order of accuracy in this work we are also concerned with reducing \textit{spectral error} which can be characterized based on Fourier analysis \cite{Lele:1992cd,Deshpande:2019uf,leveque2007finitedifferencemethods}. This will furnish us with a functional that can be directly optimized numerically for the stencil coefficients subject to the linear constraints previously discussed.

Consider $\Omega=\mathbb{T}(:=\mathbb{R}/2\pi\mathbb{Z})$ where the (periodic) interval $[0,\,2\pi)$ serves as a model for $\mathbb{T}$. Recall that family of plane-waves $(P_m)_{m\in\mathbb{Z}}$ constitute an orthonormal basis for $L^2(\mathbb{T})$ \cite{hesthaven2007spectralmethodstimedependent}. 
Suppose that $g\in L^2(\mathbb{T})$ is smooth and consider the truncated expansion:
\begin{equation}\label{eq:truncSum}
  g_N(x):=\sum_{|n|\leq N/2} \hat{g}_n \exp(i nx).
\end{equation}
Derivative approximants may be constructed utilizing Eq.\eqref{eq:truncSum} i.e.~by evaluating $\partial{}_x^{d}[g_N(x)]$
\begin{equation}\label{eq:truncDerivSum}
  \partial_x^{d}[g_N(x)]
  = \sum_{|n|\leq N/2}
  (i n)^{d} \hat{g}_n \exp(i n x)
  = \sum_{|n|\leq N/2} \hat{g}_{\hpb{}n}^{(d)} \exp(i n x),
\end{equation}
where for later convenience we have defined $\hat{g}_{\hpb{}n}^{(d)}:=(i n)^{d} \hat{g}_n$. An important property is that for $g{}^{(d)}$ and its truncated expansion we have
$\lim_{N\rightarrow\infty} \left\Vert %
g^{(d)}(x) - g^{(d)}_N(x) %
\right\Vert_2 = 0$ \cite{hesthaven2007spectralmethodstimedependent}. Suppose now that $x\in[0,\,2\pi)$ is discretized through introduction of a sequence of samples $(x_j)_{j=0}^{N-1}$ of uniform spacing $\delta x=2\pi / N$. The discrete analogue of Eq.\eqref{eq:truncSum}:
\begin{equation}\label{eq:truncSumD}
  g_{N,k}:=\sum_{|n|\leq N/2} \hat{g}_n \exp(i k n \delta x),
\end{equation}
shares similar approximation properties to the continuous expansion for smooth functions provided that $N$ is selected large enough such that aliasing error is suppressed \cite{hesthaven2007spectralmethodstimedependent,boyd2001chebyshevfourierspectral}. With discrete orthogonality focus may be restricted to a single mode where $h(x):=\exp(i n x)$:
\begin{equation}
  \label{eq:shiftingprop}
  h^{(d_K)}_{\hpb k + m} = \left.
    \partial^{d_K}_x[h(x)]
  \right|_{x=(k+m)\delta x}
  = \left(i n\right)^{d_K} \exp(i m \eta) h^{(0)}_{\hpb k}
  = \frac{1}{\delta x ^{d_K}}\left(i \eta\right)^{d_K} \exp(i m \eta) h^{(0)}_{\hpb k}
  = \exp(i m \eta) h^{(d_K)}_{\hpb k},
\end{equation}
where we have defined the normalized wave number $\eta:=n\delta x$. Set:
\begin{equation}
  \label{eq:phase_tup}
  \mathbf{E}^{(d_K)} := \left(\exp(i m \eta)\right)_{m=-L^{(d_K)}}^{R^{(d_K)}}.
\end{equation}
The selected form of $h$ together with Eq.\eqref{eq:shiftingprop}, Eq.\eqref{eq:phase_tup} and Eq.\eqref{eq:fcnTupDefn} allows for Eq.\eqref{eq:mdAnsatz} to be rewritten as:
\begin{align}
  \label{eq:fdpwerrdef}
  h^{(d_r)}_{\hpb k} &= \frac{1}{\delta x^{d_r}} (i\tilde{\eta})^{d_r} h^{(0)}_{\hpb k}, &
  (i\tilde{\eta})^{d_r} &:=
  \left(\boldsymbol{\alpha}^{(d_r)}\cdot\mathbf{E}^{(d_r)} \right)^{-1}
  \sum_{I\in\mathcal{I}}
  (i\eta)^{d_I}
  \boldsymbol{\alpha}^{(d_I)}\cdot\mathbf{E}^{(d_I)}.
\end{align}
Thus Eq.\eqref{eq:fdpwerrdef} provides for a characterization of the spectral error associated with a finite-difference scheme as we may compare the modified, normalized wave number $\tilde{\eta}$ to the analytically expected $\eta$. To this end we introduce:
\begin{equation}
  \label{eq:sp_error}
  \tilde{e}(\eta) :=
  \sum_{I\in\mathcal{I}} (i \eta)^{d_I} \mathbf{E}^{(d_I)}\cdot \boldsymbol{\alpha}^{(d_I)}
  - (i\eta)^{d_r} \mathbf{E}^{(d_r)}\cdot \boldsymbol{\alpha}^{(d_r)}.
\end{equation}
Which allows for consideration of an optimization problem based on the functional:
\begin{equation}
  \label{eq:optfunc}
  \varepsilon[\tilde{e};\,\gamma] := \int_0^\pi
  \gamma(\eta) \tilde{e}(\eta) \tilde{e}^*(\eta) \,\d \eta,
\end{equation}
where $\gamma(\eta)\geq 0$ is a non-negative weight function that allows to preferentially tune over prescribed ranges of $\eta$. The choices of Eq.\eqref{eq:sp_error} and Eq.\eqref{eq:optfunc} are motivated by the requirement of a convex optimization problem (see discussion in \cite{Deshpande:2019uf} for $|\mathcal{I}|=1$ and prescribed values $\mathbf{f}^{(0)}_{\hpb k}$). In order to construct $\boldsymbol{\alpha}{}^{(d_K)}$ when $\mathcal{C}_S>\mathcal{C}_T+1$ we solve (numerically) $\min_{\boldsymbol{\alpha}{}^{(d_K)}}\varepsilon[\tilde{e};\,\gamma]$ subject to the normalization condition $\alpha{}^{(d_r)}_{\hpb 0} = 1$ and linear constraints arising from Taylor series matching 
using \textsc{IPOPT} \cite{wachter2006implementationinteriorpointfilter}. To achieve this we refactor the functional $\varepsilon[\tilde{e};\,\gamma]$ such that quadrature may be performed numerically so as to yield an objective function involving only the sought after $\boldsymbol{\alpha}^{(d_K)}$. In brief, some choice of $d_K$, $L^{(d_K)}$, and $R^{(d_K)}$ is made. Then we set:
\begin{equation}
  \label{eq:Wdefn}
  W := \max_{\tilde{I}\in \mathcal{I}\cup\{r\}} \left\{
    L^{(d_{\tilde{I}})},\,R^{(d_{\tilde{I}})}
  \right\}
\end{equation}
and take $L^{(d_{\tilde{I}})}=R^{(d_{\tilde{I}})}=W$. The enlarged stencil coefficients we denote $\tilde{\boldsymbol{\alpha}}$. Next we construct $\tilde{\varepsilon}[\tilde{e};\,\gamma]$ with respect to $\tilde{\boldsymbol{\alpha}}$ in factored form:
\begin{align}
  \label{eq:optfuncfac}
  \tilde{\varepsilon}[\tilde{e};\,\gamma] &=
  \tilde{\mathbf{A}}
  \cdot
  \left\{
    \int_0^\pi
    \gamma(\eta)
    \boldsymbol{\epsilon}
      \left[
        \tilde{\mathbf{E}}^{(d_0)},\,\ldots,\,
        \tilde{\mathbf{E}}^{(d_{|\mathcal{I}|-1})},\,
        \tilde{\mathbf{E}}^{(d_r)};\, \eta
      \right]
    \,\d\eta
  \right\}
  \cdot
  \tilde{\mathbf{A}}^T, &
  \tilde{\mathbf{A}} &:=
  \begin{bmatrix}
    \tilde{\boldsymbol{\alpha}}^{(d_0)} & \cdots &
    \tilde{\boldsymbol{\alpha}}^{(d_{|\mathcal{I}|-1})} & \tilde{\boldsymbol{\alpha}}^{(d_r)}
  \end{bmatrix};
\end{align}
where $\tilde{\mathbf{A}}$ is a row-vector assembled from the unknown $\tilde{\boldsymbol{\alpha}}$ of size $(|\mathcal{I}|+1)(2W+1)$ and $\boldsymbol{\epsilon}$ is a square matrix formed from the enlarged $\tilde{\mathbf{E}}^{(d_K)}$ to be integrated element-wise. We use CAS to automate this factorization process. To seek a solution based on the initial target stencil sizes further auxiliary linear constraints are imposed:
\begin{equation}
  \label{eq:Zcons}
  \begin{aligned}
    0 &= \tilde{\alpha}{}^{(d_{\tilde{I}})}_{\hpb -n}\quad (n > L{}^{d_{\tilde{I}}}),
    &
    0 &= \tilde{\alpha}{}^{(d_{\tilde{I}})}_{\hpb n}\quad (n > R{}^{d_{\tilde{I}}}),
    &
    (\tilde{I}\in\mathcal{I}\cup\{r\}).
  \end{aligned}
\end{equation}
Working as above allows for rapid derivation and experimentation with a variety of schemes -- we have prepared a public notebook \cite{notebook_repo} that treats construction of the Taylor constraints, associated error functional construction and factorization, numerical quadrature evaluation, and solution for the resulting scheme coefficients at arbitrary precision.

\subsection{Dispersion and dissipation in wave propagation}
\label{ssec:phacc}
Our intention is to apply tuned, derivative approximant schemes to wave propagation problems as described by discretized hyperbolic partial differential equations and consequently we briefly recall dispersion and dissipation properties. This will allow for providing an assessment on potential phase accuracy achieved. For in-depth treatments of this together with stability and convergence properties see \cite{leveque2007finitedifferencemethods,gustafsson2013timedependentproblemsdifference}. For simplicity we focus on the linear advection equation as it is sufficient to fix conventions and illustrate concepts required later. To this end define $\Omega_T:=[0,\,T]\times \Omega$ with $\Omega$ as in \S\ref{ssec:sperr} and consider smooth $U:\Omega_T \rightarrow \mathbb{C}$ satisfying:
\begin{equation}\label{eq:adv1prot}
  \begin{cases}
    \partial_t[U] = -c_x \partial_x[U], & \Omega_T;\\
    U(t,\,0) = U(t,2\pi), & t\in[0,\,T];\\
    U(0,\,x) = u(x), & \{t=0\}\times \Omega;
  \end{cases}
\end{equation}
with $c_x>0$ or $c_x<0$ respectively representing a constant speed right-ward or left-ward propagation of the initial condition (IC) $u$. Due to linearity it is sufficient to focus on a single Fourier mode such as $u=\exp(i k x)$ which in turn leads to the time-dependent, translated profile $U(t,\,x)=\exp(ik(x-c_x t))$. In order to pass to the semi-discrete setting consider the equidistant discretization $x_k:=k\delta x=2\pi k/N$ where $k\in\{0,\,\dots,\, N-1\}$.
From Eq.\eqref{eq:adv1prot} the usual method of lines description now follows:
\begin{equation}\label{eq:adv1prot_disc}
  \begin{cases}
    \frac{\mathrm{d} \tilde{U}_k}{\mathrm{d} t}
    = -c_x \sum^{N-1}_{l=0} D{}_{kl} \tilde{U}_l, & t\in[0,\,T];\\
    \tilde{U}_k(0) = u(x_k), & \{t=0\};
  \end{cases}
\end{equation}
where the components $D{}_{kl}$ of $\mathbf{D}\in \mathbb{R}^{N\times N}$ represent values of a discrete derivative stencil in matricial form based on Eq.\eqref{eq:mdAnsatz} that embeds the periodic boundary conditions. Formal solution of Eq.\eqref{eq:adv1prot_disc} can be immediately provided through direct matrix exponentiation $\tilde{\mathbf{U}}=\exp(-t c_x \mathbf{D}) \mathbf{u}$. Whether the amplitude of a mode remains bounded under this prescription as is expected analytically depends sensitively on the spectrum of $\mathbf{D}$. To see this directly one can work in the modal representation which for an initial condition comprised of a fixed Fourier mode becomes:
\begin{equation}\label{eq:ffmadv}
  \begin{cases}
    \frac{\mathrm{d} \hat{U}_k}{\mathrm{d} t}
    = -i c_x k\frac{\tilde{\eta}}{\eta} \hat{U}_k, & t\in[0,\,T];\\
    \hat{U}_k(0) = 1, & \{t=0\};
  \end{cases}
\end{equation}
with the discrete derivative $\mathbf{D}$ appearing in Eq.\eqref{eq:adv1prot_disc} now described through the associated modified wavenumber $\tilde{\eta}$ of the selected scheme \cite{Deshpande:2019uf}. It follows that $\tilde{U}(t,\,x)=\exp(ik(x-\tilde{c}_x(k)t))$ where $\tilde{c}_x:=k c_x \tilde{\eta}/\eta$. Comparing to the analytical $U(t,\,x)$ we see that $\Re[\tilde{\eta}]$ modifies the speed of propagation i.e. the dispersion whereas non-zero $\Im[\tilde{\eta}]$ introduces amplitude attentuation (dissipation) or amplification depending on sign.

Following \cite{hesthaven2007spectralmethodstimedependent} a characterization of the number of points required to achieve an error tolerance for the phase associated with a given Fourier mode in solution of Eq.\eqref{eq:adv1prot} may now be provided. If it is the case that $\Im[\tilde{\eta}]=0$ for a selected spatial derivative approximant then there is no difference in the amplitude of the two solutions $U(t,\,x)$ and $\tilde{U}(t,\,x)$ at the semi-discrete level. Indeed we take the phase error $\varepsilon_\phi$ as the leading contribution in the relative error:
\begin{equation}
  \label{eq:phase_err_motiv}
  \left|
    \frac{U(t,\,x) - \tilde{U}(t,\,x)}{U(t,\,x)}
  \right|
  =
  \left|
    1 - \exp(ik(c_x - \tilde{c}_x)t)
  \right|
  \simeq
  \left|
    k(c_x - \tilde{c}_x)t
  \right|=:\varepsilon_\phi.
\end{equation}
If we introduce the relative error of the modified, normalized wavenumber as:
\begin{equation}\label{eq:relmodnormwavnum}
  \varepsilon_{\tilde{\eta}}(\eta) := \tilde{\eta}^{d_r}/\eta^{d_r} - 1,
\end{equation}
and we define the number of periods (in time) the solution has propagated as $N_T:=k|c_x|t/(2\pi)$ then Eq.\eqref{eq:phase_err_motiv} becomes:
\begin{equation}\label{eq:err_phi}
  \varepsilon_\phi = 2\pi N_T \left| \varepsilon_{\tilde{\eta}}(\eta) \right|.
\end{equation}
Given an IC $u=\exp(ikx)$ we have $k$ waves in the domain $\Omega$ and consequently the number of points per wavelength is $N_\nu:=N/k=2\pi/(k\delta x)=2\pi/\eta$. In order to resolve $u$ without aliasing we require that $N_\nu\geq 2$. Of interest is the number of samples required per wavelength to attain a specified phase error $\varepsilon_\phi$. For a scheme with formal order of accuracy $\mathcal{O}(\delta x^p)$ we may expand $\varepsilon_{\tilde{\eta}}(\eta) \simeq C_p \eta^p$ and consequently:
\begin{equation}\label{eq:adv_ppw_estimate}
  N_\nu \geq 2\pi \left(
    \frac{2\pi N_T |C_p|}{\varepsilon_\phi}
  \right)^{1/p},
\end{equation}
provides an estimate on the number of points required per wavelength required satisfy a phase error tolerance of $\varepsilon_\phi$.

In practice numerical solutions of systems such as Eq.\eqref{eq:adv1prot} are often constructed based on time-marching schemes involving discretization in time. Explicit Runge-Kutta (ERK) methods are the most widely used example of this. Linear stability and accuracy for an s-stage ERK method is characterized completely by the so-called stability polynomial $R$ defined by the method \cite{ketcheson2012optimalstabilitypolynomials,butcher2008numerical,hairer2010solving}. Given a constant-coefficient linear system $\partial_t[U](t,\,x)=\mathcal{L}[U](t,\,x)$ subject to initial conditions, full discretization can be reduced to an iteration involving $R$. We have that $\tilde{U}^{n+1} = R(\delta t L) \tilde{U}^n$ where $\delta t$ is a time-step, $\tilde{U}^n$ is an approximation to $U(n\delta t,\,x_k)$, and $L$ is a discretized approximation to $\mathcal{L}$ \cite{ketcheson2012optimalstabilitypolynomials}. Based on the iteration it is clear that the propagation of errors is controlled by $\Vert R(\delta t L)\Vert$. For ERK methods $R(z)=1$ defines a closed curve in the complex plane the interior of which forms the so-called absolute stability region $\mathcal{S}:=\{z\in\mathbb{C}\,:\,|R(z)\leq 1\}$. The iteration in time is considered absolutely stable if $\delta t \lambda \in \mathcal{S}$ for all eigenvalues $\lambda \in \mathrm{spec}(L)$. Tuning $\delta t$ to be sufficiently small thus implies $\tilde{U}^{n}$ remains bounded under iteration and forms a necessary\footnote{This is also a sufficient condition for stability if $L$ is a normal matrix \cite{hesthaven2007spectralmethodstimedependent}. For the more general case see e.g.~\cite{ketcheson2012optimalstabilitypolynomials} and references therein.} condition for stable propagation of errors. Typically in numerical solution of a fully-discretized system the Courant-Friedrich-Lewy (CFL) number $\mathcal{C}_{\mathrm{d}} := |c_x| \delta t / \delta x^{\mathrm{d}}\leq \mathcal{C}_{\mathrm{max}}$ is taken as one necessary criterion for stability, where $\mathcal{C}_{\mathrm{max}}$ depends on the details of $L$ and the ERK scheme applied.

\subsection{Comparison of select schemes}
\label{ssec:scmp}
The purpose of this section is to illustrate spectral properties of derivative approximant schemes that arise upon full Taylor matching in Eq.\eqref{eq:mdAnsatz} and furthermore tuning based on exploiting underdeterminedness as described in \S\ref{ssec:sperr}. For the latter we make use of a cutoff in Eq.\eqref{eq:optfunc}:
\begin{equation}\label{eq:gammacut}
  \begin{aligned}
    \gamma(\eta;\,\eta_c) &:=
    \begin{cases}
      1 & \eta\in[0,\,\eta_c],\\
      0 & \eta>\eta_c;
    \end{cases}
  \end{aligned}
\end{equation}
which is motivated by seeking to reduce spectral error at low to moderate wavenumber. Scheme stencils we have generated that are later utilized for wave-propagation problems are summarized in \S\ref{app:sch_tuned_coeff}. A general scheme is here denoted ${}^{(d_r)}S{}^{L^{(d_r)},\,R{}^{(d_r)}}_{L^{(d_0)},\,R{}^{(d_0)};\cdots}[d_I=(d_0,\,\dots),\,\eta_c]$ where indices up correspond to the left-hand-side of Eq.\eqref{eq:mdAnsatz}. Without ambiguity we can condense the notation such that for centered explicit schemes we write ${}^{(d_r)}E{}_{M}$. In the case of explicit, one point lop-siding we write ${}^{(1)}L{}_{M\pm1,M\mp1}$. The classical (fully order-matched) Pad\'{e} \cite{Lele:1992cd} we denote with ${}^{(d_r)}P{}^1_{M}$. When tuning based on Eq.\eqref{eq:gammacut} is performed $\eta_c$ is explicitly indicated. Thus ${}^{(d_r)}Q{}^1_{M}[\eta_c=1]$ takes $L^{(d_r)}=R^{(d_r)}=1$, $L^{(0)}=R^{(0)}=M$, and $\eta_c=1$ in Eq.\eqref{eq:gammacut}. Hermite schemes involve multiple derivatives which are indicated explicitly.

\begin{figure}[htbp]
  \subfloat[\label{subfig:eta_schemes_a}]{%
    \includegraphics[width=0.49\textwidth]{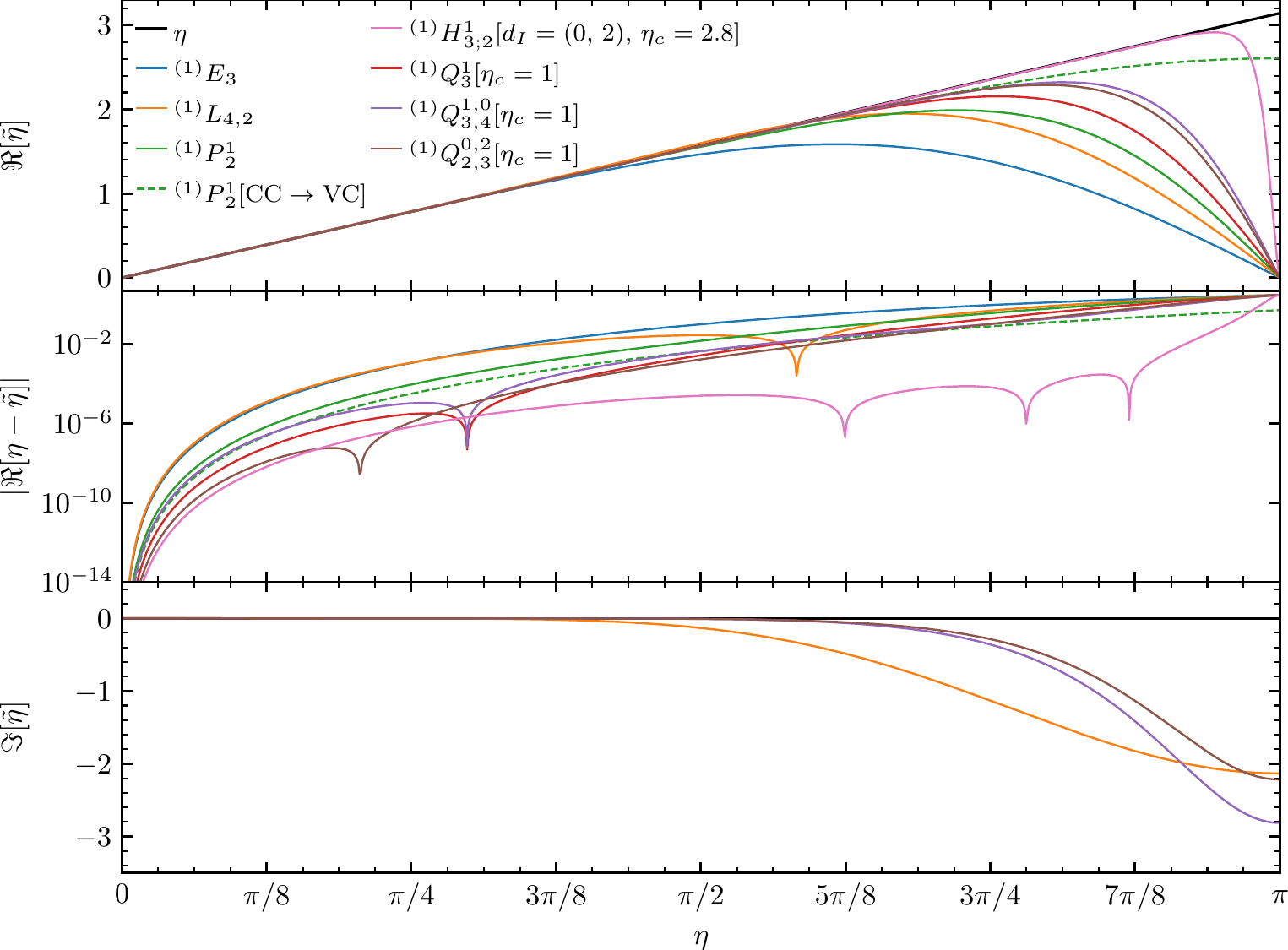}%
  }\hfill
  \subfloat[\label{subfig:eta_schemes_b}]{%
    \includegraphics[width=0.49\textwidth]{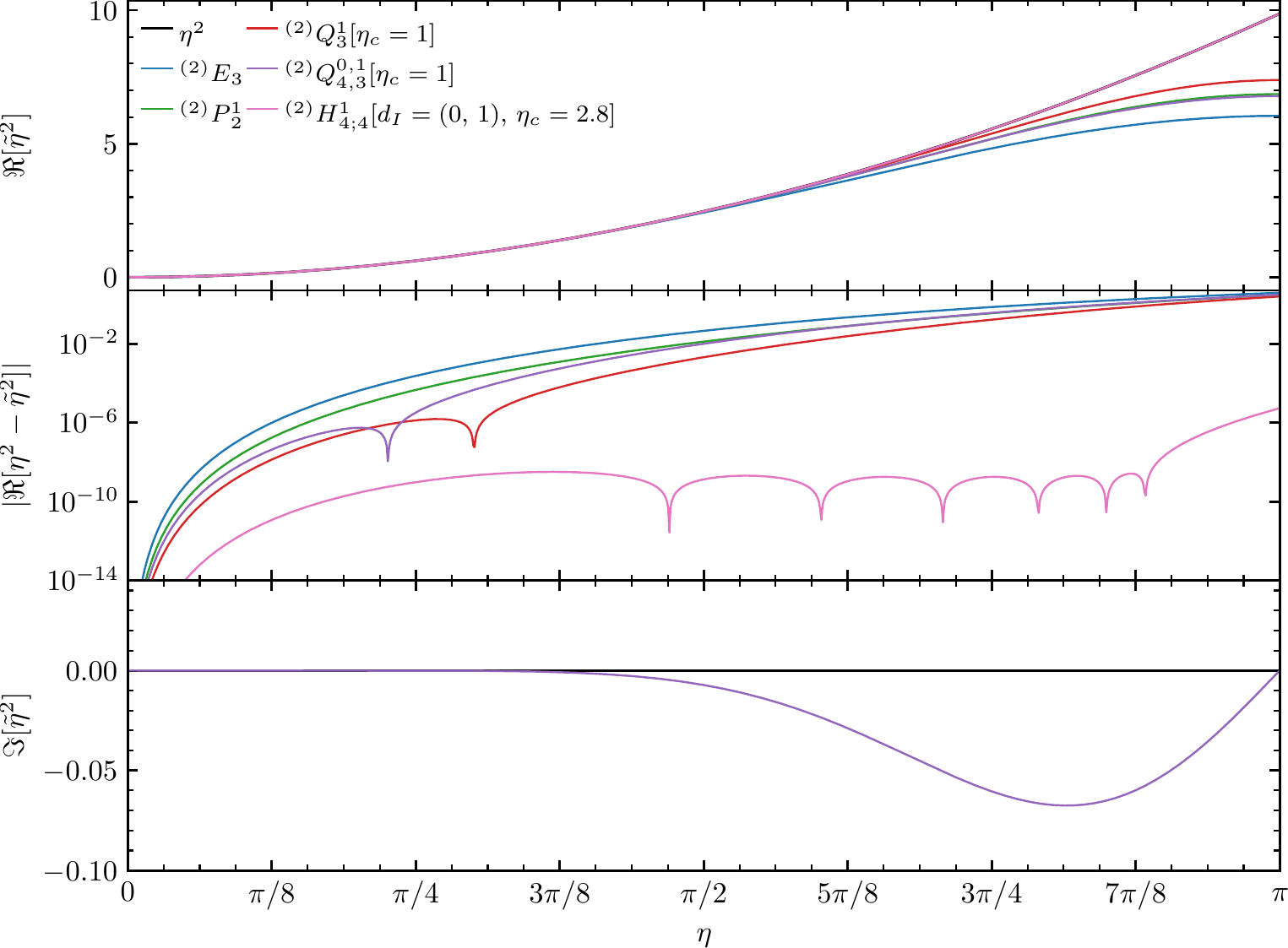}%
  }\hfill
  \caption{The modified wavenumber $\tilde{\eta}$ of various sixth-order finite-difference derivative approximants constructed with our proposed stencil generation approach. In (a) and (b) derivatives of degree one and two respectively are considered. Middle panels show the dispersion properties on a logarithmic scale; it can be seen that implicit schemes generically out-perform explicit. Bottom panels show that in wave-propagation problems biased schemes lead to non-zero mode amplification (attenuation) as $\eta\rightarrow\pi$ whereas centered schemes do not (see \S\ref{ssec:phacc}). See text for further discussion.}\label{fig:eta_schemes}
\end{figure}

Using our CAS notebook \cite{notebook_repo} we generate a variety of example stencils satisfying a formal order of accuracy of $\mathcal{O}(\delta x^6)$. In order to assess spectral properties we plot the real and imaginary parts of the modified wavenumber $\tilde{\eta}$ as defined by Eq.\eqref{eq:fdpwerrdef} over $\eta\in[0,\,\pi]$ in Fig.\ref{fig:eta_schemes}. These plots may be extended to $\eta\in[-\pi,0)$ through the parity conditions $\Re[\tilde{\eta}^d](\eta)=(-1)^d\Re[\tilde{\eta}^d](-\eta)$ whereas $\Im[\tilde{\eta}^d](\eta)=-(-1)^d\Im[\tilde{\eta}^d](-\eta)$.

As can be seen from Fig.\ref{subfig:eta_schemes_a} the standard explicit scheme ${}^{(1)}E{}_3$ has a more accurate dispersion than ${}^{(1)}L_{4,2}$ at low $\eta$, however the situation reverses as $\eta$ is increased. This is consistent with \cite{chirvasa2010finitedifferencemethods} which analyzes the use of lop-sided stencils when treating advective terms. We also observe such an advantage in numerical experiments based on setup of the aforementioned work performed in \S\ref{ssec:shifted_wave}. On the other hand it is clear that implicit schemes have better dispersion properties than their explicit counterparts at equivalent formal order of accuracy as can be seen comparing ${}^{(d_r)}E_3$ and ${}^{(d_r)}P{}^1_2$ in the middle panels of Fig.\ref{fig:eta_schemes}. Enlarging the number of function samples coupled by a stencil and leveraging underdeterminedness as described in \S\ref{ssec:sperr} allows us to construct ${}^{(d_r)}Q{}^1_3[\eta_c=1]$ which further improves over ${}^{(d_r)}P{}^1_2$. In \cite{Deshpande:2019uf} it was observed that increasing the implicit bandwidth further is also advantageous which we also observe based on e.g. ${}^{(1)}Q{}^{0,2}_{2,3}[\eta_c=1]$. If there is freedom to incorporate function derivative data we find that the implicit Hermite schemes outperform all other choices. Unfortunately a scheme such as ${}^{(1)}H{}^1_{3;2}[d_I=(0,\,2),\,\eta_c=2.8]$ requires knowledge of the second derivative of a function to compute the first which may not be available without reformulation of a given problem of interest \cite{Fornberg:2020ch}. It has been observed in \cite{Lele:1992cd} that when first degree derivatives are required on a staggered grid a stencil may be devised that does not suffer $\tilde{\eta}\rightarrow 0$ as $\eta\rightarrow \pi$. We verify this directly by generating ${}^{(1)}P{}^1_2[\mathrm{CC}\rightarrow\mathrm{VC}]$ where CC and VC have the meaning of Eq.\eqref{eq:gridDisc}.

To provide another comparison of first degree schemes we consider the semi-discretized form of the one-dimensional advection equation (see \S\ref{ssec:phacc}) and the number of points required to resolve a mode at a fixed tolerance on phase error in Tab.\ref{tab:phi_estimate}.
\begin{table}[htbp]
  \centering
  \begin{tabular}{c|l|l|l|l}
    \hline
    Scheme &
    $|C_6|$ &
    $N_\nu(\varepsilon_\phi=10^{-1})$ &
    $N_\nu(\varepsilon_\phi=10^{-2})$ &
    $N_\nu(\varepsilon_\phi=10^{-6})$ \\
    \hline
    \hline
    ${}^{(1)}E{}_{2}$ &
    $1/140$ &
    $18$ &
    $26$ &
    $119$ \\
    ${}^{(1)}L{}_{4,2}$ &
    $1/105$ &
    $19$ &
    $27$ &
    $125$ \\
    ${}^{(1)}P{}^1_{2}$ &
    $1/2100$ &
    $12$ &
    $17$ &
    $76$ \\
    ${}^{(1)}Q{}^1_{3}[\eta_c=1]$ &
    $5.3\times10^{-5}$ &
    $8$ &
    $12$ &
    $53$ \\
    ${}^{(1)}Q{}^{1,0}_{3,4}[\eta_c=1]$ &
    $2.4\times10^{-4}$ &
    $10$ &
    $15$ &
    $68$ \\
    ${}^{(1)}H{}^1_{3;2}[d_I=(0,\,2),\,\eta_c=1]$ &
    $4.9\times10^{-6}$ &
    $6$ &
    $8$ &
    $36$ \\
    \hline
  \end{tabular}
 \caption{Number of points required per wavelength $N_\nu$ for resolving a mode under one dimensional advection with fixed phase error tolerance $\varepsilon_\phi$ for a variety of $\mathcal{O}(\delta x^6)$ schemes. Number of crossing-times is $N_T=1000$. The estimate based on Eq.\eqref{eq:adv_ppw_estimate} shows that ${}^{(1)}Q{}^1_{3}[\eta_c=1]$ requires approximately half the number of points of standard explicit finite-difference ${}^{(1)}E{}_{2}$ in order to achieve the same $\varepsilon_\phi$. The error constants $C_6$ are computed from Taylor expansion in $\eta$ of the relative error of the modified, normalized wavenumber $\varepsilon_{\tilde{\eta}}(\eta)$ in Eq.\eqref{eq:relmodnormwavnum}.}
 \label{tab:phi_estimate}
\end{table}
%

\subsection{Domain decomposition: dispersion relation preservation}
\label{ssec:domdec_drp}
For large-scale problems a high degree of parallelism is required and the implicit nature of compact stencils may seem as a potential obstruction to achieve this. In this work we consider a strategy based on domain-decomposition inspired by \cite{chen2021novelparallelcomputing}. Suppose we have a decomposed and discretized domain $\Omega:=\sqcup_I \Omega_I$. Instead of solving a global, implicit system sequentially where data is communicated between neighboring $\Omega_I$ and $\Omega_J$ in succession a specification of a tuned complementary scheme on $\partial \Omega_I$ is made such that decoupled calculations may be performed in parallel that are local to each $\Omega_I$.

\begin{figure}[htbp]
  \includegraphics[width=0.6\textwidth]{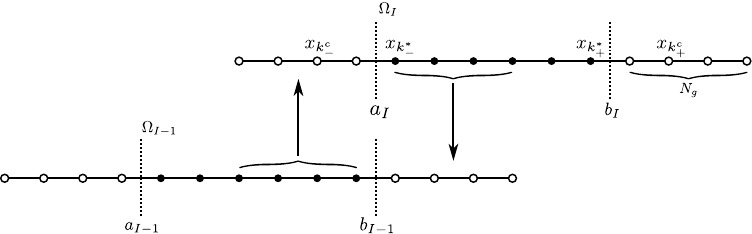}%
  \hfill
  \caption{Sampling for a sub-domain $\Omega_{I}$ and left nearest-neighbour $\Omega_{I-1}$. Cell-centered sampling $\mathcal{G}_{\mathrm{CC}}$ of Eq.\eqref{eq:gridDisc} is depicted. Indicated in shaded black circles are nodes interior to $\Omega_{I-1}$ and $\Omega_I$ where $N_B=6$. Vertical dashed lines delimit sub-domain extent. Nodes to the exterior (open circles) are ghosts filled through nearest-neighbour communication -- indicated with arrows. Here we have selected the number of ghosts to be $N_g=4$. Also indicated are the extremal interior nodes $k^*_\pm$. For later convenience (\S\ref{ssec:domdec_iter}) ghost zone closure nodes $x_{k^c_\pm}$ with $k^c_\pm=k^*_\pm \pm 2$ are also depicted. In the case of vertex-centered sampling we perform an additional averaging procedure on extremal, interior nodes as discussed in \cite{Daszuta:2021ecf}.  }
  \label{fig:stencil_drp_sketch}
\end{figure}

To illustrate this consider a one-dimensional $\Omega$ that has been discretized with uniformly spaced samples. Suppose that $\partial \Omega$ is periodically identified. A parameter $N_M$ is selected which controls the number of samples globally. Under domain-decomposition we fix $N_B$ on each $\Omega_{I}$ such that $N_B$ exactly divides $N_M$. For $\Omega_I=[a_I,\,b_I]$ we can consider cell-centered (CC) and vertex-centered (VC) sampling that are respectively defined through:
\begin{equation}
\label{eq:gridDisc}
\begin{aligned}
  \mathcal{G}_{\mathrm{CC}}[\Omega_I] &:=
  \Bigm\{
    a_I + \left(k + \frac{1}{2}\right) \frac{b_I-a_I}{N_B} \Bigm | k \in \{0,\,\ldots,\, N_B - 1\}
  \Bigm\},\\
  \mathcal{G}_{\mathrm{VC}}[\Omega_I] &:=
  \Bigm\{
    a_I + k \frac{b_I-a_I}{N_B} \Bigm | k \in \{0,\,\ldots,\, N_B\}
  \Bigm\};
\end{aligned}
\end{equation}
and the grid spacing we denote by $\delta x$. In the case of $\mathcal{G}_{\mathrm{VC}}$ the sub-domain $\Omega_I$ shares points with its nearest neighbours $\Omega_{I\pm 1}$. To the exterior of $\Omega_{I}$ a thin layer of ghost nodes $N_g$ with the same uniform spacing are appended (where $N_g<N_B$). This facilitates communication of data between sub-domains of the decomposition and evaluation of e.g. centered stencils for all $I$ in $\mathcal{G}[\Omega_I]$. We illustrate schematically in Fig.\ref{fig:stencil_drp_sketch} the grid structure together with boundary communication strategy.

Consider now evaluation of a centered derivative approximant for a scheme that implicitly couples nearest-neighbor nodes such as ${}^{(d_r)}C^1_M$. Such a scheme provides the approximation $\tilde{f}{}^{(d_r)}_{\hpb k}$ at a point $x_k\in\Omega:=\sqcup_I \Omega_I$ by coupling $\{\tilde{f}{}^{(d_r)}_{k-1},\,\tilde{f}{}^{(d_r)}_{\hpb k},\,\tilde{f}{}^{(d_r)}_{k+1}\}$ and relating to a linear combination of function samples $(f^{(0)}_{k+m})_{m=-M}^M$. In the case that we treat the domain globally then the derivative approximant under discussion may rewritten as a tridiagonal system which can be directly inverted utilizing the well-known Thomas Algorithm (TDMA) at an algorithmic complexity $\mathcal{O}(N_M)$ \cite{higham2002accuracy,golub2013matrix}. On the other hand when working with individual sub-domains from Fig.\ref{fig:stencil_drp_sketch} we see that a closure relation is required at the points with local indices $k^*_\pm$ if we are to apply TDMA with data locally available to $\Omega_I$.

Motivated by solution of wave-like propagation problems it is crucial to match both the formal order of accuracy of derivative approximants together with (approximate) preservation of the modified wavenumber \S\ref{ssec:phacc}. Given the relative error of the modified, normalized wavenumber $\varepsilon_{\tilde{\eta}}$ of Eq.\eqref{eq:relmodnormwavnum} recall that $\Re[\varepsilon_{\tilde{\eta}}]$ and $\Im[\varepsilon_{\tilde{\eta}}]$ are related to the dispersion and dissipation properties of a scheme respectively. For centered stencils such as ${}^{(d_r)}C^N_M$ the coefficient tuples $\boldsymbol{\alpha}{}^{(0)}$ are (skew)-symmetric about the central element for (odd)-even $d_r$ respectively \cite{Deshpande:2019uf}. It therefore follows from Eq.\eqref{eq:fdpwerrdef} that $\Im[\tilde{\eta}]=0$ and consequently $\Im[\varepsilon_{\tilde{\eta}}]=0$. Suppose that ${}^{(d_r)}C^N_M$ is $\mathcal{O}(\delta x^{2p})$ then:
\begin{align}
  \label{eq:errrelmodwavcent}
  \varepsilon_{\tilde{\eta}}\left[{}^{(d_r)}C^N_M\right]
  &=
  C_0 \eta^{2p} +
  C_2 \eta^{2p+2} + C_4 \eta^{2p+4} + \cdots;
\end{align}
where only even powers of $\eta$ appear, and the constants $C_{2n}$ $(n\in\mathbb{N}_0)$ are real. On $\Omega_I$ as the nodes with indices $k^*_\pm$ are approached from the interior the approximant ${}^{(d_r)}C^N_M$ can eventually no longer be applied if we wish to treat the evaluation locally in a decoupled fashion using a banded linear solver. Instead we construct a (sequence of) tuned, centered scheme(s) with fixed formal order of accuracy $\mathcal{O}(\delta x^{2p})$ where the implicit bandwidth (i.e. $2N+1$) is reduced as $k^*_\pm$ is approached. These closing schemes are instead imposed on nodes where it is not possible to impose ${}^{(d_r)}C^N_M$. If $N$ is reduced but simultaneously $M$ may increase then there is sufficient freedom to match the lowest order error constants of the closing schemes to those appearing in Eq.\eqref{eq:errrelmodwavcent}. This procedure can be tuned to \textit{approximately} preserve dispersion. In particular for $N=1$ we can apply ${}^{(d_r)}C^1_M$ on $x_k$ with $k\in\{k^*_-+1,\,\dots,\, k^+_+-1\}$ and at $k{}^*_\pm$ impose explicit, centered closures.
For ${}^{(d_r)}C^1_M[\mathcal{O}(\delta x^{2p})]$ an explicit scheme ${}^{(d_r)}\overline{C}_{p}[\mathcal{O}(\delta x^{2p})]$ can be constructed to match the formal order of accuracy. In general the constants appearing in the expansion of $\varepsilon_{\tilde{\eta}}$ for each scheme will not match. Therefore we increase the explicit stencil size and fix the number of equations arising from inserting the Taylor approximant of Eq.\eqref{eq:Tcons} into Eq.\eqref{eq:mdAnsatz} which fixes the order. This however results in an underdetermined system for the scheme coefficients. Due to underdeterminedness we may impose additional constraints that tune $\varepsilon_{\tilde{\eta}}$. We have:
\begin{equation}
  \label{eq:errelmodwavecentmatch}
  \varepsilon_{\tilde{\eta}}\left[{}^{(d_r)}C^1_M[\mathcal{O}(\delta x^{2p})]\right] -
  \varepsilon_{\tilde{\eta}}\left[{}^{(d_r)}\overline{C}_{p+1}[\mathcal{O}(\delta x^{2p})]\right]
  =
  (C_0-\overline{C}_0) \eta^{2p} +
  (C_2-\overline{C}_2) \eta^{2p+2} +
  (C_4-\overline{C}_4) \eta^{2p+4} + \cdots;
\end{equation}
where the error constants are real and all odd powers of $\eta$ must vanish. Therefore to construct ${}^{(d_r)}\overline{C}_{p+1}[\mathcal{O}(\delta x^{2p})]$ we specify $C_T=2p-1$ Taylor conditions, a unit normalization condition on $\overline{\alpha}^{(d_r)}_{\hpb 0}$ of ${}^{(d_r)}\overline{C}_{p+1}$, and also impose:
\begin{align}
  C_0 &= \left.\partial_\eta^{2p}\left[
    \varepsilon_{\tilde{\eta}}\left[
      {}^{(d_r)}\overline{C}_{p+1}
    \right]
  \right]\right|_{\eta=0}, &
  0 &= \left.\partial_\eta^{2p+1}\left[
    \varepsilon_{\tilde{\eta}}\left[
      {}^{(d_r)}\overline{C}_{p+1}
    \right]
  \right]\right|_{\eta=0}.
\end{align}
This gives us $2(p+1)$ linear conditions on the $2(p+1)$ unknown $\overline{\boldsymbol{\alpha}}$ coefficients which may be solved for directly. Hence we find a composite scheme over $\Omega_I$ at consistent formal order of approximation together with consistent lowest order $\varepsilon_{\tilde{\eta}}$ expansion constants. This allows us to treat each $\Omega_I$ independently once the ghost nodes have been populated with known function data. Once this is the case we can solve the banded system:
\begin{equation}\label{eq:sch_cent_om}
  \begin{cases}
    \tilde{f}{}^{(d_r)}_{\hpb{} k}
    =
    \frac{1}{(\delta x)^{d_r}} \sum_{m=-(p+1)}^{p+1}
    \overline{\alpha}{}^{(0)}_{\hpb{} m}
    f{}^{(0)}_{\hpb{} k + m},
    &
    k = k^*_{\pm};\\
    \alpha{}^{(d_r)}_{-1} \tilde{f}{}^{(d_r)}_{k-1} +
    \alpha{}^{(d_r)}_{\hpb{} 0} \tilde{f}{}^{(d_r)}_{k} +
    \alpha{}^{(d_r)}_{\hpb{}1} \tilde{f}{}^{(d_r)}_{k+1}
    =
    \frac{1}{(\delta x)^{d_r}} \sum_{m=-M}^{M}
    \alpha{}^{(0)}_{\hpb{} m}
    f{}^{(0)}_{\hpb{} k + m},
    &
    k \in \{k^*_- + 1,\, k^*_- + 2,\, \dots,\, k^*_+ - 1\};
  \end{cases}
\end{equation}
where $\tilde{f}{}^{(d_r)}_{\hpb{} k^*_\pm}$ are fixed by the explicit closure ${}^{(d_r)}\overline{C}_{p+1}[\mathcal{O}(\delta x^{2p})]$ whereas the remaining $\left\{\tilde{f}{}^{(d_r)}_{\hpb{} k^*_-+1},\,\dots,\,\tilde{f}{}^{(d_r)}_{\hpb{} k^*_{+}-1}\right\}$ are to be solved for. For the present closure we require that $N_g\geq p+1$ on $\Omega_I$ where in Eq.\eqref{eq:sch_cent_om} data at node indices $k+i<k^*_-$ and $k+i>k^*_+$ is supplied through communication from neighbouring sub-domains (see Fig.\ref{fig:stencil_drp_sketch}). Summarized in Tab.\ref{tab:pade_deg_one} and Tab.\ref{tab:pade_deg_two} of the appendix \S\ref{app:sch_tuned_coeff} are coefficients for ${}^{(d_r)}P^{1}_{M}$ with matched, explicit closures ${}^{(d_r)}\overline{P}_{M+2}$ for $d_r\in\{1,2\}$ and $M\in\{1,\,2,\,3\}$. The numerically optimized, centered schemes ${}^{(d_r)}Q^1_3$ of \S\ref{ssec:scmp} may be treated in a similar fashion and are summarized in Tab.\ref{tab:num_deg_one} and Tab.\ref{tab:num_deg_two}.

We now turn our attention to degree one, biased schemes as it is possible to close stencils on sub-domain boundaries in a fashion that preserves the dispersion relation \textit{exactly} \cite{chen2021novelparallelcomputing}. We demonstrate directly that this can be done for higher order derivative approximants by extending the $4^{\mathrm{th}}$ order CCU$(4,5)$ scheme of \cite{qin2014highaccuracynumerical}.
To this end we particularize Eq.\eqref{eq:mdAnsatz} as:
\begin{equation}
  \label{eq:defChenSchemes}
  {}^{\pm}\alpha{}^{(1)}_{\pm1} \tilde{f}^{(1)}_{\hpb k\pm 1} +
  {}^{\pm}\alpha{}^{(1)}_{\hpb 0} \tilde{f}^{(1)}_{\hpb k}
  =
  \frac{1}{\delta x} \sum_{m=-M}^M
  {}^{\pm}\alpha{}^{(0)}_{\pm1} f^{(0)}_{\hpb k + m}
  + \delta x {}^{\pm}\alpha{}^{(2)}_{\hpb 0} f^{(2)}_{\hpb k},
\end{equation}
where we seek even-order schemes with $\mathcal{O}(\delta x^{2M+2})$ which will be denoted ${}^{(1)}U{}_M^\pm$. We will also impose Eq.\eqref{eq:defChenSchemes} with ${}^{\pm}\alpha{}^{(2)}_{\hpb 0} = 0$ to construct odd-order schemes ${}^{(1)}V{}_M^\pm$ with $\mathcal{O}(\delta x^{2M+1})$. We refer to schemes with stencils as in Eq.\eqref{eq:defChenSchemes} involving the term $\tilde{f}^{(1)}_{\hpb k- 1}$ as left biased and $\tilde{f}^{(1)}_{\hpb k+1}$ as right biased. Given a left biased scheme of the form of Eq.\eqref{eq:defChenSchemes} then the analogous right biased scheme is given by ${}^{+}\alpha{}^{(1)}_{\hpb m} = {}^{-}\alpha{}^{(1)}_{-m}$, ${}^{+}\alpha{}^{(0)}_{\hpb m} =-{}^{-}\alpha{}^{(0)}_{-m}$, and ${}^{+}\alpha{}^{(2)}_{\hpb 0} = -{}^{-}\alpha{}^{(2)}_{0}$ \cite{chen2021novelparallelcomputing,Deshpande:2019uf}.

In \cite{fu1997highorderaccurate,qin2014highaccuracynumerical} the so-called UCD-style schemes ${}^{(1)}U{}_1^\pm$ and ${}^{(1)}V{}_2^\pm$ have been presented however here we use the flexibility of our method in allowing for automatic generation and solution of Taylor constraints to easily extend to higher order approximants. Stencil coefficient tuples together with expansions of the modified wavenumber for a variety of orders are provided in Tab.\ref{tab:ucd_slow} and Tab.\ref{tab:ucd_fast}.
\begin{table}[htbp]
  \centering
  \begin{tabular}{c|c|lll|l}
    \hline
    Scheme &
    Order &
    ${}^{-}\boldsymbol{\alpha}^{(1)}$ &
    ${}^{-}\boldsymbol{\alpha}^{(0)}$ &
    ${}^{-}{\alpha}^{(2)}_{\hphantom{(}0}$ &
    $\varepsilon_{\tilde{\eta}}$ \\
    \hline
    \hline
    ${}^{(1)}U_{1}^{-}$ & 4 &
    $\left(\frac{1}{2},\, 1 \right)$ &
    $\frac{1}{4}\left(-7,\, 8,\, -1 \right)$ &
    $\frac{1}{2}$ &
    $-\frac{1}{180}\eta^4 - \frac{i}{1080}\eta^5 + \mathcal{O}(\eta^6)$ \\
    ${}^{(1)}U_{2}^{-}$ & 6 &
    $\left(\frac{2}{3},\, 1 \right)$ &
    $\frac{1}{72}\left(-3,\, -136,\, 162,\, -24,\, 1 \right)$ &
    $\frac{1}{2}$ &
    $-\frac{1}{2100}\eta^6 - \frac{11i}{84000}\eta^7 - \mathcal{O}(\eta^8)$ \\
    ${}^{(1)}U_{3}^{-}$ & 8 &
    $\left(\frac{3}{4},\, 1 \right)$ &
    $\frac{1}{25200}\left(70,\, -1890,\, -48825,\, 59500,\, -9450,\,630,\,-35 \right)$ &
    $\frac{1}{2}$ &
    $-\frac{1}{17640}\eta^8 - \frac{23i}{1234800}\eta^9 + \mathcal{O}(\eta^{10})$ \\
    ${}^{(1)}U_{4}^{-}$ & 10 &
    $\left(\frac{4}{5},\, 1 \right)$ &
    $\frac{1}{30}\left(-\frac{1}{112},\, \frac{4}{21},\, -3,\, -\frac{294}{5},\, \frac{1745}{24},\,-12,\,1,\,-\frac{2}{21},\,\frac{3}{560} \right)$ &
    $\frac{1}{2}$ &
    $-\frac{1}{124740}\eta^{10} - \frac{13i}{4490640}\eta^{11} + \mathcal{O}(\eta^{12})$ \\
    \hline
  \end{tabular}
 \caption{Coefficient tuples and associated relative error in the normalized wavenumber $\varepsilon_{\tilde{\eta}}$ for a variety of left biased UCD$(2M+2)$-style schemes we have derived. The analogous ${}^{(1)}U{}_M^{+}$ coefficients are given by ${}^+\alpha{}^{(0)}_{\hpb m} = - {}^-\alpha{}^{(0)}_{-m}$, ${}^+\alpha{}^{(1)}_{\hpb m} = {}^-\alpha{}^{(1)}_{-m}$, and ${}^+\alpha{}^{(2)}_{\hpb 0} = - {}^-\alpha{}^{(2)}_{0}$. The formal order of a given scheme is reflected in $\varepsilon_{\tilde{\eta}}$. Given $\varepsilon_{\tilde{\eta}}$ for ${}^{(1)}U{}_M^{-}$ that of ${}^{(1)}U{}_M^{+}$ may be found through complex conjugation as $\varepsilon_{\tilde{\eta}}\left[{}^{(1)}U{}_M^{+}\right]=\varepsilon_{\tilde{\eta}}\left[{}^{(1)}U{}_M^{-}\right]^*$. The scheme ${}^{(1)}U{}_1^{\pm}$ coincides with UCD$(4)$ of \cite{qin2014highaccuracynumerical}.}
 \label{tab:ucd_slow}
\end{table}
\begin{table}[htbp]
  \centering
  \begin{tabular}{c|c|ll|l}
    \hline
    Scheme &
    Order &
    ${}^{-}\boldsymbol{\alpha}^{(1)}$ &
    ${}^{-}\boldsymbol{\alpha}^{(0)}$ &
    $\varepsilon_{\tilde{\eta}}$ \\
    \hline
    \hline
    ${}^{(1)}V{}_1^{-}$ & 3 &
    $\left(\frac{1}{2},\, 1 \right)$ &
    $\frac{1}{4}\left(-5,\,4,\,1  \right)$ &
    $-\frac{i}{36}\eta^3 + \frac{1}{270}\eta^4 - \mathcal{O}(\eta^5)$ \\
    ${}^{(1)}V{}_2^{-}$ & 5 &
    $\left(\frac{2}{3},\, 1 \right)$ &
    $\frac{1}{36}\left(
      -3,\,-44,\,36,\,12,\,-1
    \right)$ &
    $-\frac{i}{300}\eta^5 + \frac{3}{3500}\eta^6 + \mathcal{O}(\eta^7)$ \\
    ${}^{(1)}V{}_3^{-}$ & 7 &
    $\left(\frac{3}{4},\, 1 \right)$ &
    $\frac{1}{240}\left(
      2,\,-36,\,-285,\,240,\,90,\,-12,\,1
    \right)$ &
    $-\frac{i}{1960}\eta^7 + \frac{1}{6174}\eta^8 + \mathcal{O}(\eta^9)$ \\
    ${}^{(1)}V{}_4^{-}$ & 9 &
    $\left(\frac{4}{5},\, 1 \right)$ &
    $\frac{1}{4200}\left(
      -5,\, 80,\, -840,\, -4872,\, 4200,\, 1680,\, -280,\, 40,\, -3
    \right)$ &
    $-\frac{i}{11340}\eta^9 + \frac{1}{32076}\eta^{10} + \mathcal{O}(\eta^{11})$ \\
    \hline
  \end{tabular}
 \caption{Coefficient tuples and associated relative error in the normalized wavenumber $\varepsilon_{\tilde{\eta}}$ for a variety of UCD$(2M+1)$-style schemes. The analogous ${}^{(1)}V{}_M^{+}$ coefficients are given by taking ${}^+\alpha{}^{(0)}_{\hpb m} = - {}^-\alpha{}^{(0)}_{-m}$ and ${}^+\alpha{}^{(1)}_{\hpb m} = {}^-\alpha{}^{(1)}_{-m}$ in the above. The formal order of a given scheme is reflected in $\varepsilon_{\tilde{\eta}}$. Given $\varepsilon_{\tilde{\eta}}$ for ${}^{(1)}V{}_M^{-}$ that of ${}^{(1)}V{}_M^{+}$ may be found through complex conjugation as $\varepsilon_{\tilde{\eta}}\left[{}^{(1)}V{}_M^{+}\right]=\varepsilon_{\tilde{\eta}}\left[{}^{(1)}V{}_M^{-}\right]^*$. The result of our derivation approach for the scheme ${}^{(1)}V{}_2^{\pm}$ gives an equivalent stencil to UCD$(5)$ of \cite{fu1997highorderaccurate}. }
 \label{tab:ucd_fast}
\end{table}
An interesting observation made in \cite{chen2021novelparallelcomputing} that is compatible with the $\varepsilon_{\tilde{\eta}}$ expansions we have provided and verified is that switching between left and right bias does not alter the dispersion for ${}^{(1)}U{}_M^\pm[\mathcal{O}(\delta x^{2M+2})]$ and ${}^{(1)}V{}_M^\pm[\mathcal{O}(\delta x^{2M+1})]$ schemes derived based on Eq.\eqref{eq:defChenSchemes} and Taylor matching. In particular we have:
\begin{align}
  \label{eq:ucd_slow_rel_err}
  \varepsilon_{\tilde{\eta}}[{}^{(1)}U{}_M^\pm]
  &= U_0 \eta^{2M+2} \pm i U_1 \eta^{2M+3} + U_2 \eta^{2M+4} \pm i U_3 \eta^{2M+5} + \cdots,\\
  \label{eq:ucd_fast_rel_err}
  \varepsilon_{\tilde{\eta}}[{}^{(1)}V{}_M^\pm]
  &= \pm i V_1 \eta^{2M+1} + V_2 \eta^{2M+2} \pm i V_3 \eta^{2M+3} + V_4 \eta^{2M+4} + \cdots;
\end{align}
where the error constants $U_j$ and $V_k$ are real. Due to the bias of these schemes closure is only required on a single point and therefore ${}^{(1)}U{}_M^{\pm}$ may be closed at $k^*_\pm$ utilizing ${}^{(1)}U{}_M^{\mp}$ and similarly for ${}^{(1)}V{}_M^{\pm}$ without modifying dispersion. However stability properties for a wave propagation problem depend on the dissipation. To modify this latter without affecting dispersion the closure for ${}^{(1)}U_M^\pm$ is instead based on a modification to ${}^{(1)}U{}_M^{\mp}$ where at $k^*_\pm$ one imposes a stencil of the form \cite{chen2021novelparallelcomputing}:
\begin{equation}
  \label{eq:ucd_slow_closure_chi}
  \delta x\,\boldsymbol{\alpha}^{(1)}[{}^{(1)}U_M^{\mp}]
  \cdot \tilde{\mathbf{f}}_{\hpb k^*_{\pm}}^{(1)}
  =
  \boldsymbol{\alpha}^{(0)}[{}^{(1)}U_M^{\mp}] \cdot \mathbf{f}_{\hpb k^*_{\pm}}^{(0)}
  +\delta x^2 \alpha{}^{(2)}_{\hpb 0}[{}^{(1)}U_M^{\mp}] f^{(2)}_{\hpb k^*_{\pm}}
  +\xi_{\pm} \delta x^{2M+3}
  \boldsymbol{\alpha}^{(1)}[{}^{(1)}U_M^{\mp}] \cdot \mathbf{f}_{\hpb k^*_{\pm}}^{(2M+2)},
\end{equation}
where the values of $\mathbf{f}_{\hpb k^*_{\pm}}^{(2M+2)}$ can be approximated using ${}^{(2M+2)}E{}_{M+1}$ of Tab.\ref{tab:diss_scheme} and once $\xi_{\pm}$ are determined we have closures ${}^{(1)}\overline{U}{}_M^{\pm}$ for the schemes ${}^{(1)}U{}_M^{\pm}$ to be applied at $k^*_\pm$. Incorporating the high-degree derivative as in Eq.\eqref{eq:ucd_slow_closure_chi} means that Eq.\eqref{eq:ucd_slow_rel_err} is modified to:
\begin{equation}
  \varepsilon_{\tilde{\eta}}[{}^{(1)}\overline{U}{}_M^\pm]
  =
  U_0 \eta^{2M+2} \mp i \overline{U}_1(\xi_{\pm}) \eta^{2M+3} + U_2 \eta^{2M+4} \mp i \overline{U}_3(\xi_{\pm}) \eta^{2M+5} + \cdots,
\end{equation}
where for $k\in\mathbb{N}_0$ the constants $U_{2k}$ are left unmodified whereas $\overline{U}_{2k+1}$ now depend on $\xi_\pm$. In order to determine $\xi_\pm$ we match the next to leading order error term:
\begin{equation}
  \left.
  \partial_{\eta}^{2M+3}\left[
    \varepsilon_{\tilde{\eta}}[{}^{(1)}U{}_M^\pm]
  \right]
  \right|_{\eta=0}
  =
  \left.\partial_{\eta}^{2M+3}\left[
    \varepsilon_{\tilde{\eta}}[{}^{(1)}\overline{U}{}_M^\pm]
  \right]\right|_{\eta=0},
\end{equation}
which gives rise to a linear relation for $\xi_\pm$. The result of this is summarized for the schemes of Tab.\ref{tab:ucd_slow} in Tab.\ref{tab:ucd_slow_closure}.
\begin{table}[htbp]
  \centering
  \begin{tabular}{c|l|l}
    \hline
    Scheme &
    $\xi_-$ &
    $\varepsilon_{\tilde{\eta}}$ \\
    \hline
    \hline
    ${}^{(1)}\overline{U}{}_1^{-}$ &
    $-1/540$ &
    $-\frac{1}{180}\eta^4 - \frac{i}{1080}\eta^5 - \mathcal{O}(\eta^6)$ \\
    ${}^{(1)}\overline{U}{}_2^{-}$ &
    $11/42000$ &
    $-\frac{1}{2100}\eta^6 - \frac{11i}{84000}\eta^7 - \mathcal{O}(\eta^8)$ \\
    ${}^{(1)}\overline{U}{}_3^{-}$ &
    $-23/617400$ &
    $-\frac{1}{17640}\eta^8 - \frac{23i}{1234800}\eta^9 + \mathcal{O}(\eta^{10})$ \\
    ${}^{(1)}\overline{U}{}_4^{-}$ &
    $13/2245320$ &
    $-\frac{1}{124740}\eta^{10} - \frac{13i}{4490640}\eta^{11} + \mathcal{O}(\eta^{12})$ \\
    \hline
  \end{tabular}
 \caption{Constants $\xi_-$ entering Eq.\eqref{eq:ucd_slow_closure_chi} for the closures ${}^{(1)}\overline{U}^-_M$ for schemes of Tab.\ref{tab:ucd_slow} induced through the stencils ${}^{(1)}U^+_M$ to be applied at $k^*_-$ as described in the text. The analogous coefficients for ${}^{(1)}\overline{U}^+_M$ closures are $\xi_+=-\xi_-$. As in Tab.\ref{tab:ucd_fast} given $\varepsilon_{\tilde{\eta}}$ for ${}^{(1)}\overline{U}{}_M^{-}$ that of of ${}^{(1)}\overline{U}{}_M^{+}$ may be found through complex conjugation as $\varepsilon_{\tilde{\eta}}\left[{}^{(1)}\overline{U}{}_M^{+}\right]=\varepsilon_{\tilde{\eta}}\left[{}^{(1)}\overline{U}{}_M^{-}\right]^*$. }
 \label{tab:ucd_slow_closure}
\end{table}
In the case of ${}^{(1)}V{}_M^{\pm}$ the closure is based on an analogous modification to ${}^{(1)}V{}_M^{\mp}$ where we consider:
\begin{equation}
  \label{eq:ucd_fast_closure_chi}
  \delta x\,\boldsymbol{\alpha}^{(1)}[{}^{(1)}V_M^{\mp}]
  \cdot \tilde{\mathbf{f}}_{\hpb k^*_{\pm}}^{(1)}
  =
  \boldsymbol{\alpha}^{(0)}[{}^{(1)}V_M^{\mp}] \cdot \mathbf{f}_{\hpb k^*_{\pm}}^{(0)}
  +\xi_- \delta x^{2M+2}
  \boldsymbol{\alpha}^{(1)}[{}^{(1)}V_M^{\mp}] \cdot \mathbf{f}_{\hpb k^*_{\pm}}^{(2M+2)},
\end{equation}
and $\mathbf{f}_{\hpb k^*}^{(2M+2)}$ may again be approximated with ${}^{(2M+2)}E{}_{M+1}$ of Tab.\ref{tab:diss_scheme}. Once $\xi_{\pm}$ is determined we have closures ${}^{(1)}\overline{V}{}_M^{\pm}$ for ${}^{(1)}V{}_M^{\pm}$ to be applied at $k^*_{\pm}$. For the closure specification of Eq.\eqref{eq:ucd_fast_closure_chi} it is the case that Eq.\eqref{eq:ucd_fast_rel_err} is modified to:
\begin{equation}
  \varepsilon_{\tilde{\eta}}[{}^{(1)}\overline{V}{}_M^\pm]
  =
  \pm i \overline{V}_1(\xi_{\pm}) \eta^{2M+1} + V_2 \eta^{2M+2} \pm i \overline{V}_3(\xi_{\pm}) \eta^{2M+3} + V_4 \eta^{2M+4} + \cdots;
\end{equation}
where for $k\in\mathbb{N}_0$ the constants $V_{2k}$ are left unmodified whereas $\overline{V}_{2k+1}$ now depend on $\xi_\pm$. In order to determine $\xi_\pm$ we match the leading order error term:
\begin{equation}
  \left.
  \partial_{\eta}^{2M+1}\left[
    \varepsilon_{\tilde{\eta}}[{}^{(1)}V{}_M^\pm]
  \right]
  \right|_{\eta=0}
  =
  \left.\partial_{\eta}^{2M+1}\left[
    \varepsilon_{\tilde{\eta}}[{}^{(1)}\overline{V}{}_M^\pm]
  \right]\right|_{\eta=0},
\end{equation}
The result of this is summarized for the schemes of Tab.\ref{tab:ucd_fast} in Tab.\ref{tab:ucd_fast_closure}.
\begin{table}[htbp]
  \centering
  \begin{tabular}{c|l|l}
    \hline
    Scheme &
    $\xi_-$ &
    $\varepsilon_{\tilde{\eta}}$ \\
    \hline
    \hline
    ${}^{(1)}\overline{V}{}_1^{-}$ &
    $1/18$ &
    $-\frac{i}{36}\eta^3 + \frac{1}{270}\eta^4 + \mathcal{O}(\eta^5)$ \\
    ${}^{(1)}\overline{V}{}_2^{-}$ &
    $-1/150$ &
    $-\frac{i}{300}\eta^5 + \frac{3}{3500}\eta^6 + \mathcal{O}(\eta^7)$ \\
    ${}^{(1)}\overline{V}{}_3^{-}$ &
    $1/980$ &
    $-\frac{i}{1960}\eta^7 + \frac{1}{6174}\eta^8 + \mathcal{O}(\eta^9)$ \\
    ${}^{(1)}\overline{V}{}_4^{-}$ &
    $-1/5670$ &
    $-\frac{i}{11340}\eta^9 + \frac{1}{32076}\eta^{10} + \mathcal{O}(\eta^{11})$ \\
    \hline
  \end{tabular}
 \caption{Constants $\xi_-$ entering Eq.\eqref{eq:ucd_fast_closure_chi} for the closures ${}^{(1)}\overline{V}^-_M$ for schemes of Tab.\ref{tab:ucd_fast} induced through the stencils ${}^{(1)}V^+_M$ to be applied at $k^*_-$ as described in the text. The analogous coefficients for ${}^{(1)}\overline{V}^+_M$ closures are $\xi_+=-\xi_-$. As in Tab.\ref{tab:ucd_fast} given $\varepsilon_{\tilde{\eta}}$ for ${}^{(1)}\overline{V}{}_M^{-}$ that of of ${}^{(1)}\overline{V}{}_M^{+}$ may be found through complex conjugation as $\varepsilon_{\tilde{\eta}}\left[{}^{(1)}\overline{V}{}_M^{+}\right]=\varepsilon_{\tilde{\eta}}\left[{}^{(1)}\overline{V}{}_M^{-}\right]^*$. }
 \label{tab:ucd_fast_closure}
\end{table}
\begin{figure}[htbp]
  \subfloat[\label{subfig:eta_schemes_drp_a}]{%
    \includegraphics[width=0.49\textwidth]{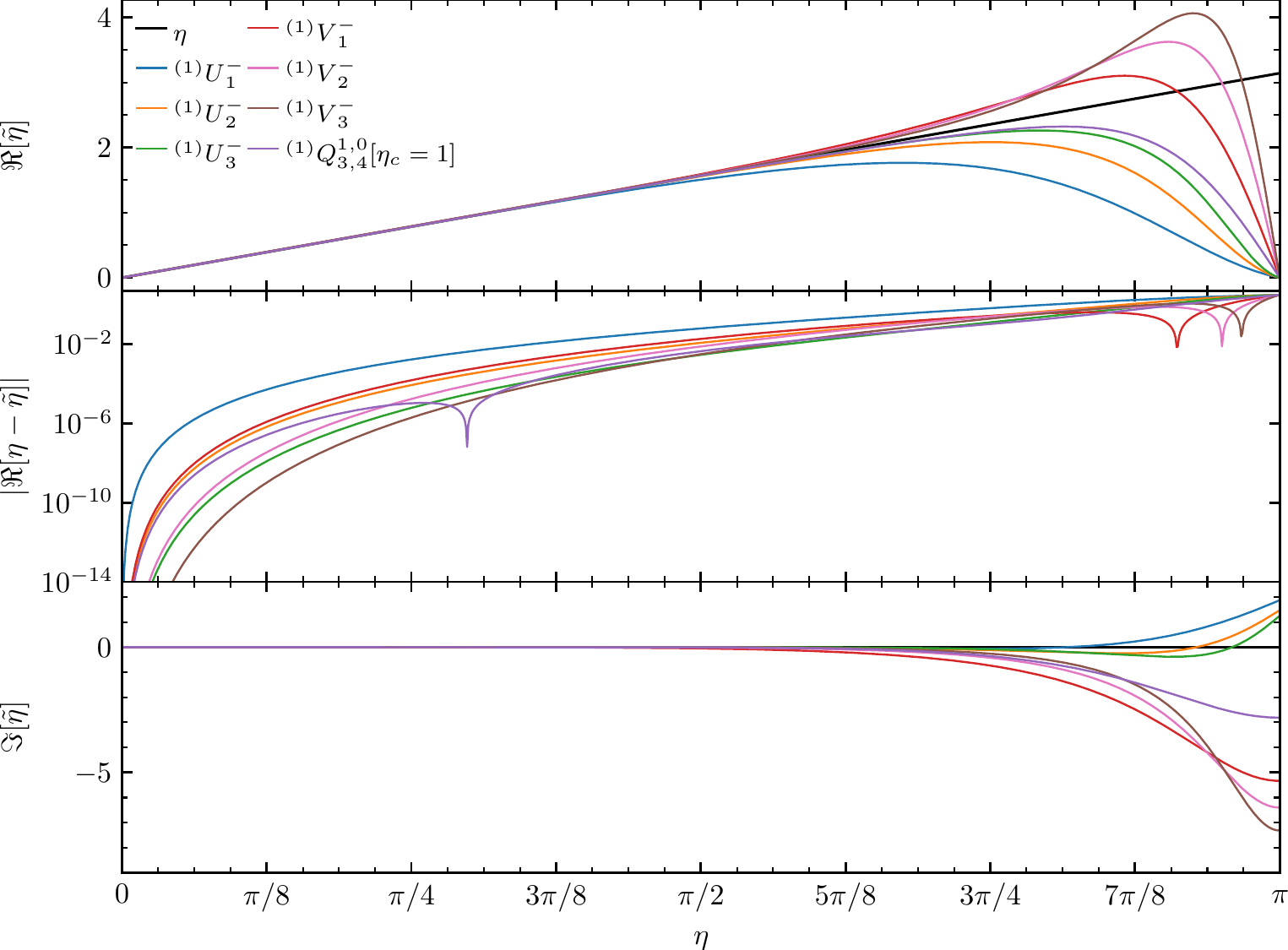}%
  }\hfill
  \subfloat[\label{subfig:eta_schemes_drp_b}]{%
    \includegraphics[width=0.49\textwidth]{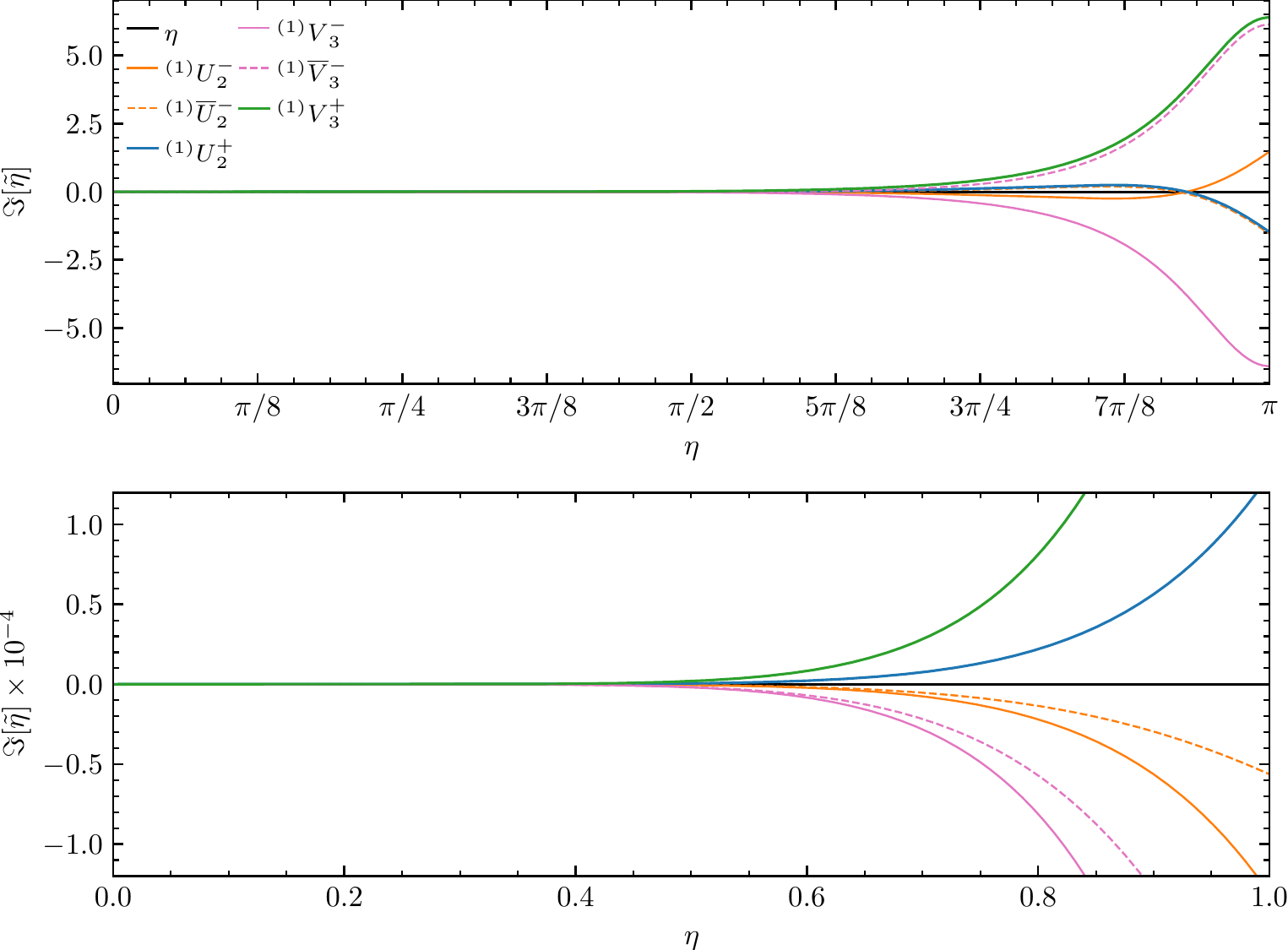}%
  }\hfill
  \caption{The modified, normalized wavenumber $\tilde{\eta}$ for generalizations of UCD style schemes to higher formal order of accuracy. In (a) the so-called "slow" ${}^{(1)}U{}^-_{M}$ and "fast" ${}^{(1)}V{}^-_{M}$ schemes are shown. To facilitate comparison the $\mathcal{O}(\delta x^6)$ scheme ${}^{(1)}Q{}_{3,4}^{1,0}[\eta_c=1]$ is also shown. In the top panel the respective over and under-estimation of $\eta$ can be seen for the two UCD varieties. In (a, bottom) we observe that $\Im[\tilde{\eta}[{}^{(1)}U{}_M^-]]\leq0$ whereas for the fast scheme the sign is indefinite. In (b) closures for ${}^{(1)}U{}^-_{2}$ and ${}^{(1)}V{}^-_{3}$ are depicted. We only show $\Im[\tilde{\eta}]$ as $\Re[\tilde{\eta}]$ is matched exactly by construction. It can be seen as $\eta\rightarrow\pi$ that $\Im[\tilde{\eta}[{}^{(1)}\overline{U}{}^-_{2}]]\rightarrow\Im[\tilde{\eta}[{}^{(1)}U{}^+_{2}]]$. This is expected as only the leading term of the expansion in $\Im[\tilde{\eta}]$ has been matched. In (b, bottom) we see this latter reflected in qualitative matching of the $\Im[\tilde{\eta}]$ trend for small $\eta$. The same behaviour is observed for the fast scheme. Similar conclusions can also be drawn for the other UCD schemes presented in the text (not depicted).
  See text for further discussion.}\label{fig:eta_schemes_drp}
\end{figure}
The schemes ${}^{(1)}U{}_M^\pm$ and ${}^{(1)}V{}_M^\pm$ that arise are termed so-called slow and fast schemes due to their respective under- and over-estimation of $\Re(\tilde{\eta})$ and its role in wave propagation (see \S\ref{ssec:phacc}). We illustrate in Fig.\ref{fig:eta_schemes_drp}. Consequently it has been suggested \cite{qin2014highaccuracynumerical,chen2021novelparallelcomputing} to linearly combine schemes with a weighting parameter. To this end we consider:
\begin{equation}
  \label{eq:LCslowfast}
  \begin{aligned}
    \left(
      \lambda \boldsymbol{\alpha}[{}^{(1)}U{}_M^\pm]
      + (1-\lambda) {}^{(\pm)}\boldsymbol{\alpha}[{}^{(1)}V{}_{M+1}^\pm]
    \right)
      \cdot\left(
        \delta x\,\tilde{\mathbf{f}}{}^{(1)}_{\hpb k}
      \right) =&
      \left(
        \lambda \boldsymbol{\alpha}[{}^{(0)}U{}_M^\pm]
        + (1-\lambda) {}^{(\pm)}\boldsymbol{\alpha}[{}^{(0)}V{}_{M+1}^\pm]
      \right)
      \cdot
      \mathbf{f}{}^{(0)}_{\hpb k}\\
      &+
      \lambda \boldsymbol{\alpha}[{}^{(2)}U{}_M^\pm]
      \cdot\left(
        \delta x^2\,\mathbf{f}{}^{(2)}_{\hpb k}
      \right),
  \end{aligned}
\end{equation}
where $\lambda\in[0,\,\lambda_\mathrm{max}]$, and analogously for the closure relations. In prescribing Eq.\eqref{eq:LCslowfast} we have formal order of accuracy of at least $\mathcal{O}(\delta x^{2M+2})$. The resultant schemes that combine ${}^{(1)}U{}_M^\pm$ and ${}^{(1)}V{}_{M+1}^\pm$ are generalizations of CCU$(4,5)$ \cite{qin2014highaccuracynumerical,chen2021novelparallelcomputing}.

In many problems of interest $f{}^{(2)}_{\hpb k}$ may not be available therefore this term must be approximated. One may consider utilizing e.g.~${}^{(2)}P{}^1_M$ unfortunately within the closure approach described this only approximately preserves dispersion under domain decomposition. Such CCU$(M,M+1)$ schemes we will denote ${}^{(1)}W{}^\pm_{M}$. Construction of ${}^{(1)}W{}^\pm_{M}$ is more involved as the now embedded, implicit specification of $\tilde{f}{}^{(2)}_{\hpb k}$ requires not only a closure ${}^{(1)}\overline{W}{}^\pm_{M}$ at at $k^*_\pm$ for the first derivative approximant but also ${}^{(1)}\widetilde{W}{}^\pm_{M}$ at $k^*_\mp$ for the second. Alternatively one can make use of explicit, derivative approximants such as ${}^{(2)}E_{M+2}$ for $\tilde{f}{}^{(2)}_{\hpb k}$ (see Tab.\ref{tab:misc_explicit}). This results in a deformed $\Im[\tilde{\eta}]$ for ${}^{(1)}U{}_M^\pm$ and ${}^{(1)}\overline{U}{}_M^\pm$ which exactly preserves dispersion between them.
Such CCU$(M,M+1)$ schemes we will denote ${}^{(1)}X{}^\pm_{M}$. In both approaches we select $\lambda$ with a simple strategy by starting at $\lambda_\mathrm{max}=1$ and reducing until $\tilde{\eta}\leq \eta$ over $\eta\in[0,\pi]$. In the case of ${}^{(1)}U{}_3^\pm$ when an embedded, explicit, second degree derivative approximation is made for $f{}^{(2)}_{\hpb k}$ it turns out that $\tilde{\eta}$ overestimates $\eta$ and so instead we set $\lambda_\mathrm{max}=2$.
\begin{figure}[htbp]
  \subfloat[\label{subfig:eta_schemes_ccu_a}]{%
    \includegraphics[width=0.49\textwidth]{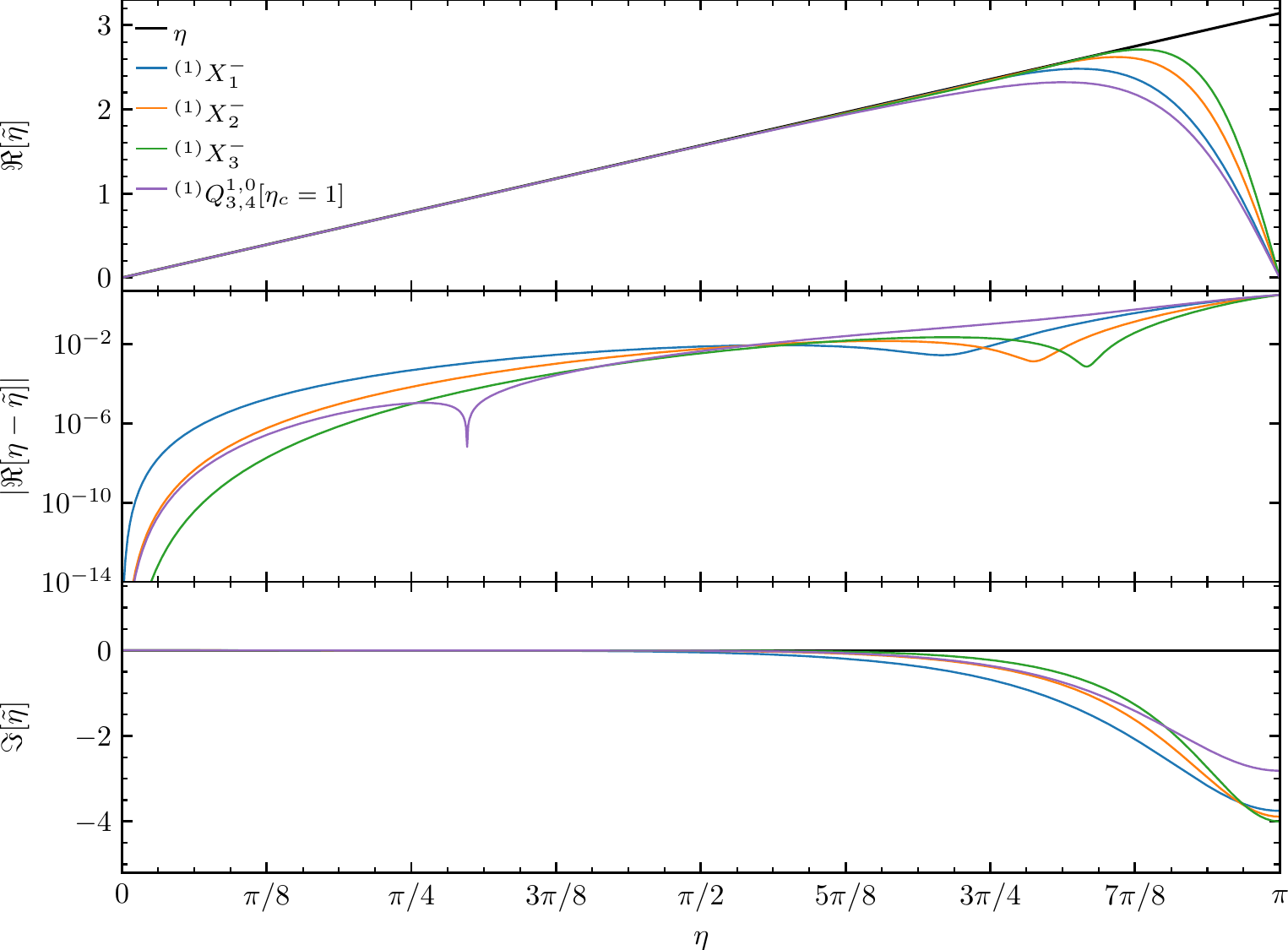}%
  }\hfill
  \subfloat[\label{subfig:eta_schemes_ccu_b}]{%
    \includegraphics[width=0.49\textwidth]{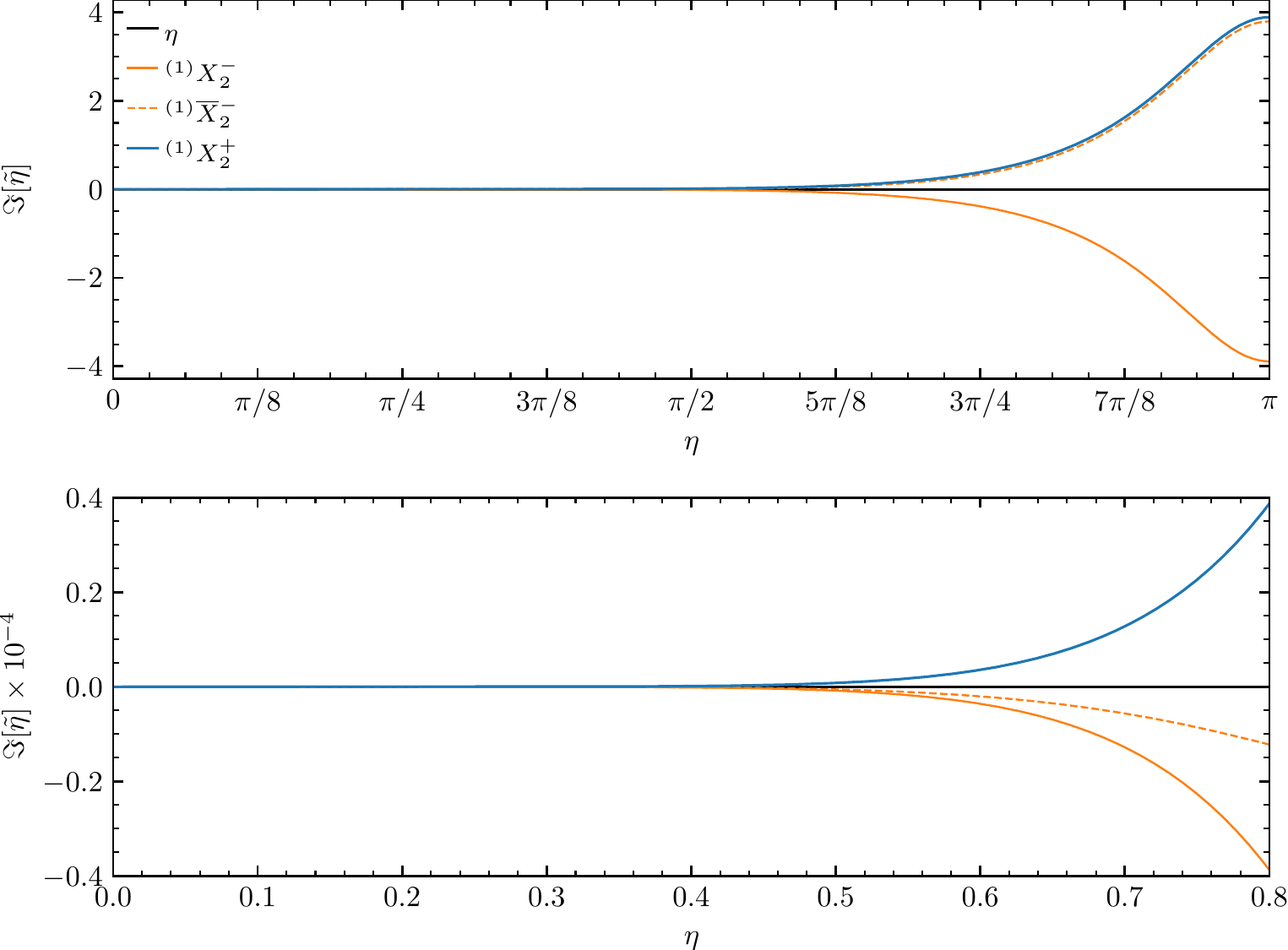}%
  }\hfill
  \caption{The modified normalized wavenumber $\tilde{\eta}$ for generalizations of CCU style schemes to higher formal order of accuracy. In (a) ${}^{(1)}X{}^\pm_{M}$ schemes are shown. To facilitate comparison the $\mathcal{O}(\delta x^6)$ scheme ${}^{(1)}Q{}_{3,4}^{1,0}[\eta_c=1]$ is also shown. It is clear that $\tilde{\eta}$ approximates $\eta$ well over a broad range of $\eta$. In (a, middle) we see that ${}^{(1)}Q{}_{3,4}^{1,0}[\eta_c=1]$ is a better approximation than the $\mathcal{O}(\delta x^6)$ scheme ${}^{(1)}X{}^\pm_{2}$ on $\eta\in[0,27/50\pi]$. This is to be expected as ${}^{(1)}Q{}_{3,4}^{1,0}[\eta_c=1]$ is numerically tuned on $\eta\in[0,1]$. In (a, bottom) we observe that $\Im[\tilde{\eta}]\leq0$ for all schemes depicted. In (b) closure for ${}^{(1)}X{}^\pm_{2}$ is depicted. We only show $\Im[\tilde{\eta}]$ as $\Re[\tilde{\eta}]$ is matched exactly by construction. It can be seen as $\eta\rightarrow\pi$ that $\Im[\tilde{\eta}[{}^{(1)}\overline{X}{}^-_{2}]]\rightarrow\Im[\tilde{\eta}[{}^{(1)}X{}^+_{2}]]$. This is expected as only the leading term of the expansion in $\Im[\tilde{\eta}]$ has been matched. In (b, bottom) we see this latter reflected in qualitative matching of the $\Im[\tilde{\eta}]$ trend for small $\eta$. Similar conclusions can also be drawn for the other CCU schemes presented in the text (not depicted).
  See text for further discussion.}\label{fig:eta_schemes_ccu}
\end{figure}
For convenience scheme ${}^{(1)}W{}^\pm_{M}$ coefficients and closures are provided in Tab.\ref{tab:ccuw_deg_one_col} whereas for ${}^{(1)}X{}^\pm_{M}$ see Tab.\ref{tab:ccux_deg_one_col}. The normalized, modified wavenumber for ${}^{(1)}X{}^\pm_{M}$ is shown in Fig.\ref{fig:eta_schemes_ccu}. On a sub-domain $\Omega_I$ in order to compute derivative data $\tilde{f}{}^{(1)}$ for the left biased schemes such as ${}^{(1)}X{}^-_{M}$ as closed by ${}^{(1)}\overline{X}{}^-_{M}$ we solved the banded linear system (cf. Eq.\eqref{eq:sch_cent_om}):
\begin{equation}\label{eq:closure_left_X}
  \begin{cases}
    {}^-\overline{\alpha}{}^{(1)}_{\hpb{}0} \tilde{f}{}^{(1)}_{\hpb{} k} +
    {}^-\overline{\alpha}{}^{(1)}_{\hpb{}1} \tilde{f}{}^{(1)}_{k+1} +
    = \frac{1}{\delta x} \sum_{m=-(M+2)}^{M+3}
    {}^-\overline{\alpha}{}^{(0)}_{\hpb{}m} f{}^{(0)}_{k+m},
    & k=k^*_-;\\
    {}^-\alpha{}^{(1)}_{-1} \tilde{f}{}^{(1)}_{k-1} +
    {}^-\alpha{}^{(1)}_{\hpb{}0} \tilde{f}{}^{(1)}_{\hpb{}k} +
    = \frac{1}{\delta x} \sum_{m=-(M+2)}^{M+2}
    {}^-\alpha{}^{(0)}_{\hpb{}m} f{}^{(0)}_{k+m},
    & k\in\left\{
      k^*_- + 1,\, k^*_- + 2,\,\dots,\, k^*_+
    \right\};
  \end{cases}
\end{equation}
similarly for the right biased cases with coefficients now those of ${}^{(1)}X{}^+_{M}$ as closed by ${}^{(1)}\overline{X}{}^+_{M}$:%
\begin{equation}
  \begin{cases}
    {}^+\alpha{}^{(1)}_{\hpb{}0} \tilde{f}{}^{(1)}_{\hpb{}k} +
    {}^+\alpha{}^{(1)}_{\hpb{}1} \tilde{f}{}^{(1)}_{k+1} +
    = \frac{1}{\delta x} \sum_{m=-(M+2)}^{M+2}
    {}^+\alpha{}^{(0)}_{\hpb{}m} f{}^{(0)}_{k+m},
    & k\in\left\{
      k^*_-,\, k^*_- + 1,\,\dots,\, k^*_+ - 1
    \right\};\\
    {}^+\overline{\alpha}{}^{(1)}_{-1} \tilde{f}{}^{(1)}_{k-1} +
    {}^+\overline{\alpha}{}^{(1)}_{\hpb{}0} \tilde{f}{}^{(1)}_{\hpb{}k} +
    = \frac{1}{\delta x} \sum_{m=-(M+3)}^{M+2}
    {}^+\overline{\alpha}{}^{(0)}_{\hpb{}m} f{}^{(0)}_{k+m},
    & k=k^*_+;
  \end{cases}
\end{equation}
and in both cases we take the size of the ghost node layer to be $N_g\geq M+2$.

\subsection{Domain decomposition: iteration of closures}
\label{ssec:domdec_iter}
Having investigated construction of dispersion relation preservation (DRP) in \S\ref{ssec:domdec_drp} we now turn our attention to a combined, hybrid strategy which features a DRP prescription and further iteratively updates values employed in the implicit system closure through additional communication of points.

Under domain-decomposition application of derivative scheme closures on a sub-domain $\Omega_I$ leads to numerical approximants with error that accumulates towards $\partial \Omega_I$. While this error is convergent at the constructed formal order, when compared to a unpartitioned, global scheme, an artifact of the decomposition can often be observed. This motivates us to consider construction of a sequence of approximants ${}^{[i]}\tilde{\mathbf{f}}{}^{(d_r)}$ that iteratively refine ${}^{[0]}\tilde{\mathbf{f}}{}^{(d_r)}$ as found using the closure prescription of \S\ref{ssec:domdec_drp} and consequently better capture the global scheme.

To this end, consider again ${}^{(1)}X{}^-_M$ with the matched closure ${}^{(1)}\overline{X}{}^-_M$. We solve Eq.\eqref{eq:closure_left_X} which provides us with ${}^{[0]}\tilde{\mathbf{f}}{}^{(1)}$. For $i>0$ we then solve:
\begin{equation}\label{eq:iter_left_X}
  \begin{cases}
    {}^{[i]}\tilde{f}{}^{(1)}_{\hpb{}k}
    = {}^{[i-1]}\tilde{f}{}^{(1)}_{\hpb{}k},
    & k=k^c_-;\\
    {}^-\alpha{}^{(1)}_{-1} {}^{[i]}\tilde{f}{}^{(1)}_{k-1} +
    {}^-\alpha{}^{(1)}_{\hpb{}0} {}^{[i]} \tilde{f}{}^{(1)}_{\hpb{}k} +
    = \frac{1}{\delta x} \sum_{m=-(M+2)}^{M+2}
    {}^-\alpha{}^{(0)}_{\hpb{}m} f{}^{(0)}_{k+m},
    & k\in\left\{
      k^c_- + 1,\, k^c_- + 2,\,\dots,\, k^*_+
    \right\};
  \end{cases}
\end{equation}
where during discretization on $\Omega_I$ the number of ghosts $N_g$ is selected such that ${}^{(1)}\overline{X}{}^-_M$ can be imposed and we can select $k^c_-$ such that $k^c_-<k^*_-$. In particular $N_g=M+2$ allows us to take $k^c_-=k^*_- - 1$. In practice the relation of Eq.\eqref{eq:iter_left_X} may be iterated until a desired tolerance is attained. While this requires additional communication only a single value ${}^{[i]}\tilde{f}{}^{(1)}_{\hpb{} k^c_-}$ needs updating prior to solution of Eq.\eqref{eq:iter_left_X} as the sampled function $\mathbf{f}^{(0)}$ remains unchanged. Furthermore the linear system when considered as a matrix equation appears with diagonal and single sub-diagonal and consequently ${}^{[i]}\tilde{\mathbf{f}}^{(1)}$ can be cheaply solved for at $\mathcal{O}(N_B)$ algorithmic complexity.

In a similar manner we may treat centered implicit schemes such as ${}^{(d_r)}P_M^1$ with matched closures ${}^{(d_r)}\overline{P}_{M+2}$ which fashion us with ${}^{[0]}\tilde{\mathbf{f}}{}^{(d_r)}$ based on Eq.\eqref{eq:sch_cent_om}. For $i>0$ one then solves:
\begin{equation}\label{eq:iter_center_P}
  \begin{cases}
    {}^{[i]}\tilde{f}{}^{(d_r)}_{\hpb{}k}
    = {}^{[i-1]}\tilde{f}{}^{(d_r)}_{\hpb{}k},
    & k=k^c_\pm;\\
    {}^-\alpha{}^{(d_r)}_{-1} {}^{[i]}\tilde{f}{}^{(d_r)}_{k-1} +
    {}^-\alpha{}^{(d_r)}_{\hpb{}0} {}^{[i]} \tilde{f}{}^{(d_r)}_{\hpb{}k} +
    = \frac{1}{(\delta x)^{d_r}} \sum_{m=-M}^{M}
    {}^-\alpha{}^{(0)}_{\hpb{}m} f{}^{(0)}_{k+m},
    & k\in\left\{
      k^c_- + 1,\, k^c_- + 2,\,\dots,\, k^c_+ - 1
    \right\};
  \end{cases}
\end{equation}
where for each additional iteration derivative values at $k^c_\pm$ must now be communicated. For ${}^{(d_r)}P_M^1$ we utilize $N_g=M+2$ and impose Eq.\eqref{eq:iter_center_P} with $k^c_-=k^*_--2$ and $k^c_+=k^*_++2$.

\begin{figure}[htbp]
  \subfloat[\label{subfig:der_01_decomp_a}]{%
    \includegraphics[width=0.49\textwidth]{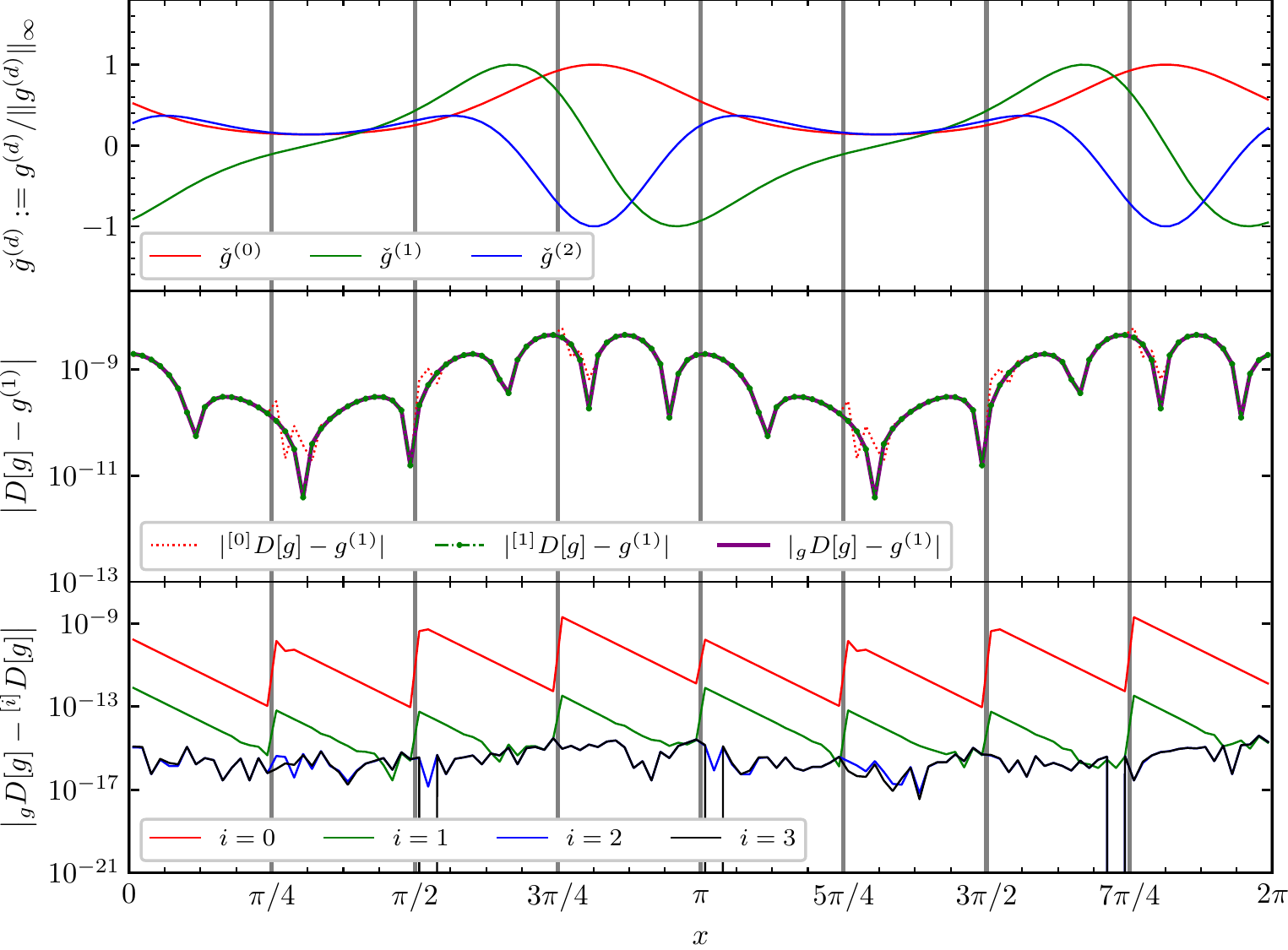}%
  }\hfill
  \subfloat[\label{subfig:der_01_decomp_b}]{%
    \includegraphics[width=0.49\textwidth]{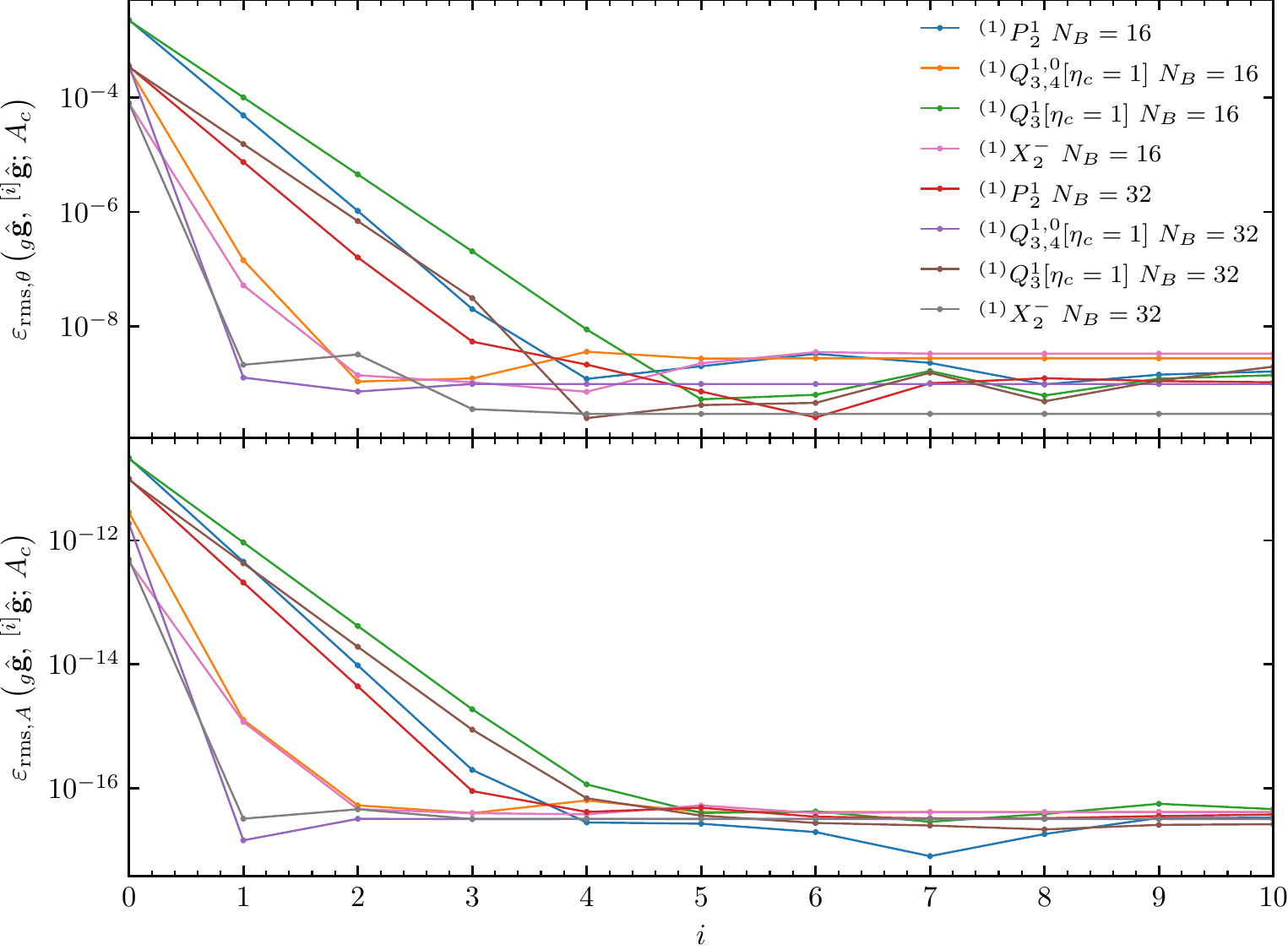}%
  }\hfill
  \caption{Effect of iterating the closure for first degree derivatives of $g$ as defined in Eq.\eqref{eq:test_grid_conv_g} at fixed spatial resolution $N_M=128$ on decomposed domains with $N_g=4$. Function parameters are $A=5/100$, $S=1$, $N_1=2$, and $x_0=2/10$ throughout. In (a, top) we depict the function $g^{(d)}$ for $d\in\{0,\,1,\,2\}$. In (a, middle) the ${}^{(1)}Q^{1,0}_{3,4}[\eta_c=1]$ scheme is utilized. Error is compared pointwise against $g^{(1)}$ for the domain-decomposed approximants ${}^{[0]}D[g]$ and ${}^{[1]}D[g]$. The latter involves a single iteration on the closure value and can be seen to coincide well with the global approximant ${}_g D[g]$ while mitigating edge artifacts. Deviation from the global scheme is greatest at the sub-domain edge used for the closure. In (a, bottom) convergence under iteration is demonstrated. Common to (a): $N_B=16$, and $k^c_\pm=k^*_\pm \pm 2$ with CC sampling as in Eq.\eqref{eq:gridDisc}. Sub-domains are delimited with vertical gray lines.
  In (b) closure is iterated for a variety of derivative approximants utilizing VC sampling. For $i=0$ no iteration occurs and only the salient closure is employed as constructed according to \S\ref{ssec:domdec_drp}. Under iteration biased schemes converge more rapidly than centered. At fixed $i$ error is smaller for larger ($N_B=32$) sub-domains; VC and CC (not depicted) sampling perform comparably. In (b, upper) and (b, lower) we show in the modal representation the RMS of differences between domain-decomposed and global derivative approximants for phases and amplitudes respectively. An ampltitude cut $A_c=10^{-8}$ was chosen in Eq.\eqref{eq:err_met_rms_sp} to select relevant modes.
    See text for further discussion.}\label{fig:der_01_decomp}
\end{figure}

We now turn to tests where prescribed functions are numerically differentiated globally and compared against the decomposed, local method. Subsequently we verify through grid convergence tests that compare against numerical sampling of the analytically expected result that error converges away at the design order of the underlying schemes. This will allow us to assess the performance of the explicit closures with (approximate) dispersion relation preservation under domain decompopsition introduced in \S\ref{ssec:domdec_drp} and the iterated closures described by Eq.\eqref{eq:iter_left_X} and Eq.\eqref{eq:iter_center_P}. To this end, define the smooth functions:
\begin{align}
  \label{eq:test_grid_conv_g}
  g(x) &= A \exp(-S \sin(N_1(x - x_0))),\\
  \label{eq:test_grid_conv_h}
  h(x) &= \exp(-S(x-x_0)^2) \left(
    c_1 \sin(N_2 x) + c_2 \cos(N_2 x)
  \right) + \sin(N_1 x- \phi_S).
\end{align}
As a first direct numerical probe of the effect of decomposition we fix the number of samples on $\Omega$ as $N_M$ and partition to sub-domains $\Omega_I$ with $N_B$ samples each. To inspect the effect on the spectrum directly upon numerical differentiation with our proposed compact finite-difference schemes we consider Fourier decomposition of the result and quantify error in the amplitude and phase respectively according to:
\begin{equation}
\label{eq:err_met_rms_sp}
\begin{aligned}
  \varepsilon_{\mathrm{rms},A}(
    {}_g\hat{\mathbf{a}}{}^{(d_r)},\,{}^{[i]}\hat{\mathbf{a}}{}^{(d_r)};\,A_c
  ) &:=
  \left(
  \frac{1}{|\mathcal{N}|}
  \sum_{n\in\mathcal{N}}
  \left|
    {}_g \hat{a}_n - {}^{[i]} \hat{a}_n
  \right|^2
  \right)^{1/2},\\
  \varepsilon_{\mathrm{rms},\theta}(
    {}_g\hat{\mathbf{a}}{}^{(d_r)},\,{}^{[i]}\hat{\mathbf{a}}{}^{(d_r)};\,A_c
  ) &:=\left(
    \frac{1}{|\mathcal{N}|}
    \sum_{n\in\mathcal{N}}
    \left(
      \arg\left({}_g \hat{a}_n\right) -
      \arg\left({}^{[i]} \hat{a}_n\right)
    \right)^2
    \right)^{1/2};
\end{aligned}
\end{equation}
where ${}_g\hat{\mathbf{a}}{}^{(d_r)}$ is a vector of coefficients (appearing in Eq.\eqref{eq:truncSumD}) as found from the global scheme, ${}^{[i]}\hat{\mathbf{a}}{}^{(d_r)}$ are the coefficients of the analogous domain-decomposed approach, and we select coefficients according to a relative amplitude threshold\footnote{In the modal representation smooth functions as discussed here have coefficients with amplitudes that decay exponentially \cite{boyd2001chebyshevfourierspectral,hesthaven2007spectralmethodstimedependent} which for a sufficiently high band-limit yields negligible contribution to description of the underlying function. Thus we impose a cut when computing $\varepsilon_{\mathrm{rms},\theta}$ to avoid polluting the error analysis with spurious $\arg{\hat{a}_n}$ values arising from coefficients with exponentially small amplitude.} $\mathcal{N}=\{n\,|\, |\hat{a}_n| / \max_m |\hat{a}_m| \leq A_c\}$ taken with respect to the global scheme. We select $g$ of Eq.\eqref{eq:test_grid_conv_g} with parameters $A=5/100$, $S=1$, $N_1=2$, and $x_0=2/10$ such that the amplitude of the normalized, modal coefficients of the Fourier decomposition (Eq.\eqref{eq:truncSumD}) are at numerical round-off i.e. $|\hat{g}_n| / \max_m |\hat{g}_m| \sim 10^{-16}$ for $|n| \gtrsim 32$. In Fig.\ref{fig:der_01_decomp} we investigate first degree numerical differentiation of $g$ with the (iterated) schemes: ${}^{(1)}P{}^1_2$ and ${}^{(1)}Q{}^1_3[\eta_c=1]$ with iteration on $k^c_\pm=k^*_\pm \pm 2$, whereas our CCU(6,7) generalization ${}^{(1)}X{}^-_2$ and ${}^{(1)}Q{}^{1,0}_{3,4}[\eta_c=1]$ are iterated with $k^c_-=k^*_--1$ and $k^c_-=k^*_--2$ respectively. It is clear that the implicit, interior compact finite-difference and the complimentary, explicit, tuned closures lead to error that has comparable order of magnitude (see Fig.\ref{subfig:der_01_decomp_a}). The decomposition induced error (near the boundaries of the sub-domains) is mitigated through the iterative prescription of this section. This is particularly effective for the biased schemes as can be seen in Fig.\ref{subfig:der_01_decomp_b}.
\begin{figure}[htbp]
  \subfloat[\label{subfig:der_02_decomp_a}]{%
    \includegraphics[width=0.49\textwidth]{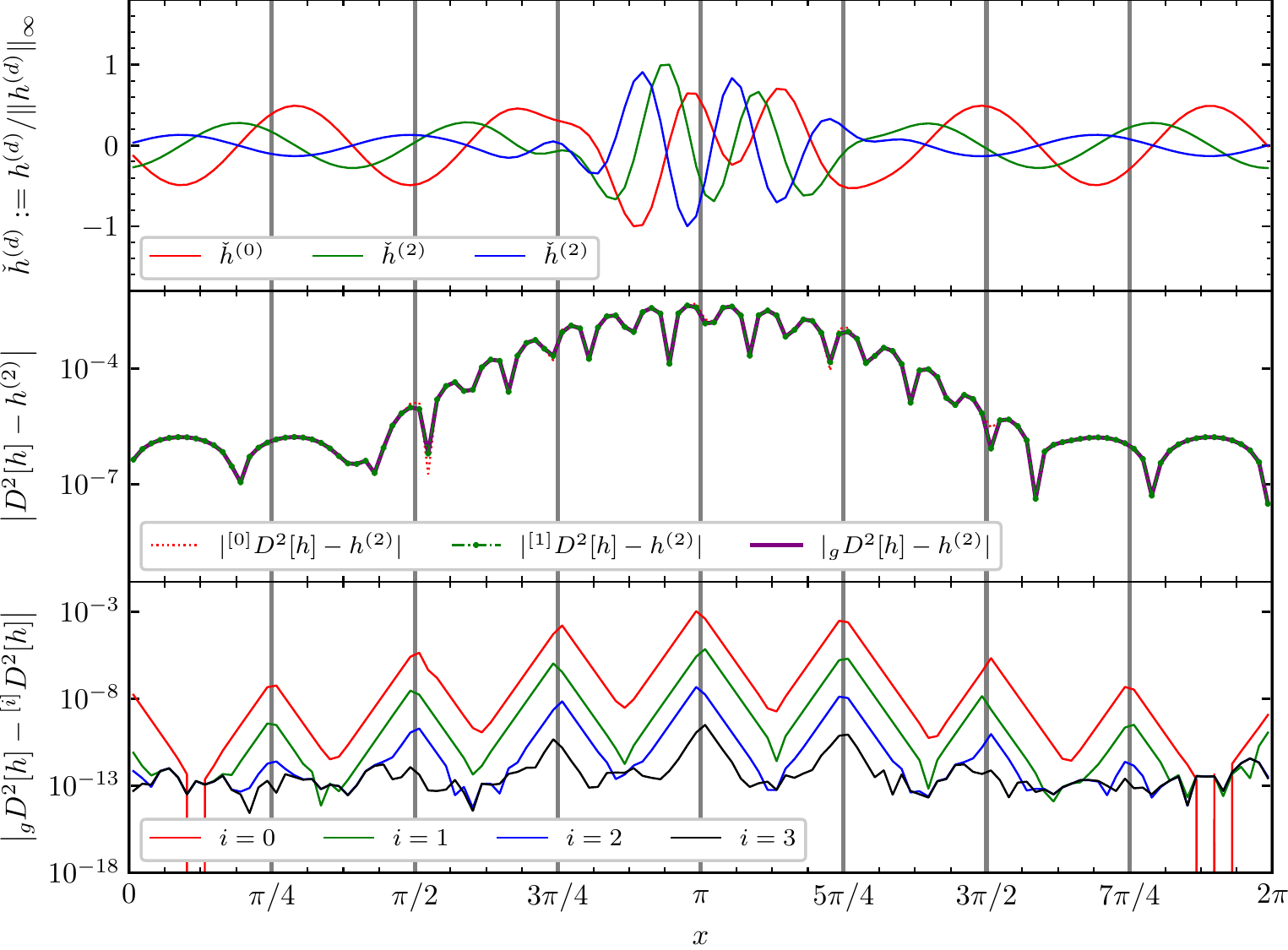}%
  }\hfill
  \subfloat[\label{subfig:der_02_decomp_b}]{%
    \includegraphics[width=0.49\textwidth]{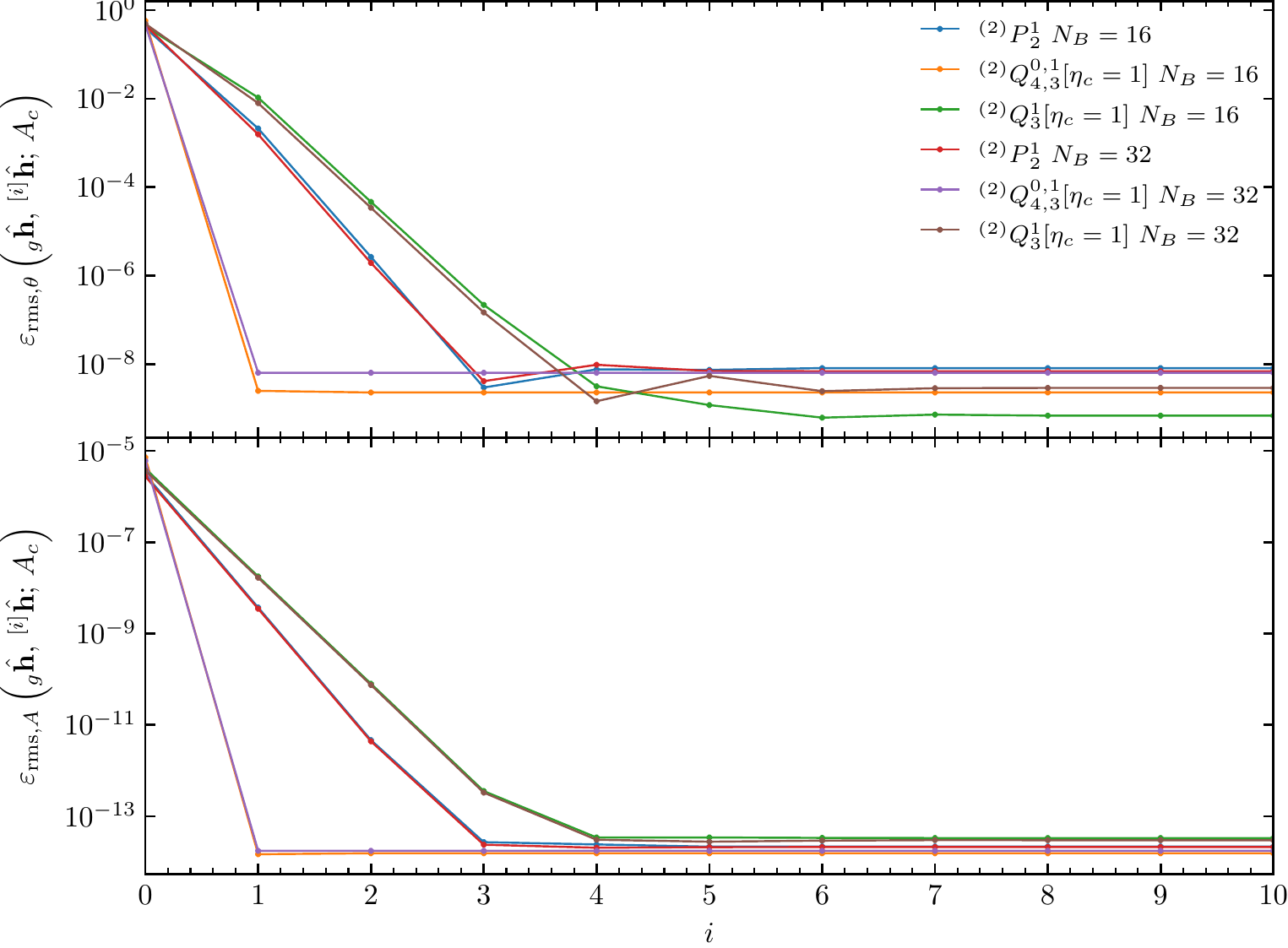}%
  }\hfill
  \caption{Effect of iterating the closure for second degree derivatives of $h$ as defined in Eq.\eqref{eq:test_grid_conv_h} at fixed spatial resolution $N_M=128$ on decomposed domains with $N_g=4$. Function parameters are $S=3$, $x_0=\pi$, $N_1=5$, $N_2=11$, $\phi_S=3$, $c_1=12/10$, and $c_2=-1$ throughout. In (a, top) we depict the function $h^{(d)}$ for $d\in\{0,\,1,\,2\}$.
  In (a, middle) the ${}^{(2)}P{}^{1}_{2}$ scheme of Tab.\ref{tab:pade_deg_two} is utilized. Deviation from the global scheme is greatest at both sub-domain edges used for the closure -- cf.~Fig.\ref{subfig:der_01_decomp_a}. In (a, bottom) convergence towards the global scheme is demonstrated. Common to (a): $N_B=16$, and $k^c_\pm=k^*_\pm \pm 2$ with CC sampling.
  In (b) same setup as~Fig.\ref{subfig:der_01_decomp_b} and similar conclusions hold however for the function examined sub-domain sampling of $N_B=16$ and $N_B=32$ appears to perform comparably with respect to error.
      See text for further discussion.}\label{fig:der_02_decomp}
\end{figure}
In a similar fashion we take $h$ of Eq.\eqref{eq:test_grid_conv_h} and select parameters $S=3$, $x_0=\pi$, $N_1=5$, $N_2=11$, $\phi_S=3$, $c_1=12/10$, and $c_2=-1$ such that $|\hat{h}_n| / \max_m |\hat{h}_m| \sim 10^{-16}$ for $|n| \gtrsim 32$. We investigate second degree numerical differentiation of $h$ with the (iterated) schemes: ${}^{(2)}P{}^1_2$ and ${}^{(2)}Q{}^1_3[\eta_c=1]$ with iteration on $k^c_\pm=k^*_\pm \pm 2$, whereas ${}^{(2)}Q{}^{0,1}_{4,3}[\eta_c=1]$ is iterated with $k^c_+=k^*_++2$. Results are depicted in Fig.\ref{fig:der_02_decomp}.

The number of samples (i.e.~size) of a sub-domain has an influence on the accuracy when compared with the equivalent global scheme. At fixed sampling for $\Omega$ where $N_M=128$ taking more samples per $\Omega_I$ that is $N_B=32$ cf. $N_B=16$ is favoured which is particularly apparent in the degree one case. This is compatible with the expectation that introducing fewer sub-domain boundaries leads to fewer artifacts due to them; however, it may not always be practical to select larger $N_B$ due to e.g. concerns involving partitioning a problem to utilize parallelism efficiently. During tests we have found that CC and VC sampling (Eq.\eqref{eq:gridDisc}) performs comparably. Due to the higher rate at with which biased schemes converge under iteration -- for the schemes we have investigated a single iteration is required to reach saturation -- a symmetrization procedure may be considered. To this end we compute derivatives utilizing ${}^{(d_r)}Q{}^{1,0}_{3,4}[\eta_c=1]$ and ${}^{(d_r)}Q{}^{0,1}_{4,3}[\eta_c=1]$ and then take as the final result after iteration as the average of the two. Such symmetrized schemes we denote as $\mathfrak{S}[{}^{(d_r)}Q{}^{1,0}_{3,4}[\eta_c=1]]$.

\begin{figure}[htbp]
  \subfloat[\label{subfig:der_grid_conv_a}]{%
    \includegraphics[width=0.49\textwidth]{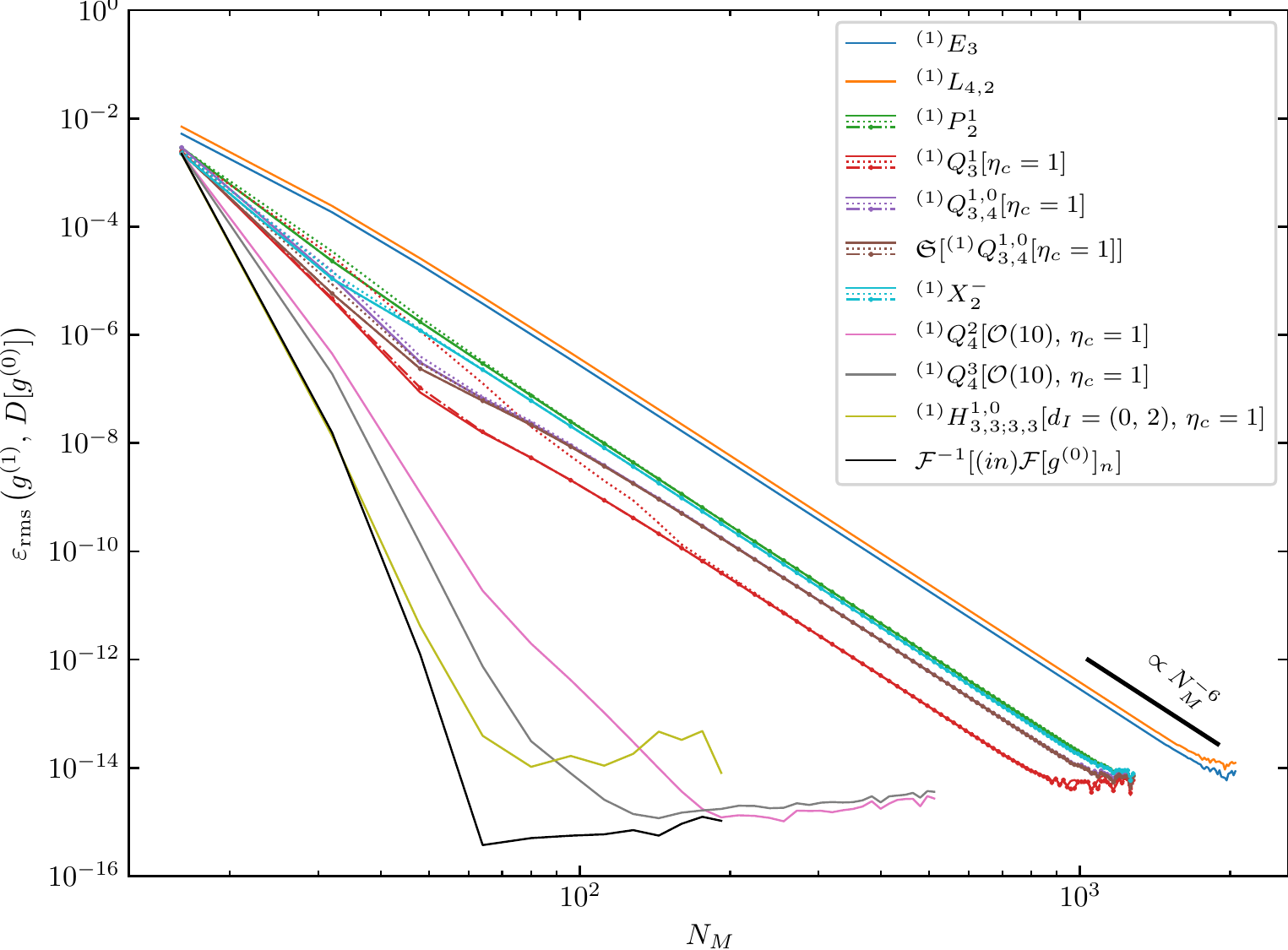}%
  }\hfill
  \subfloat[\label{subfig:der_grid_conv_b}]{%
    \includegraphics[width=0.49\textwidth]{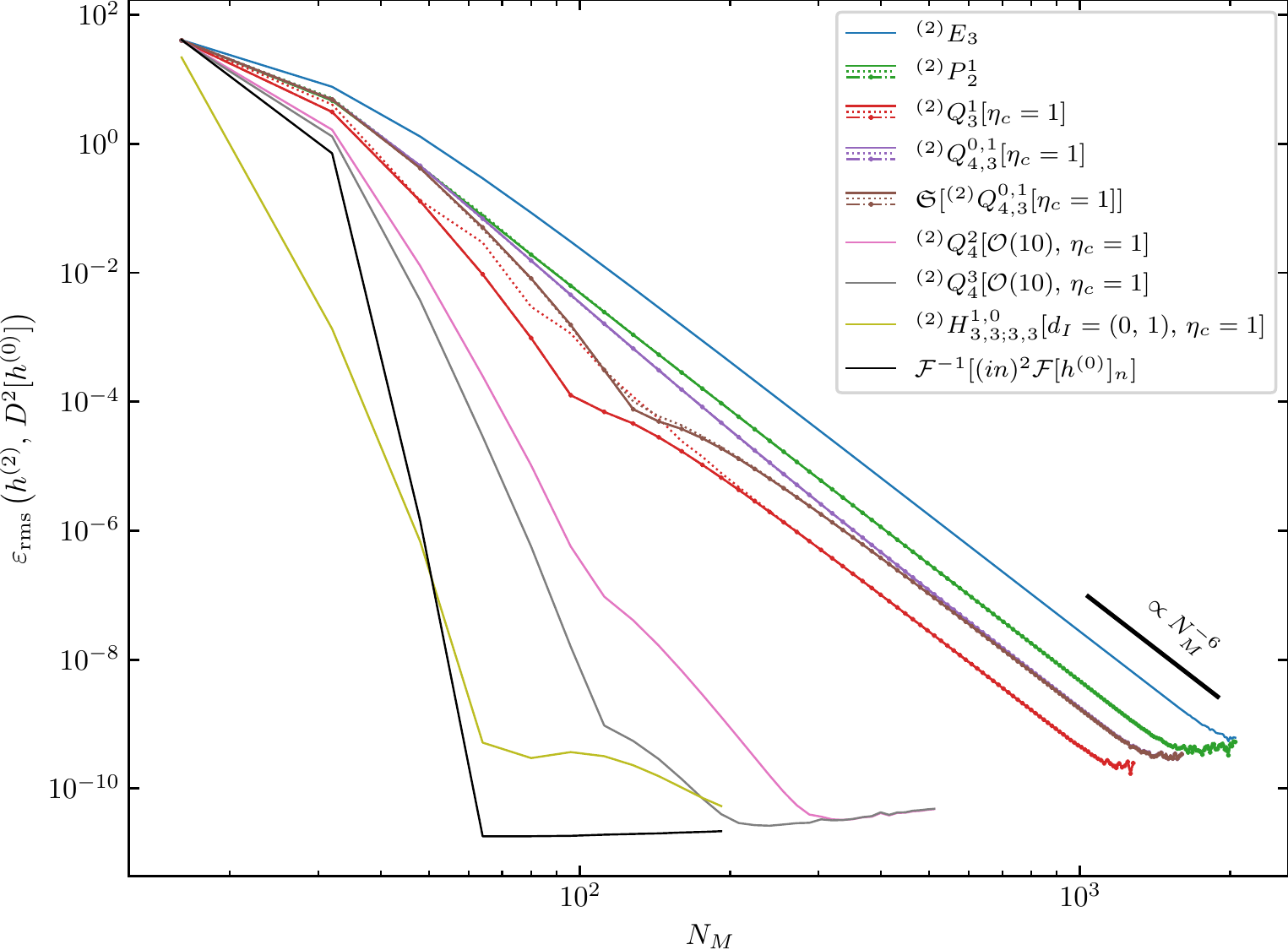}%
  }\hfill
  \caption{Grid convergence test for derivative approximants to $g$ of Eq.\eqref{eq:test_grid_conv_g} and $h$ of Eq.\eqref{eq:test_grid_conv_h} with function parameters as in Fig.\ref{fig:der_01_decomp} and Fig.\ref{fig:der_02_decomp}. A variety of schemes that we numerically constructed are compared against standard, explicit finite differencing. For several cases we compare domain decomposed analogues. A scheme with a legend involving multiple line-styles has depicted in: solid, global approximant; dashed, tailored sub-domain closure; dashed-dot, single iteration.
  The examples here are sampled on VC grids though CC sampling (not-depicted) gives comparable trends. For convergence tests involving sub-domain decomposition we take $N_B=16$. For $6^{\mathrm{th}}$ order schemes we observe a $6^{\mathrm{th}}$ order trend prior to saturation in convergence as expected.
  See text for further discussion.}\label{fig:der_grid_conv}
\end{figure}

In Fig.\ref{fig:der_grid_conv} we verify that the error of our proposed schemes converges as resolution is increased at the formal order anticipated from design choices. We find that symmetrization of the biased scheme ${}^{(d_r)}Q{}^{1,0}_{3,4}[\eta_c=1]$ leads to a reduction in error at intermediate $N_M$ and $\mathfrak{S}[{}^{(d_r)}Q{}^{1,0}_{3,4}[\eta_c=1]]\rightarrow {}^{(d_r)}Q{}^{1,0}_{3,4}[\eta_c=1]$ as $N_M$ is increased. To demonstrate the generality of our approach we also generate a variety of $\mathcal{O}(\delta x^{10})$ schemes tailored to the full domain $\Omega$; it can be seen that for higher implicit bandwidth error is more efficiently reduced at equal order, which is in agreement with general conclusions of \cite{Deshpande:2019uf}. For reference with more well-known schemes we also compare explicit finite differencing together with a Fourier based spectral method.
%

\subsection{Embedded operations}
\label{ssec:embedded}
As discussed in \S\ref{ssec:phacc} given a discretized wave propagation problem the spectrum of the associated discretized operator governing the time-evolution controls stability properties. With the introduction of domain decomposition and tuned closures for derivative schemes the aforementioned spectrum will be modified cf. the analogous global scheme. It is thus useful to have a practical method for inspecting the spectrum. Consequently we summarize a simple strategy that embeds into matrices a representation of sub-domain communication and imposition of boundary conditions together with the iterated closures discussed in \S\ref{ssec:domdec_iter}.

Consider a partitioned, periodic, one-dimensional domain $\Omega:=\sqcup_I \Omega_I$ where we select $N_M$ as the sampling parameter over $\Omega$ and $N_B$ for $\Omega_I$ where $N_B$ divides $N_M$ and the number of sub-domains is $N_S:=N_M/N_B$. Thus $I\in\{0,\,\dots,\,N_S\}$. Suppose that $f{}^{(0)}$ is sampled on each $\Omega_I$ using $\mathcal{G}_{\mathrm{CC}}$ of Eq.\eqref{eq:gridDisc} and we extend each $\Omega_I$ by $N_g$ ghost nodes in each direction. Collectively this data may be concatenated into the partitioned vector:
\begin{equation}\label{eq:concat_domdec}
  \mathbf{F}{}^{(0)}
  =
  \left[
    \begin{array}{c|c|c|c}
      f{}^{(0)}_{0,k} & f{}^{(0)}_{1,k} & \cdots & f{}^{(0)}_{N_S-1,k}
    \end{array}
  \right]^T,
\end{equation}
where each sub-domain (fixed $I$) has associated to it the $2N_g+N_B$ elements $f{}^{(0)}_{I,k}$. To capture communication operations between sub-domains together with imposition of boundary conditions we consider the introduction of a block-partitioned matrix $\mathbf{G}$ with $N_s\times N_s$ blocks where each block $G_{IJ}$ is of shape $(2N_g+N_B)\times(2N_g+N_B)$. The diagonal blocks $G_{II}$ are comprised of unit entries\footnote{In the case of VC the approach is similar but for consistency shared nodes are averaged \cite{Daszuta:2021ecf}.} which in indices local to the block satisfy\footnote{The standard definition of the Kronecker delta $\delta{}_{ij}=1$ if $i=j$ and $0$ otherwise is extended to logical conditionals where $1$ is taken if satisfied.} $\delta{}_{jk}\delta{}_{k\geq k^*_-}\delta{}_{k\leq k^*_+}$. To populate the ghost layer of $\Omega_I$ we utilize data data from $\Omega_{I_\pm}$ where $I_\pm:=(I\pm1)\mod(N_S)$. Thus $G_{II_-}$ has entries with local indices $\delta{}_{j,k-N_B}\delta{}_{j<k^*_-}$ whereas $G_{II_+}$ has entries $\delta{}_{j-N_B,k}\delta{}_{j>k^*_+}$. As an example suppose $N_S=3$ the overall structure of $\mathbf{G}$ becomes:
\begin{equation}
  \label{eq:example_G}
  \mathbf{G}:=
  \left[
    \begin{array}{ccc|ccc|ccc}
      0_{N_g} & 0_H              & 0_{N_g} & 0_{N_g} & 0_H                                       & 0_{N_g} & 0_{N_g} & 0_{N_g\times(N_B-N_g)}\;\mathbb{I}_{N_g}  & 0_{N_g} \\
      0_V     & \mathbb{I}_{N_B} & 0_V     & 0_V     & 0_{N_B}                                   & 0_V     & 0_V     & 0_{N_B}                                   & 0_V     \\
      0_{N_g} & 0_H              & 0_{N_g} & 0_{N_g} & \mathbb{I}_{N_g}\; 0_{N_g\times(N_B-N_g)} & 0_{N_g} & 0_{N_g} & 0_H                                       & 0_{N_g} \\
      \hline
      0_{N_g} & 0_{N_g\times(N_B-N_g)}\;\mathbb{I}_{N_g} & 0_{N_g} &  0_{N_g} & 0_H              & 0_{N_g} & 0_{N_g} & 0_H                                       & 0_{N_g} \\
      0_V     & 0_{N_B}                                  & 0_V     & 0_V      & \mathbb{I}_{N_B} & 0_V     & 0_V     & 0_{N_B}                                   & 0_V     \\
      0_{N_g} & 0_H                                      & 0_{N_g} &  0_{N_g} & 0_H              & 0_{N_g} & 0_{N_g} & \mathbb{I}_{N_g}\; 0_{N_g\times(N_B-N_g)} & 0_{N_g} \\
      \hline
      0_{N_g} & 0_H                                       & 0_{N_g} & 0_{N_g} & 0_{N_g\times(N_B-N_g)}\;\mathbb{I}_{N_g}  & 0_{N_g} & 0_{N_g} & 0_H  & 0_{N_g} \\
      0_V     & 0_{N_B}                                   & 0_V     & 0_V     & 0_{N_B}                                   & 0_V     & 0_V     & \mathbb{I}_{N_B}                          & 0_V     \\
      0_{N_g} & \mathbb{I}_{N_g}\; 0_{N_g\times(N_B-N_g)} & 0_{N_g} & 0_{N_g} & 0_H                                       & 0_{N_g} & 0_{N_g} & 0_H                                       & 0_{N_g}
    \end{array}
  \right],
\end{equation}
where $0_N$, $0_V$, and $0_H$ are $N\times N$, $N_B\times N_g$, and $N_g\times N_B$ zero matrices, and  $\mathbb{I}_{N}$ is an $N\times N$ identity matrix. 
In practice, during calculation of an implicit, centered numerical derivative on a sub-domain $\Omega_I$ we form a relation as in Eq.\eqref{eq:sch_cent_om} which would yield $\tilde{f}{}^{(d_r)}_{\hpb{} k}$ on nodes $k\in\{k^*_-,\,\dots,\,k^*_+\}$ through the use of a banded solver. Alternatively we may form the inverse matrix to solve Eq.\eqref{eq:sch_cent_om} directly. In order to work with square matrices that have shapes compatible with the blocks appearing in $\mathbf{G}$ and $\mathbf{F}{}^{(0)}$ we define:
\begin{align}
  \label{eq:diff_mat_subL}
  \overline{D}{}^{(d_r)}_{\hpb{}L}[\boldsymbol{\alpha}^{(d_r)}]
  &:=
  \left[
  \begin{array}{c|ccccc|c}
    \mathbb{I}_{N_g} &                       &      &  & & & \vphantom{\Big(}\\
    \hline
                               & 1                     &      &  & & & \vphantom{\Big(}\\
                               & \alpha{}^{(d_r)}_{-1} & \alpha{}^{(d_r)}_{\hpb{} 0} & \alpha{}^{(d_r)}_{\hpb{} 1} & & & \vphantom{\Big(}\\
                               &                       & \ddots & \ddots & \ddots & \vphantom{\Big(}\\
                              & & & \alpha{}^{(d_r)}_{-1} & \alpha{}^{(d_r)}_{\hpb{} 0} & \alpha{}^{(d_r)}_{\hpb{} 1} & \vphantom{\Big(}\\
                              &                       &      &  & & 1 & \vphantom{\Big(}\\
  \hline
                               &                       &      &  & & & \mathbb{I}_{N_g} \vphantom{\Big(}
  \end{array}
  \right], \\
  \label{eq:diff_mat_subR}
  \overline{D}{}^{(d_r)}_{\hpb{}R}[\boldsymbol{\alpha}^{(0)},\,\overline{\boldsymbol{\alpha}}^{(0)}]
  &:=
  \left[
  \begin{array}{ccc|ccccc|ccc}
    \hphantom{\cdots} & \mathbb{I}_{(N_g-(p+1))}           &                     &                                     &                           &                                       &                           &                           &                           &                                 & \vphantom{\Big(}\\
    \hline
    & \overline{\alpha}{}^{(0)}_{-(p+1)} & \cdots              & \overline{\alpha}{}^{(0)}_{\hpb{}0} & \cdots                    & \overline{\alpha}{}^{(0)}_{\hpb{}p+1} &                           &                           &                           &                                 & \vphantom{\Big(}\\
    &                                    & \alpha{}^{(0)}_{-M} & \cdots                              & \alpha{}^{(0)}_{\hpb{} 0} & \cdots                                & \alpha{}^{(0)}_{\hpb{} M} &                           &                           &                                 & \vphantom{\Big(}\\
    &                                    &                     & \ddots                              & \ddots                    & \ddots                                & \ddots                    & \ddots                    &                           &                                 & \vphantom{\Big(}\\
    &                                    &                     &                                     & \alpha{}^{(0)}_{-M}       & \cdots                                & \alpha{}^{(0)}_{\hpb{} 0} & \cdots                    & \alpha{}^{(0)}_{\hpb{} M} &                                 & \vphantom{\Big(}\\
    &                                    &                     &                                     &                           & \overline{\alpha}{}^{(0)}_{-(p+1)}    & \cdots                    & \overline{\alpha}{}^{(0)} & \cdots                    & \overline{\alpha}{}^{(0)}_{p+1} & \vphantom{\Big(}\\
    \hline
    &                                    &                     &                                     &                           &                                       &                           &                           &                           & \mathbb{I}_{(N_g-(p+1))}        & \hphantom{\cdots} \vphantom{\Big(}
  \end{array}
  \right];
\end{align}
where $\overline{D}{}^{(d_r)}_{\hpb{}L}[\boldsymbol{\alpha}^{(d_r)}]$ and $\overline{D}{}^{(d_r)}_{\hpb{}R}[\boldsymbol{\alpha}^{(0)},\,\overline{\boldsymbol{\alpha}}^{(0)}]$ have shapes $(N_B + 2 N_g)\times (N_B + 2 N_g)$, and zero entries we now suppress. In Eq.\eqref{eq:diff_mat_subL} and Eq.\eqref{eq:diff_mat_subR} elements are partitioned according to whether they act on data of $\Omega_I$ or the ghost layer. On each $\Omega_I$ we may thus evaluate derivatives using ${}^{[0]}D^{(d_r)}:=(\delta x)^{-d_r}(\overline{D}{}^{(d_r)}_{\hpb{}L})^{-1} \overline{D}{}^{(d_r)}_{\hpb{}R}$. For collective application to each sub-domain described by $\mathbf{F}{}^{(0)}$ we need $N_S$ copies of ${}^{[0]}D^{(d_r)}$ embedded as: ${}^{[0]}\mathbf{D}^{(d_r)}:=\mathbf{G}\,\mathrm{diag}({}^{[0]}D^{(d_r)},\,\dots,\,{}^{[0]}\,D^{(d_r)})\,\mathbf{G}$. This describes an initial communication populating ghost layer data, followed by differentiation, and finalized by another communication. To provide ${}^{[i]}\mathbf{D}^{(d_r)}$ with $i>0$ we need to incorporate the iterated scheme of Eq.\eqref{eq:iter_center_P}. To this end define $D{}^{(d_r)}_{\hpb{}L}[\boldsymbol{\alpha}^{(d_r)}]$ by replacing $N_g$ with $N_c:=k^c_-$ in $\overline{D}{}^{(d_r)}_{\hpb{}L}[\boldsymbol{\alpha}^{(d_r)}]$ of Eq.\eqref{eq:diff_mat_subL} whereas $\overline{D}{}^{(d_r)}_{\hpb{}R}[\boldsymbol{\alpha}^{(0)},\,\overline{\boldsymbol{\alpha}}^{(0)}]$ of Eq.\eqref{eq:diff_mat_subR} becomes:
\begin{align}
  \label{eq:diff_mat_subMR}
  D{}^{(d_r)}_{\hpb{}R}[\boldsymbol{\alpha}^{(0)}]
  &:=
  \left[
  \begin{array}{ccc|ccccc|ccc}
    \hphantom{\cdots} & \mathbb{I}_{N_c}           &                     &                                     &                           &                                       &                           &                           &                           &                                 & \vphantom{\Big(}\\
    \hline
    &                                    &                     & 1                                   &                           &                                       &                           &                           &                           &                                 & \vphantom{\Big(}\\
    &                                    & \alpha{}^{(0)}_{-M} & \cdots                              & \alpha{}^{(0)}_{\hpb{} 0} & \cdots                                & \alpha{}^{(0)}_{\hpb{} M} &                           &                           &                                 & \vphantom{\Big(}\\
    &                                    &                     & \ddots                              & \ddots                    & \ddots                                & \ddots                    & \ddots                    &                           &                                 & \vphantom{\Big(}\\
    &                                    &                     &                                     & \alpha{}^{(0)}_{-M}       & \cdots                                & \alpha{}^{(0)}_{\hpb{} 0} & \cdots                    & \alpha{}^{(0)}_{\hpb{} M} &                                 & \vphantom{\Big(}\\
    &                                    &                     &                                     &                           &                                       &                           & 1                         &                           &                                 & \vphantom{\Big(}\\
    \hline
    &                                    &                     &                                     &                           &                                       &                           &                           &                           & \mathbb{I}_{N_c}        & \hphantom{\cdots} \vphantom{\Big(}
  \end{array}
  \right].
\end{align}
By introducing $(E^c_\pm){}_{jk}:=\delta{}_{jk}\delta{}_{kk^c_\pm}$ and $(Z^c_\pm){}_{jk}:=\delta{}_{jk}\delta{}_{k\geq k^c_-}\delta{}_{k\leq k^c_+} - \delta{}_{kk^c_\pm}$ matrix multiplication will allow for extraction and fixing of values at nodes $k^c_\pm$ of $\Omega_I$. We introduce the block-diagonal matrices $\mathbf{E}^c_\pm:=\mathrm{diag}(E^c_\pm,\,\dots,\,E^c_\pm)$, $\mathbf{Z}^c_\pm:=\mathrm{diag}(Z^c_\pm,\,\dots,\,Z^c_\pm)$, $(\mathbf{D}{}^{(d_r)}_{\hpb{}L})^{-1}=\mathrm{diag}((D{}^{(d_r)}_{\hpb{}L})^{-1},\,\dots,\,(D{}^{(d_r)}_{\hpb{}L})^{-1})$, and $\mathbf{D}{}^{(d_r)}_{\hpb{}R}=\mathrm{diag}(D{}^{(d_r)}_{\hpb{}R},\,\dots,\,D{}^{(d_r)}_{\hpb{}R})$. Finally this allows for:
\begin{equation}
  \label{eq:iter_scheme}
  {}^{[i]}\mathbf{D}^{(d_r)} =
  \begin{cases}
    \mathbf{G}
    \left(
      (\mathbf{D}{}^{(d_r)}_{\hpb{}L})^{-1}
      \left(
        \frac{1}{(\delta x)^{d_r}}\mathbf{Z}^c_- \mathbf{Z}^c_+
        \mathbf{D}{}^{(d_r)}_{\hpb{}R} + (\mathbf{E}^c_- + \mathbf{E}^c_+) {}^{[i-1]}\mathbf{D}^{(d_r)}
      \right)
    \right), &
    i>0; \\
    \mathbf{G}\,\mathrm{diag}({}^{[0]}D^{(d_r)},\,\dots,\,{}^{[0]}\,D^{(d_r)})\,\mathbf{G}, & i=0.
  \end{cases}
\end{equation}
Given ${}^{[i]}\mathbf{D}^{(d_r)}$ and domain-decomposed function data $\mathbf{F}{}^{(0)}$ assembled as in Eq.\eqref{eq:concat_domdec} we may perform numerical experiments involving construction of derivative approximants based on implicit finite-difference schemes under domain-decomposition with closures iterated as described in \S\ref{ssec:domdec_iter}. Equation~\eqref{eq:iter_scheme} also allows for direct inspection of the eigenvalues which may be used to rapidly gain insight on stability properties for wave propagation problems. If biased schemes are instead considered then the iteration matrix ${}^{[i]}\mathbf{D}$ is constructed analogously and therefore we omit details. While we detailed the construction of ${}^{[i]}\mathbf{D}^{(d_r)}$ for a one-dimensional grid this may be extended to higher dimensions through use of Kronecker products.

\section{Applications}
\label{sec:res}

We have considered construction of a variety of compact finite-difference schemes together with their properties under domain-decomposition in the numerical approximation of function derivatives. The goal of this section is to demonstrate their application to wave propagation problems. In particular, numerical solution of the standard, two dimensional homogeneous advection equation \cite{evans2010partialdifferentialequations} where sensitivity to characteristics \cite{leveque2007finitedifferencemethods} will allow for careful examination of biased schemes. Thereafter the shifted wave equation \cite{chirvasa2010finitedifferencemethods} will be considered as it provides a simple model \cite{calabrese2005finitedifferencingsecond,babiuc2006testingnumericalrelativity} for first order in time, second order in space partial differential equations governing numerical relativity formulations such as BSSNOK \cite{Nakamura:1987zz,Shibata:1995we,Baumgarte:1998te} and Z4c \cite{Bernuzzi:2009ex,Ruiz:2010qj,Weyhausen:2011cg,Hilditch:2012fp}. These toy problems we take as an initial stepping-stone allowing for verification of implementation and feasibility of the approach thereafter attention is turned to the Z4c system in \S\ref{ssec:z4c}.

\subsection{Two dimensional advection}
\label{ssec:wave_eqn}
\begin{figure}[htbp]
  \subfloat[\label{subfig:eig_stab_no_iter_a}]{%
    \includegraphics[width=0.49\textwidth]{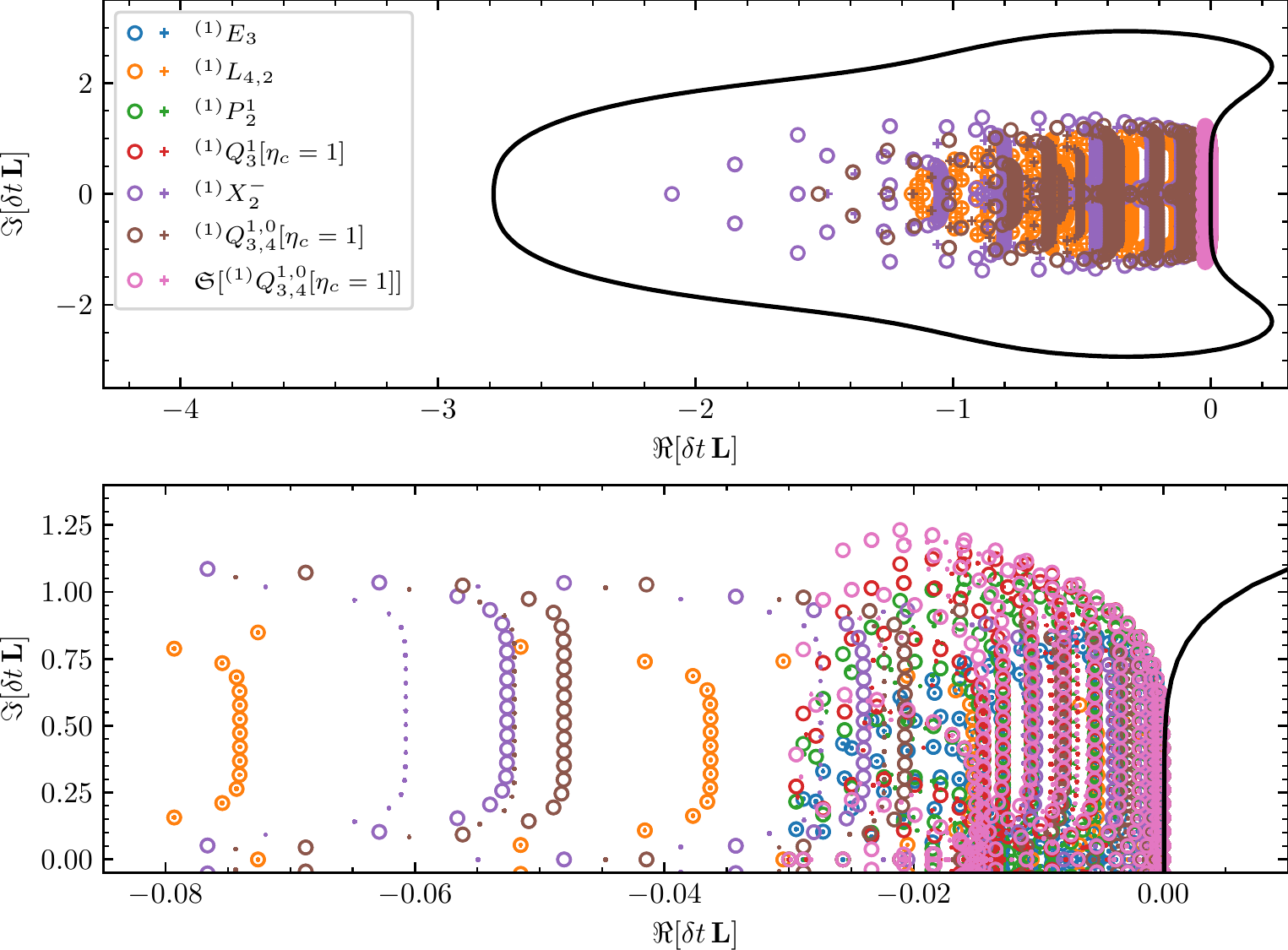}%
  }\hfill
  \subfloat[\label{subfig:eig_stab_no_iter_b}]{%
    \includegraphics[width=0.49\textwidth]{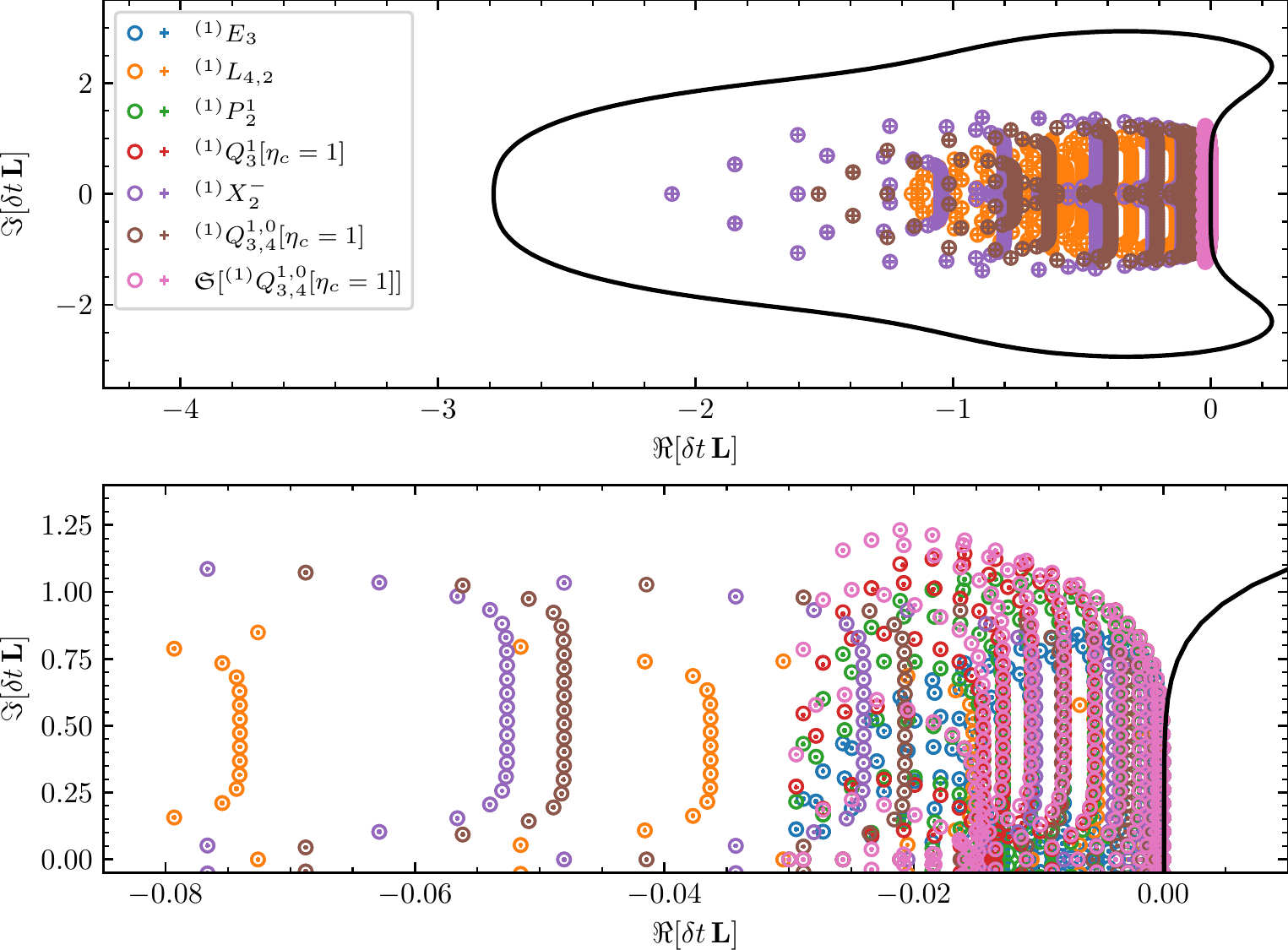}%
  }\hfill
  \caption{
    Eigenvalues of the discretized linear operator representing the two dimensional advection problem of Eq.\eqref{eq:sd2pde} scaled by $\delta t$ as determined from CFL of $\mathcal{C}_1=3/8$ for a variety of schemes. The interior of the back curve depicts the absolute stability region for classical ERK4 \cite{butcher2008numerical,hairer2010solving}. In (a,b) open circles denote $\mathrm{spec}(\delta t\,\mathbf{L})$ which corresponds to the single domain formulation whereas $\mathrm{spec}\left({}^{[i]}\mathbf{L}\right)$ is shown in crosses. In (a) $i=0$ is taken with closures for domain-decomposed derivative schemes applied as described in \S\ref{ssec:domdec_drp}. It is clear that the closure induces a deformation of the spectrum for implicit schemes. In the case of manifestly explicit schemes domain-decomposition leaves the spectrum unaffected. In (b) a single iteration is performed ($i=1$) according to the prescription of \S\ref{ssec:domdec_iter} with $\mathrm{spec}\left({}^{[1]}\mathbf{L}\right)$ evaluated based on Eq.\eqref{eq:iter_scheme}. It is clear that employing a hybrid strategy that involves an additional iteration leads to spectra that closely coincide with one another (cf.~Fig.\ref{fig:eig_stab_wave}).
    See text for further discussion.
  }\label{fig:eig_stab_advection}
\end{figure}

We begin with the two-dimensional, spatially periodic, initial value problem for the advection equation. Define $\Omega^2_T:=[0,\,T]\times\Omega\times\Omega$ with $\Omega$ as in \S\ref{ssec:sperr}. Consider smooth $U:\Omega^2_T\rightarrow \mathbb{R}$ satisfying:
\begin{equation}\label{eq:tr2pde}
  \begin{cases}
    \partial_t[U] + c_x \partial_x[U] + c_y \partial_y[U] = 0, &
    \Omega^2_T;\\
    U(t,\,x,\,0) = U(t,\,x,\,2\pi), & t\in[0,T],\,x\in\Omega; \\
    U(t,\,0,\,y) = U(t,\,2\pi,\,y), & t\in[0,T],\,y\in\Omega; \\
    U(0,\,x,\,y) = u(x,\,y), &
    \{t=0\} \times \Omega^2;
  \end{cases}
\end{equation}
with $\mathbf{c}:=(c_x,\,c_y)\in \mathbb{R}^2$ subject to the unit-speed constraint $|c_x|^2 + |c_y|^2=1$. The direction of propagation of an initial advected profile $u$ is parametrized through selection of the angle $\varphi$ that $\mathbf{c}$ forms with the $x$ axis. Under single domain, uniformly spaced, spatial discretization with $N_{M_x}$ and $N_{M_y}$ samples in $x$ and $y$ directions the system of Eq.\eqref{eq:tr2pde} becomes:
\begin{equation}\label{eq:sd2pde}
  \begin{cases}
    \frac{\mathrm{d} \tilde{U}_{ij}}{\mathrm{d}t}
    =
    -c_x \sum_{k=0}^{N_{M_x} - 1} \tilde{D}{}^{(1)}_{x,ik} \tilde{U}_{kj}(t)
    -c_y \sum_{l=0}^{N_{M_y} - 1} \tilde{D}{}^{(1)}_{y,jl} \tilde{U}_{il}(t), &
    t\in[0,\,T];\\
    \tilde{U}_{ij}(0) = u(x_i,\,y_j), &
    \{t=0\};
  \end{cases}
\end{equation}
where enforcement of the periodic boundary conditions is considered embedded within the discretized derivatives and initial conditions are compatible with periodicity. For simplicity in this section we will restrict attention to cell-centered sampling. We may also view Eq.\eqref{eq:sd2pde} as $\frac{\mathrm{d}}{\mathrm{d} t}[\tilde{\mathbf{U}}](t)=\mathbf{L} \tilde{\mathbf{U}}(t)$ where $\mathbf{L}:=(-c_x \tilde{\mathbf{D}}_x)\oplus(-c_y \tilde{\mathbf{D}}_y)=-c_x \tilde{\mathbf{D}}_x\otimes \mathbb{I}_{N_{M_y}} -c_y \mathbb{I}_{N_{M_x}}\otimes\tilde{\mathbf{D}}_y$ acts on the state vector $\mathbf{U}$ comprised of elements $\tilde{U}_{ij}$. Formal solution is provided through matrix exponentiation $\tilde{\mathbf{U}}(t)=\exp(t \mathbf{L}) \tilde{\mathbf{U}}(0)$. Consequently to describe propagating solutions to the semi-discrete problem with bounded amplitude we require that $\Lambda\in\mathrm{spec}(\mathbf{L})$ satisfies $\Re[\Lambda]\leq 0$ (see discussion of \S\ref{ssec:phacc}). Due to this when utilizing biased schemes for $\tilde{D}{}^{(1)}$ care needs to be taken to appropriately upwind. This can be achieved by ensuring that for $c>0$ we select a biased scheme\footnote{These choices can be understood through considering the propagating, single mode solution to Eq.\eqref{eq:ffmadv} and the modified wavenumber.} with $\Im[\tilde{\eta}]\leq0$ whereas for $c<0$ we require $\Im[\tilde{\eta}]\geq0$. Suppose $\mathcal{S}_-:=\{{}^{(1)}L_{2,4},\, {}^{(1)}Q{}^{0,1}_{4,3}[\eta_c=1],\, {}^{(1)}X{}^+_2\}$ and $\mathcal{S}_+:=\{{}^{(1)}L_{4,2},\, {}^{(1)}Q{}^{1,0}_{3,4}[\eta_c=1],\,{}^{(1)}X{}^-_2\}$ where $S_\pm\in\mathcal{S}_\pm$ are the previously introduced $\mathcal{O}(\delta x^6)$ biased schemes. During calculations based on Eq.\eqref{eq:sd2pde} involving approximants of $\mathcal{S}_\pm$ we replace:
\begin{equation}\label{eq:biasderapproxadv}
  \tilde{D}{}^{(1)} \rightarrow
  \frac{1}{2}(|c| - c) \tilde{D}{}^{(1)}[S_-] +
  \frac{1}{2}(|c| + c) \tilde{D}{}^{(1)}[S_+],
  \quad (S_\pm\in\mathcal{S}_\pm).
\end{equation}
In the case of decomposition of $\Omega$ into $N_{S_x}\times N_{S_y}$ sub-domains $\Omega_{IJ}$ the form of Eq.\eqref{eq:sd2pde} does not change if indices are instead viewed as local to a sub-domain. Furthermore in light of Eq.\eqref{eq:iter_scheme} a global description of the domain decomposed problem with iterated closures for implicitly specified derivative approximants is provided through ${}^{[i]}\mathbf{L}:=-c_x{}^{[i]}\tilde{\mathbf{D}}_x \otimes \mathbb{I}_{N_{S_y}(N_{B_y}+2N_g)} - c_y \mathbb{I}_{N_{S_x}(N_{B_x}+2N_g)}\otimes {}^{[i]} \tilde{\mathbf{D}}_y$. This is of particular utility during consideration of the fully discretized problem where $[0,\,T]$ is uniformly partitioned into time-steps $\delta t$ as stability properties may be assessed based on $\mathrm{spec}\left(\delta t\, {}^{[i]}\mathbf{L}\right)$. For numerical calculation of the eigenvalues we do not need to assemble the full matrix explicitly. It is the case that if $\mathbf{A}:=\mathbf{B}\otimes\mathbb{I}+\mathbb{I}\otimes\mathbf{C}$ then for $\lambda\in\mathrm{spec}(\mathbf{B})$ and $\mu\in\mathrm{spec}(\mathbf{C})$ we have that $\lambda + \mu \in \mathrm{spec}(\mathbf{A})$ \cite{schacke2004kroneckerproduct}. We fix $c_x=c_y=1/\sqrt{2}$ and compare the spectrum for $\delta t\,\mathbf{L}$ constructed with respect to a single domain $\Omega$ where $(N_{M_x},\,N_{M_y})=(32,\,32)$ with that of $\delta t\,{}^{[i]}\mathbf{L}$ where $(N_{B_x},\,N_{B_y})=(16,\,16)$ and $(N_{S_x},\,N_{S_y})=(2,\,2)$ corresponding to the domain-decomposed problem for a variety of schemes in Fig.\ref{fig:eig_stab_advection}.
We find that the spectrum is deformed during domain-decomposition on account of decoupling the implicit specification of derivative approximants with explicit closures however as is evident a single iteration is sufficient to almost entirely mitigate this effect. Additionally we observe that for the biased stencils of $\mathcal{S}_\pm$ imposing Eq.\eqref{eq:biasderapproxadv} leads to all schemes investigated satisfying $\Re\left[\mathrm{spec}\left({}^{[i]}\mathbf{L}\right)\right]\leq 0$ to numerical round-off. From Fig.\ref{fig:eig_stab_advection} we also see that selecting $\delta t$ based on a CFL of $\mathcal{C}_1=3/8$ satisfies a neccessary (and in this case sufficient) condition for stability of the fully discrete problem evolved with the classical ERK4 as all eigenvalues are contained with the stability polynomial of the method. Furthermore we have verified that these properties are robust under changes to the direction of $\mathbf{c}$ together with changes in the number of samples and sub-domains.

Having confirmed stability properties through numerical spectra we now turn to numerical solution of the evolution problem. In the semi-discrete case we make use of Eq.\eqref{eq:sd2pde} supplemented by Eq.\eqref{eq:biasderapproxadv} for biased schemes and consider formal exponential integration. Recall that $(\mathbf{A}\otimes\mathbf{B})(\mathbf{C}\otimes\mathbf{D})=(\mathbf{A}\mathbf{C})\otimes(\mathbf{B}\mathbf{D})$ and consequently the commutator $[\mathbf{A}\otimes\mathbb{I},\,\mathbb{I}\otimes\mathbf{B}]$ vanishes. Thus through the use of the Zassenhaus formula \cite{suzuki1977convergenceexponentialoperators} we have $\exp(\mathbf{A}\oplus\mathbf{B})=\exp(\mathbf{A})\otimes\exp(\mathbf{B})$ and therefore:
\begin{equation}
  \label{eq:exp_integ_adv2}
  \tilde{U}{}_{ij}(t)
  =
  \sum_{k,l}
  \exp\left(-t c_x\tilde{\mathbf{D}}^{(1)}_{\hpb{}x}\right){}_{ik}
  \exp\left(-t c_y\tilde{\mathbf{D}}^{(1)}_{\hpb{}y}\right){}_{jl}\,
  \tilde{U}{}_{kl}(0),
\end{equation}
where matrix exponentials we evaluate numerically based on \cite{almohy2010newscalingsquaring}.

In the fully-discrete case we make use of classical ERK4 which provides a time-integrator of formal order of accuracy $\mathcal{O}(\delta t^4)$ \cite{butcher2008numerical,hairer2010solving}. In the context of hyperbolic evolution for a spatial discretization of $\mathcal{O}(\delta x^{2r-2})$ it is common to add dissipation involving derivatives of degree $2r$ through the standard Kreiss-Oliger prescription \cite{gustafsson2013timedependentproblemsdifference} on each field component and in each spatial direction:
\begin{align}
  \label{eq:dissdef}
  Q{}^{(2r)}
  &:=
  -\frac{(-1)^r}{2^{2r}}
  \sigma (\delta x)^{2r-1} (D_+)^r (D_-)^r, &
  D{}_\pm[f_i] := \pm \frac{1}{\delta x}(f_{i\pm1} - f_i);
\end{align}
where $\sigma\geq0$ regulates the strength of the added dissipation. The derivative product $D{}^{(2r)}:=(D_+)^r (D_-)^r$ may be evaluated through ${}^{(2r)}E_r$ of Tab.\ref{tab:diss_scheme}. The choice of order and degree is made such that in the case of non-linear hyperbolic PDE (as tested in \S\ref{ssec:z4c}) stability in appropriate norm may be demonstrated and attained for various classes of problems \cite{gustafsson2013timedependentproblemsdifference}. While our numerical experiments show that addition of dissipation does not appear strictly necessary for full discretization of Eq.\eqref{eq:tr2pde} for the $T$ investigated our purpose here is to ensure that conventions are consistently selected in the context of this simple problem.

\begin{figure}[htbp]
  \subfloat[\label{subfig:adv2_polar_a}]{%
    \includegraphics[width=0.49\textwidth]{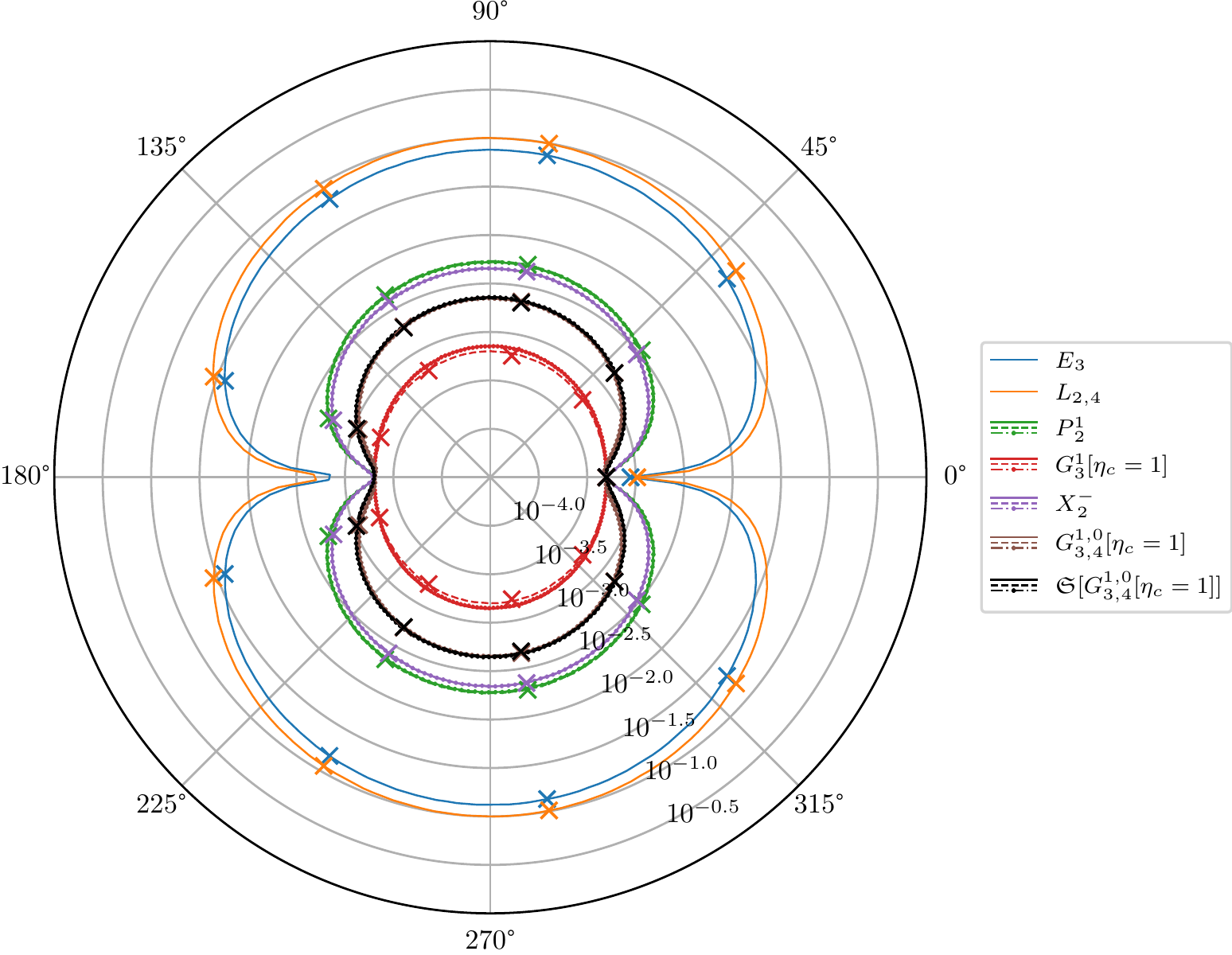}%
  }\hfill
  \subfloat[\label{subfig:adv2_polar_b}]{%
    \includegraphics[width=0.49\textwidth]{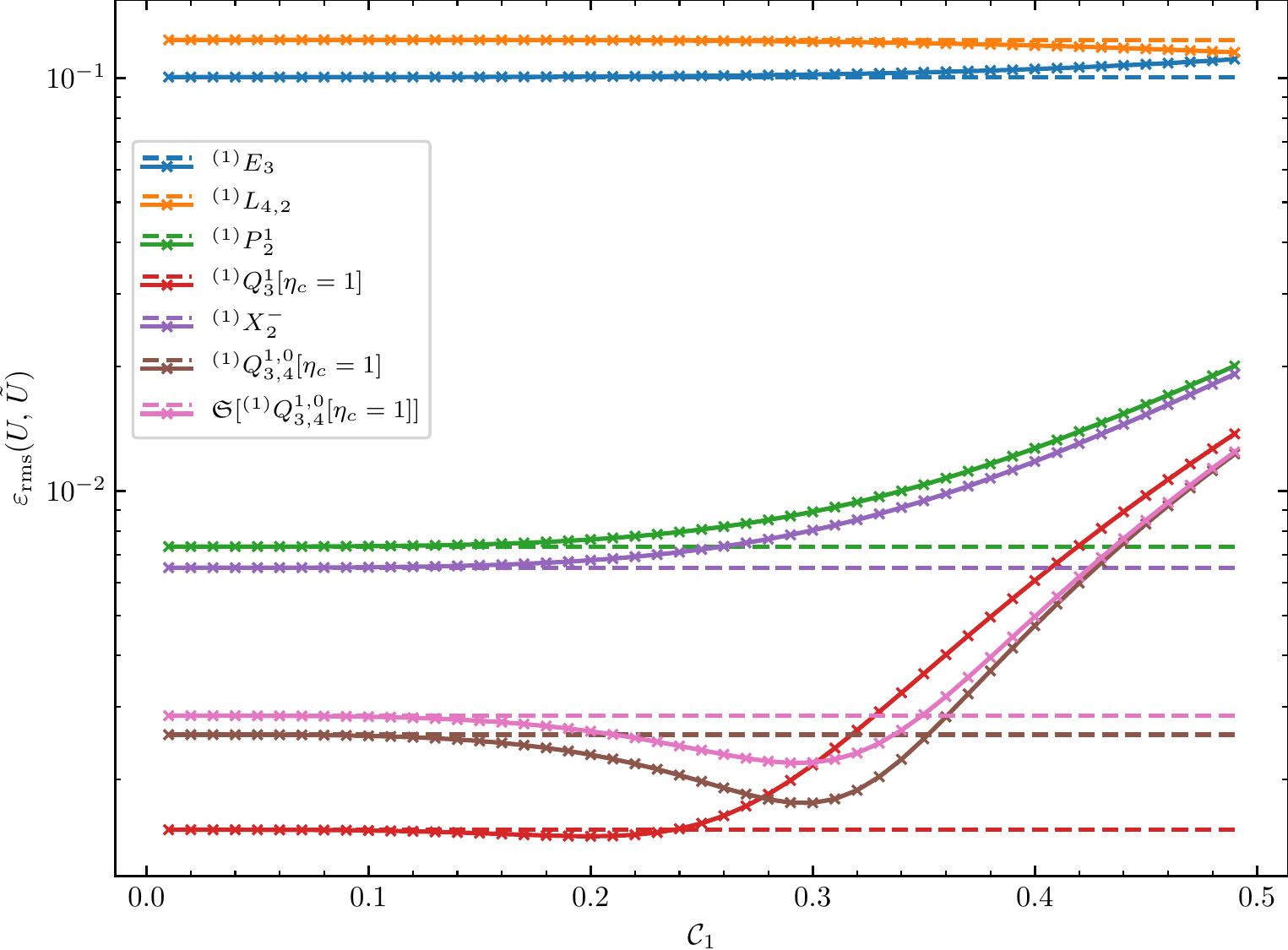}%
  }\hfill
  \caption{
    Error associated with numerical solution of Eq.\eqref{eq:tr2pde} based on a variety of derivative approximants and temporal integration schemes. In both sub-figures $(N_{M_x},\,N_{M_y})=(192,\,192)$ is selected together with a final evolution time of $T=200\pi$. During domain-decomposition sub-domains are selected with $(N_{B_x},\,N_{B_y})=(16,\,16)$ and $N_g=4$.
    In (a) $\Vert U_{ij}-\tilde{U}_{ij}\Vert_\infty$ is shown as a function of the propagation angle $\varphi$ selected by choice of $\mathbf{c}$ at time $T$.     Schemes with solid lines in the legend have $\tilde{U}$ constructed according to exponential integration on a single domain. Dashed lines indicate domain-decomposition with closure relation applied (\S\ref{ssec:domdec_drp}); dot-dashed indicate a single iteration on the closure relation ($i=1$ in Eq.\eqref{eq:iter_scheme}). Crosses show solution to full discretization of Eq.\eqref{eq:sd2pde} as based on ERK4 with $\mathcal{C}_1=1/10$. We find error for all approaches to be compatible. In (b) the resulting RMS error at time $T$ for ERK4 evolution as the choice of CFL is varied is shown in lines with crosses. Dashed lines indicate the resulting error from exponential integration. Kreiss-Oliger dissipation has been added with $\sigma=0.04$ based on $Q{}^{(8)}$ of Eq.\eqref{eq:dissdef} and $E{}^{(8)}_{\hpb{}4}$ of Tab.\ref{tab:diss_scheme}.
    See text for further discussion.
  }\label{fig:adv2_polar}
\end{figure}

For initial conditions we form $u(x,\,y)=g(x)h(y)$ with $g$ is defined in Eq.\eqref{eq:test_grid_conv_g} and parameters selected as $(A,\,x_0,\,N_1,\,S)=(5/100,\,2/10,\,2,\,1)$ whereas $h$ is defined in Eq.\eqref{eq:test_grid_conv_h} and we take $(S,\,x_0,\,c_1,\,c_2,\,N_1,\,N_2,\,\phi_S)=(15/10,\,\pi,\,12/10,\,-1,\,2,\,7,\,1)$. As the numerical solutions approximate Eq.\eqref{eq:tr2pde} we compare the sampled, advected initial condition $U_{ij}(T):=u((x_i-Tc_x)\mod(2\pi),\,(y_j-Tc_y)\mod(2\pi))$ pointwise to $\tilde{U}_{ij}(T)$ at $T=200\pi$ for the methods described. Given fixed spatial resolution taken to be uniform in $x$ and $y$ directions Fig.\ref{fig:adv2_polar} depicts the error associated with the result of exponential integration (Eq.\eqref{eq:exp_integ_adv2}) and similarly that of ERK4 based solution for the single domain and domain-decomposed approaches at a variety of angles $\varphi$ for $\mathbf{c}$.
From Fig.\ref{subfig:adv2_polar_a} we see that schemes tend to have a more pronounced error as the direction of propagation tends towards alignment along the $y$ axis. This can be understood from the comparing the spectral content of $g(x)$ and $h(y)$. We find that $|\hat{g}_n| / \max_m |\hat{g}_m| \sim 10^{-16}$ for $|n|\simeq 30$ whereas for $|\hat{h}_n| / \max_m |\hat{g}_m|\simeq 10^{-7}$ at $|n|\simeq 30$ and consequently higher resolution would be required in the $y$ direction to achieve a more uniform error as $\varphi$ is varied. Crucially we observe that biased schemes remain stable (and indeed error is symmetric under reflection about axes) when upwinding in accordance with Eq.\eqref{eq:biasderapproxadv}. As can be seen in Fig.\ref{subfig:adv2_polar_b} by judiciously selecting CFL when evolving with ERK4 the temporal error may be made comparable to that of the spatial discretization and consequently $\varepsilon_{\mathrm{rms}}$ converges to the error associated with exponential integration (EI) of the semi-discrete problem indicating consistency between approaches. Interestingly we find that for some schemes an intermediate regime of $\mathcal{C}_1$ exists where ERK4 outperforms EI. One possible explanation for this is that while in this work we exclusively focus on tuning modified wavenumber for derivative approximants based on arguments involving semi-discretization; ERK schemes propagating the fully discrete system may further modify dispersion and dissipation \cite{hu1996lowdissipationlowdispersionrunge}. For the present setup we find that utilizing implicit schemes for specification of spatial derivatives reduces maximum error by a factor of $13.6$ for the scheme ${}^{(1)}P{}^1_2$ or $65.9$ for ${}^{(1)}Q{}^1_3[\eta_c=1]$ when compared with the standard explicit finite-difference approach ${}^{(1)}E{}_3$. This maintains the trend observed in the grid convergence study of \S\ref{ssec:domdec_iter}.

\subsection{Shifted wave equation}
\label{ssec:shifted_wave}
As another example we consider the shifted wave equation \cite{chirvasa2010finitedifferencemethods}. This will also allow us to test the previously introduced second degree implicit derivative schemes. Suppose $(\mathcal{M},\,g)$ is a Lorentzian manifold endowed with metric $g$. For simplicity, suppose $g{}^{00}=-1$ and introduce the projector\footnote{Geometric quantities may feature space-time $a,\,b,\,\dots$ or spatial $i,\,j,\,\dots$ indices respectively. Juxtaposition of an index that appears raised and lowered implies summation on that index.} $\gamma{}^{ij}=g{}^{ij} + \beta{}^i\beta{}^j$ where $\beta{}^i$ corresponds to a shift vector. The homogeneous, scalar, wave equation $\Box[\Phi]=0$ may then be written as:
\begin{equation}\label{eq:toyWaveEqn}
  \partial{}_{tt}[\Phi] = 2 \beta{}^i \partial_i[\partial{}_t[\Phi]] + \left(\gamma{}^{ij}-\beta{}^i\beta{}^j\right)\partial{}_i[\partial{}_j[\Phi]].
\end{equation}
We reduce Eq.\eqref{eq:toyWaveEqn} to a first order in time system by defining the auxiliary field:
\begin{equation}
  \mathrm{K}:=\partial_t[\Phi] - \beta{}^j \partial{}_j[\Phi].
\end{equation}
The shifted wave equation can now be written as:
\begin{equation}\label{eq:toyShiftWaveEqnSys}
  \partial_t
  \begin{bmatrix}
    \Phi \\
    \mathrm{K}
  \end{bmatrix}
  =
  \begin{bmatrix}
    \beta{}^j \partial{}_j[\Phi] + \mathrm{K}\\
    \beta{}^j\partial{}_j[\mathrm{K}] + \gamma{}^{ij}\partial{}_i[\partial{}_j[\Phi]]
  \end{bmatrix}
  =
  \begin{bmatrix}
    \beta{}^j \partial{}_j                 & 1\\
    \gamma{}^{ij}\partial{}_i\partial{}_j  & \beta{}^j\partial{}_j
  \end{bmatrix}
  \begin{bmatrix}
    \Phi \\
    \mathrm{K}
  \end{bmatrix};
\end{equation}
subject to supplementation with suitable initial and boundary conditions. As our goal here is to provide a dynamical test of the second degree derivative schemes we simplify the problem by working in $(1+1)$ dimensions, impose spatial periodicity, and freeze $\beta{}^i$ to be constant with $\gamma{}^{ij}$ selected as a flat background. Under these assumptions we thus seek smooth $\Phi:\Omega_T\rightarrow\mathbb{R}$ and $\mathrm{K}:\Omega_T\rightarrow\mathbb{R}$ satisfying:
\begin{equation}\label{eq:wav2prot}
  \begin{cases}
    \partial_t
    \begin{bmatrix}
      \Phi \\
      \mathrm{K}
    \end{bmatrix}
    =
    \begin{bmatrix}
      \beta \partial{}_x  & 1 \\
      \partial{}_x^2      & \beta \partial{}_x
    \end{bmatrix}
    \begin{bmatrix}
      \Phi \\
      \mathrm{K}
    \end{bmatrix}
    , & \Omega_T; \\
    \Phi(t,\,0) = \Phi(t,\,2\pi),
    \quad
    \mathrm{K}(t,\,0) = \mathrm{K}(t,\,2\pi), & t\in[0,\,T];\\
        \Phi(0,\,x) = \phi(x),
    \quad
    \mathrm{K}(0,\,x) = \mathrm{k}(x), & \{t=0\};
      \end{cases}
\end{equation}
where $\beta\in\mathbb{R}$. In particular $\beta=0$ reduces Eq.\eqref{eq:wav2prot} to the standard un-shifted case. For the single domain, uniformly spaced, spatial discretization with $N_M$ samples on $\Omega$ we define:
\begin{align}
  \label{eq:sd_wave_ops}
  \tilde{\mathbf{U}} &:=
  \begin{bmatrix}
    \tilde{\boldsymbol{\Phi}} \\
    \tilde{\mathbf{K}}
  \end{bmatrix}, &
  \mathbf{L}&:=
  \begin{bmatrix}
    \beta \tilde{\mathbf{D}}^{(1)} & \mathbb{I} \\
    \tilde{\mathbf{D}}^{(2)} & \beta \tilde{\mathbf{D}}^{(1)}
  \end{bmatrix};
\end{align}
such that the semi-discrete formulation of Eq.\eqref{eq:wav2prot} is given by:
\begin{equation}
  \label{eq:sd_wave}
  \begin{cases}
    \frac{\mathrm{d}}{\mathrm{d} t}
    \tilde{\mathbf{U}}
    =
    \mathbf{L}
    \tilde{\mathbf{U}}, &
    t\in[0,\,T];\\
      \tilde{\boldsymbol{\Phi}}
    {}_i = \phi(x{}_i),
    \quad
      \tilde{\mathbf{K}}
    {}_i = \mathrm{k}(x{}_i), & \{t=0\};
  \end{cases}
\end{equation}
where enforcement of the periodic boundary conditions is considered embedded within the discretized derivatives appearing in $\mathbf{L}$ and initial conditions are compatible with periodicity. If biased derivative approximation schemes are selected in Eq.\eqref{eq:sd_wave_ops} we consider suitable upwinding based on Eq.\eqref{eq:biasderapproxadv} with $c$ replaced by $\beta$. For simplicity in this section we assume cell-centered sampling. As in the case of the advection problem we may view the formal solution to the semi-discrete problem as provided through matrix exponentiation. In passing to the domain-decomposed problem over $\Omega=\sqcup_{I}\Omega_I$ with $N_S$ sub-domains the general form of Eq.\eqref{eq:sd_wave_ops} and Eq.\eqref{eq:sd_wave} remains unchanged if discretized derivative operators are understood in the sense of Eq.\eqref{eq:iter_scheme}.

\begin{figure}[htbp]
  \subfloat[\label{subfig:eig_stab_a}]{%
    \includegraphics[width=0.49\textwidth]{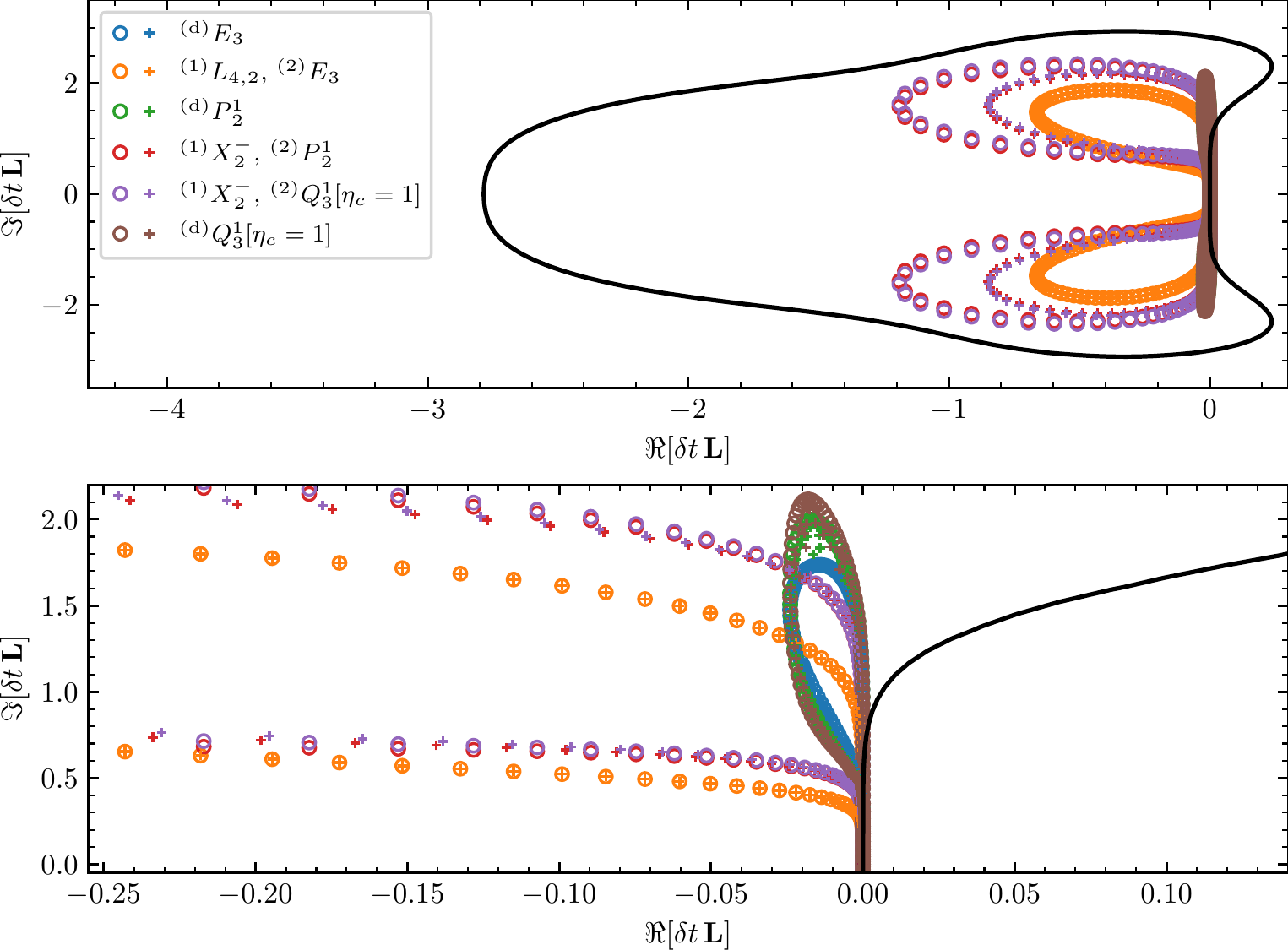}%
  }\hfill
  \subfloat[\label{subfig:eig_stab_b}]{%
    \includegraphics[width=0.49\textwidth]{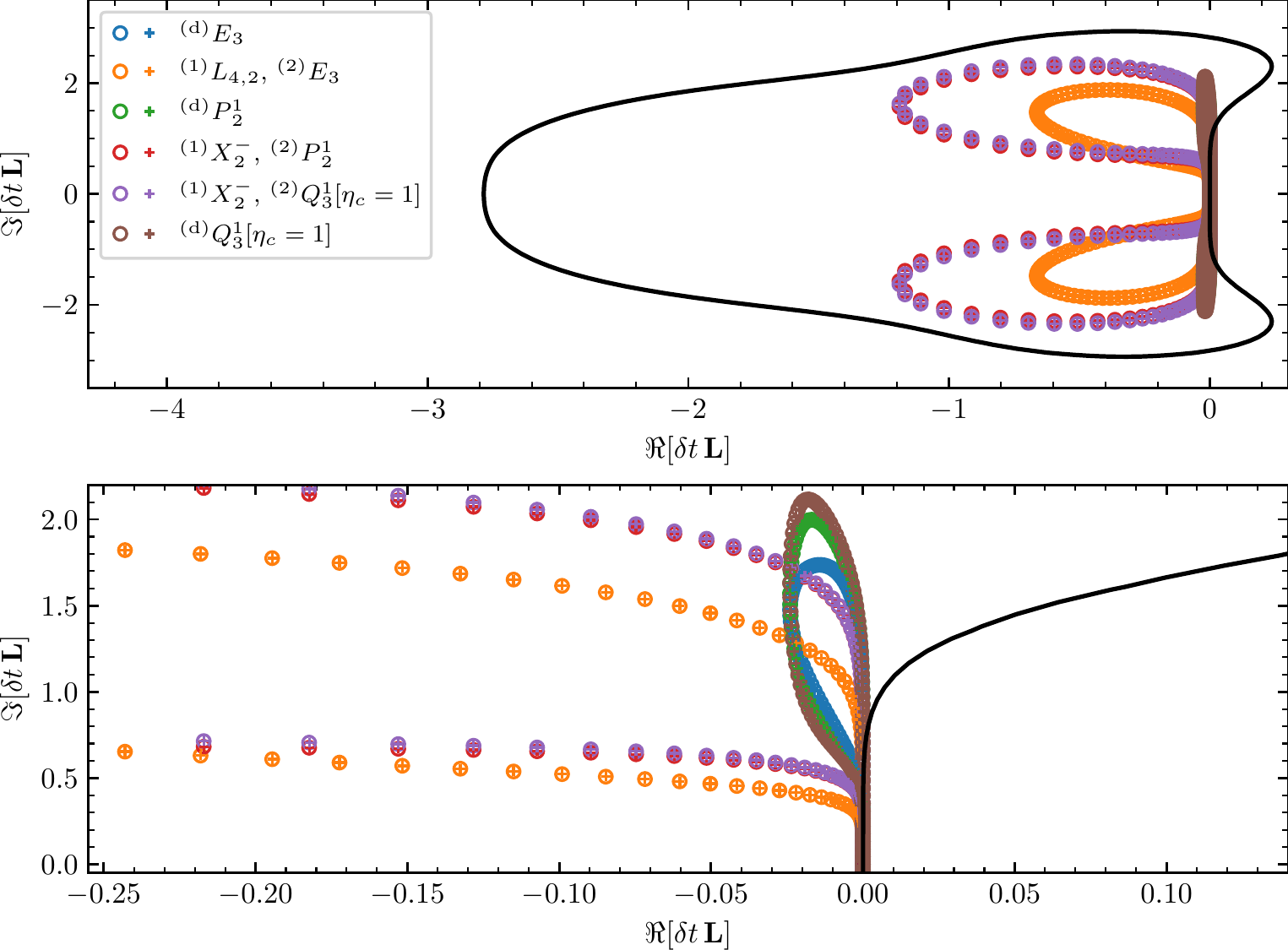}%
  }\hfill
  \caption{
    Eigenvalues of the discretized linear operator representing the shifted wave equation of Eq.\eqref{eq:sd_wave_ops} and Eq.\eqref{eq:sd_wave} with $\beta=1/2$ scaled by $\delta t$ as determined from CFL of $\mathcal{C}_1=6/10$ for a variety of schemes. The interior of the back curve depicts the absolute stability region for classical ERK4 \cite{butcher2008numerical,hairer2010solving}. In (a,b) open circles denote $\mathrm{spec}(\delta t\,\mathbf{L})$ which corresponds to the single domain formulation whereas $\mathrm{spec}\left({}^{[i]}\mathbf{L}\right)$ is shown in crosses. In (a) $i=0$ is taken with closures for domain-decomposed derivative schemes applied as described in \S\ref{ssec:domdec_drp}. It is clear that the closure induces a deformation of the spectrum for implicit schemes. In (b) a single iteration is performed ($i=1$) according to the prescription of \S\ref{ssec:domdec_iter} with $\mathrm{spec}\left({}^{[1]}\mathbf{L}\right)$ evaluated based on Eq.\eqref{eq:iter_scheme}. A single iteration on the closure leads to good agreement between spectra of global and domain decomposed operators (cf.~Fig.\ref{fig:eig_stab_advection})
  }\label{fig:eig_stab_wave}
\end{figure}

Full-discretization with $[0,\,T]$ sampled with uniform time-steps $\delta t$ entails that the eigenvalues $\lambda_n \in \mathrm{spec}(\delta t{}^{[i]}\mathbf{L})$ are to be assessed. However $\mathbf{L}$ of Eq.\eqref{eq:sd_wave_ops} is non-normal and satisfying $\Re\left[\lambda_n \right]\leq 0$ for all $n$ together with containment within the ERK4 stability region only provide a necessary condition for stability. On the other hand it is known \cite{Calabrese:2005ft,chirvasa2010finitedifferencemethods} that for this system fully-discrete stability can be established for explicit finite differencing based on a modified $\mathrm{L}^2$ norm involving additional derivative terms. We do not seek to extend this analytical result here for implicit derivative schemes but rather take it as a guide. In order to gain insight on the properties of the implicit derivative scheme closures and hybrid iteration for this system featuring multiple derivative degrees we select $N_M=128$, $N_B=16$, and $N_S=8$ and investigate $\mathrm{spec}(\delta t{}^{[i]}\mathbf{L})$ in Fig.\ref{fig:eig_stab_wave}.
As in the case of the advection problem of \S\ref{ssec:wave_eqn} we find that the spectrum of $\delta t\,{}^{[0]}\mathbf{L}$ is deformed when compared with the single domain approach. However as can be seen in Fig.\ref{subfig:eig_stab_b} this can be mitigated through the use of a single hybrid iteration.

\begin{figure}[htbp]
  \subfloat[\label{subfig:wav_sp_conv_a}]{%
    \includegraphics[width=0.49\textwidth]{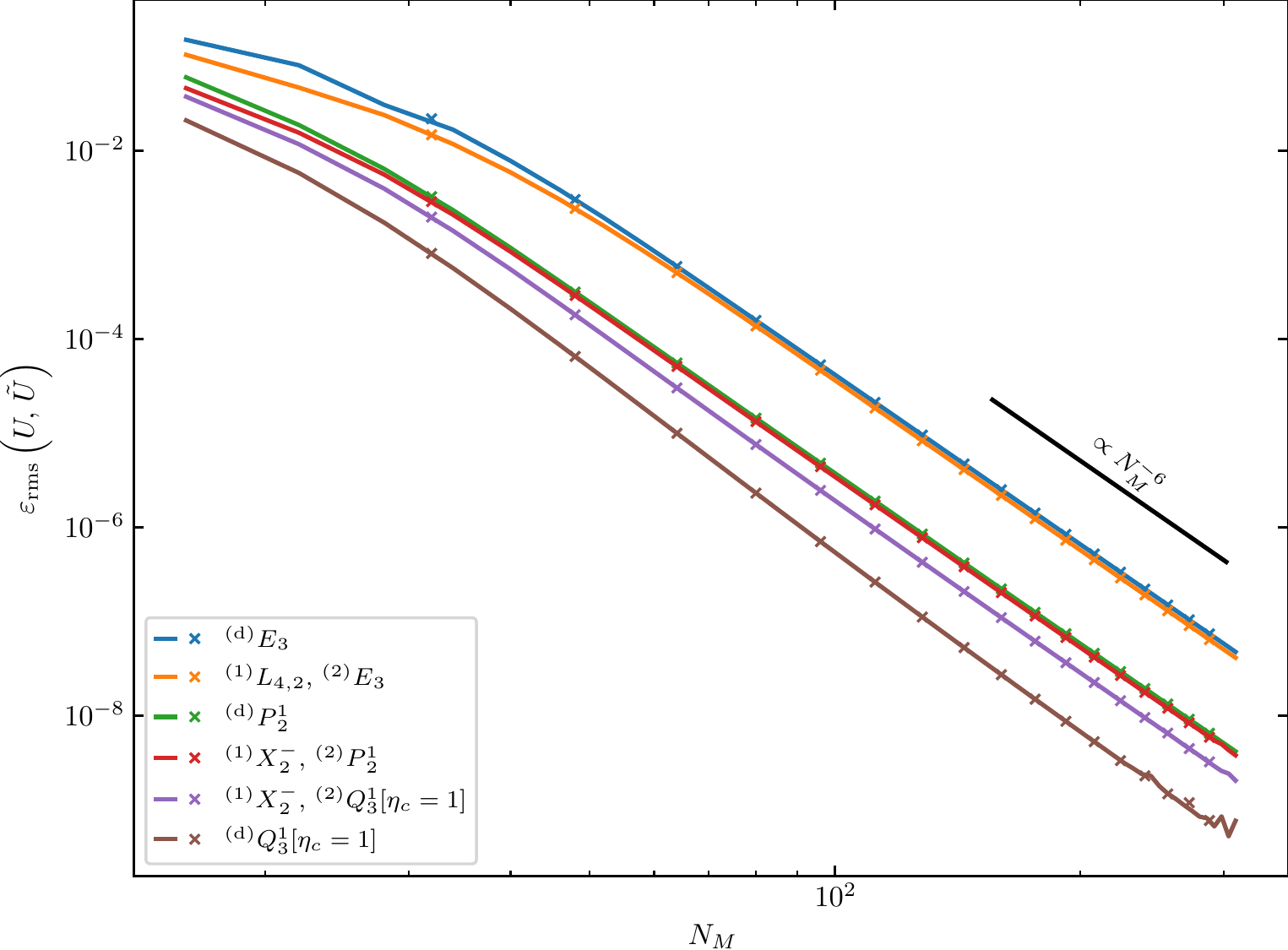}%
  }\hfill
  \subfloat[\label{subfig:wav_sp_conv_b}]{%
    \includegraphics[width=0.49\textwidth]{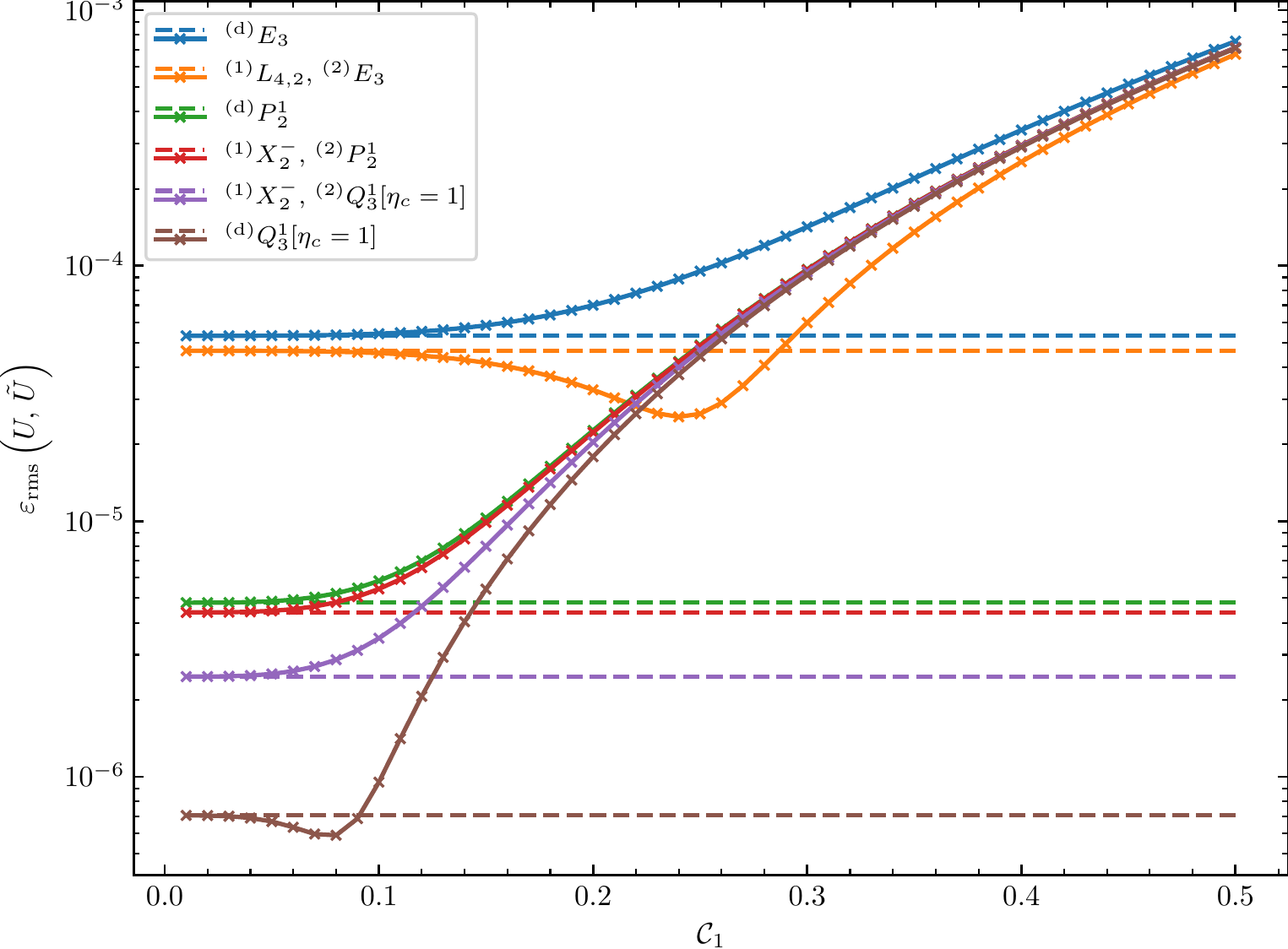}%
  }\hfill
  \caption{
    Convergence properties of numerical solutions to the periodic, shifted wave equation Eq\eqref{eq:toyShiftWaveEqnSys} subject to initial conditions of Eq.\eqref{eq:ini_wav}.  We take $T=200\pi$ which corresponds to $100$ crossing-times and examine associated error as compared to the initial condition. In (a) exponential integration of the semi-discretized system of Eq.\eqref{eq:sd_wave_ops} and Eq.\eqref{eq:sd_wave} utilizing a single domain (solid lines) and under domain-decomposition with $N_B=16$ (crosses). A single hybrid iteration is taken in the latter to close implicit derivative approximants. We find excellent agreement between the two approaches. In (b) $N_M=96$, $N_B=16$, and $N_S=6$ are fixed. In solid lines with crosses RMS error associated with ERK4 is investigated as CFL (and consequently $\delta t$) is swept; dashed lines depict RMS error from exponential integration. Kreiss-Oliger dissipation has been added with $\sigma=0.04$ based on $Q{}^{(8)}$ of Eq.\eqref{eq:dissdef} and $E{}^{(8)}_{\hpb{}4}$ of Tab.\ref{tab:diss_scheme}. See text for further discussion.
  }\label{fig:wav_sp_conv}
\end{figure}

We now consider propagating the initial condition:
\begin{align}\label{eq:ini_wav}
  \phi(x) &= \exp\left(
    -(2\pi \tau)^{-2}\sin^2\left(
      \frac{x}{2} - \frac{\pi}{2}
    \right)
  \right), &
  \mathrm{k}(x) &= a\,\partial_x[\phi(x)];
\end{align}
with $\beta=1/2$, $\tau=8/100$, and $a=1$, which describes a left-ward propagating Gaussian. With regard to the continuum problem Eq.\eqref{eq:wav2prot} after a crossing time of $2\pi$ the profile described by Eq.\eqref{eq:ini_wav} is reconstructed. This feature must appear in the discretized solution if it is accurate and consequently allows us to characterize error stroboscopically at integer multiples of the crossing time through direct comparison with the initial profile. We perform convergence testing based on solution to the semi-discrete problem posed in single and domain-decomposed form together with verification of convergence of the latter in the fully-discretized context with ERK4 in Fig.\ref{fig:wav_sp_conv}.
As is clear from Fig.\ref{subfig:wav_sp_conv_a} utilizing implicit schemes for specification of spatial derivatives at $N_M=128$ can reduce maximum error by a factor of $11.2$ for the scheme ${}^{(\mathrm{d})}P{}^1_2$ or $85.7$ for ${}^{(\mathrm{d})}Q{}^1_3[\eta_c=1]$ when compared with the standard explicit finite-difference approach ${}^{(\mathrm{d})}E{}_3$. From Fig.\ref{subfig:wav_sp_conv_b} we find agreement with observations made in \cite{chirvasa2010finitedifferencemethods} in that for non-zero $\beta$ there exists a CFL regime where utilizing a combination of ${}^{(1)}L_{4,2}$ and ${}^{(2)}E_3$ reduces error when compared to full centering (i.e.~using instead ${}^{(\mathrm{d})}E_3$ for both derivative degrees).

\subsection{Z4c: system description}
\label{ssec:z4c}

A primary target application of this work is numerical solution of the Cauchy problem for the Einstein field equations (EFE). For problems in the absence of symmetries, this requires considerable computational infrastructure and highly performant code. We therefore utilize the octree-based, adaptive mesh refinement (AMR) infrastructure offered by \GRAthena{} \cite{Daszuta:2021ecf} where hybrid MPI-OMP provides parallelism at scale. The generalized finite-difference schemes investigated in prior sections we have coupled via a header-only, templated \Cpp{} library. Before describing our numerical tests we briefly recall some formulation details. In the context of the Cauchy problem for the EFE the conformal formulations of BSSNOK \cite{Nakamura:1987zz,Shibata:1995we,Baumgarte:1998te} and \z4c{} \cite{Bernuzzi:2009ex,Ruiz:2010qj,Weyhausen:2011cg,Hilditch:2012fp}, and the generalized harmonic gauge (GHG) approach \cite{Friedrich:1985,Pretorius:2005gq,Lindblom:2005qh} have found success in simulation of a wide variety of physical problems. The former two formulations leverage consideration of a globally hyperbolic space-time $(\mathcal{M},\,g)$ as foliated by a family of non-intersecting spatial slices; initial data is provided on a selected slice $\Sigma_{t^\star}$ of the foliation and well-posed evolution equations compatible with the EFE must be prescribed and solved\footnote{For an introductory account of geometric, analytical, and numerical considerations see the textbooks \cite{Gourgoulhon:2007ue,alcubierre2008introduction,Baumgarte:2010,Shibata:2016}.} over $t\in[t^*,\,T]$. In particular the \z4{} formulation \cite{Bona:2003fj} augments the EFE through introduction of an auxiliary, dynamical vector field $Z{}^a$ and first-order covariant derivatives thereof. This results in evolution equations involving the variables
$\left(%
\gamma{}_{ij},\,K{}_{ij},\,\Theta,\,\check{Z}_i%
\right)$
where $\gamma{}_{ij}$ and $K{}_{ij}$ are the induced metric and extrinsic curvature associated with $\Sigma_t$ respectively whereas $\Theta$ and $\check{Z}{}_i$ are normal and spatial projections of $Z{}^a$. Furthermore Hamiltonian, momentum, and auxiliary vector constraints must also be satisfied $\mathcal{C}_U:=(\mathcal{H},\,\mathcal{M}_i,\,Z_a)=0$ such that a numerical space-time is faithful to a solution of the standard EFE. Importantly for a space-time without boundary if $\mathcal{C}_U=0$ for some element of the foliation (e.g.~$\Sigma_{t^*}$) then analytically this property extends for all $t$ \citep{Bona:2003fj}. This strategy leads to a framework wherein certain strengths of BSSNOK and GHG can be blended \cite{bona2010actionprinciplenumericalrelativity}. Isolating a spatial conformal degree of freedom leads to \z4c{} which is a conformal, free-evolution scheme featuring prescribable constraint damping. One defines\footnote{Existence of a global chart with Cartesian coordinatization $x{}^i\dot{=}(x^1,\,x^2,\,x^3)=(x,\,y,\,z)$ is assumed throughout.}:
\begin{align}
  \label{eq:confxform}
  \tilde{\gamma}{}_{ij} :=& \psi{}^{-4}\gamma{}_{ij}, &
  \tilde{A}{}_{ij}:=& \psi^{-4}\Big(K{}_{ij} - \frac{1}{3}K\gamma{}_{ij}\Big);
\end{align}
with $K:=K{}_{ij}\gamma{}^{ij}$ and $\psi:=\gamma^{1/12}$ where $\gamma$ is the determinant of $\gamma{}_{ij}$. Define:
\begin{align}
  &\chi:=\gamma{}^{-1/3}, &&
  &\hat{K}:= K-2\Theta;\\
  &\tilde{\Gamma}{}^i := 2\tilde{\gamma}{}^{ij} \check{Z}{}_j
  + \tilde{\gamma}{}^{ij}\tilde{\gamma}{}^{kl} \pd{}_l[\tilde{\gamma}{}_{jk}],&&
  &\defG{}^i := \tilde{\gamma}{}^{jk}\tilde{\Gamma}{}^i{}_{jk}.
  \label{eq:defn_transforms}
\end{align}
The \z4c{} system has dynamical variables
$\big(%
\chi,\,\tilde{\gamma}{}_{ij},\,\hat{K},\tilde{A}{}_{ij},\,\Theta,\,\tilde{\Gamma}{}^i%
\big)$ which in vacuum are governed by the \textit{evolution equations}:
\begin{equation}
  \label{eq:evo_chi}
  \pd{}_t[\chi] = \frac{2}{3}\chi
  \left(
    \alpha(\hat{K} + 2\Theta) - \partial_i[\beta{}^i]
  \right) + \beta{}^i\partial_i[\chi],
\end{equation}
\begin{equation}
  \label{eq:evo_conf_metr}
  \pd{}_t[\tilde{\gamma}{}_{ij}]
  =
  -2\alpha\tilde{A}{}_{ij} + \beta{}^k\partial{}_k[\tilde{\gamma}{}_{ij}]
  -\frac{2}{3}\tilde{\gamma}{}_{ij}\partial{}_k[\beta{}^k]
  +2\tilde{\gamma}{}_{k(i}\partial{}_{j)}[\beta{}^k],
\end{equation}
\begin{equation}
  \label{eq:evo_Khat}
  \pd{}_t[\hat{K}]
  =
  -\D{}^i[\D{}_i[\alpha]]
  + \alpha\left[
    \tilde{A}{}_{ij}\tilde{A}{}^{ij} + \frac{1}{3}(\hat{K}+2\Theta)^2
  \right]
  + \beta{}^i \pd{}_i[\hat{K}]
  +\alpha\kappa{}_1(1-\kappa{}_2)\Theta,
\end{equation}
\begin{equation}
  \label{eq:evo_conf_Atil}
  \pd{}_t[\tilde{A}{}_{ij}]
  =
  \chi\left\{-\D{}_i[\D{}_j[\alpha]] + R{}_{ij}\right\}^{\mathrm{tf}}
  +\alpha[
    (\hat{K} + 2\Theta)\tilde{A}{}_{ij}
    -2\tilde{A}{}^k{}_i\tilde{A}{}_{kj}
  ] + \beta{}^k \pd{}_k[\tilde{A}_{ij}]
  + 2\tilde{A}{}_{k(i} \pd{}_{j)}[\beta{}^k]
  - \frac{2}{3}\tilde{A}{}_{ij}\pd{}_k[\beta{}^k],
\end{equation}
\begin{equation}
  \label{eq:evo_Theta}
  \pd{}_t[\Theta]
  =
  \frac{\alpha}{2}
  \left[
    \tilde{\mathcal{H}} -2\kappa{}_1(2+\kappa{}_2) \Theta
  \right] + \beta{}^i \pd{}_i[\Theta],
\end{equation}
\begin{align}
  \nonumber
  \pd{}_t[\tilde{\Gamma}{}^i] =&
  -2\tilde{A}{}^{ij} \pd{}_j[\alpha]
  +2\alpha
  \Big[
    \tilde{\Gamma}{}^i{}_{jk} \tilde{A}{}^{jk}
    -\frac{3}{2}\tilde{A}{}^{ij} \pd{}_j[\ln(\chi)]
  -\kappa{}_1(\tilde{\Gamma}{}^i-\defG{}^i)
  -\frac{1}{3}\tilde{\gamma}{}^{ij}\pd{}_j[2\hat{K}+\Theta]
  \Big]\\
  &
  +\tilde{\gamma}{}^{jk}\pd{}_k[\pd{}_j[\beta{}^i]]
  +\frac{1}{3}\tilde{\gamma}{}^{ij}\pd{}_j[\pd{}_k[\beta{}^k]]
  +\beta{}^j \pd{}_j[\tilde{\Gamma}{}^i]
  -\defG{}^j \pd{}_j[\beta{}^i]
  +\frac{2}{3}\defG{}^i \pd{}_j[\beta{}^j];
  \label{eq:evo_conf_GamTil}
\end{align}
where $\D{}_i$ is the covariant derivative compatible with $\gamma{}_{jk}$, $\kappa_1$ and $\kappa_2$ are constraint damping parameters, and in Eq.\eqref{eq:evo_conf_Atil} the trace-free (tf) operation is computed with respect to
$\gamma{}_{ij}$. The intrinsic curvature is split as $R{}_{ij} = \tilde{R}^\chi{}_{ij} + \tilde{R}{}_{ij}$ and utilizing the conformal connection $\tilde{\mathrm{D}}{}_i$ compatible with $\tilde{\gamma}{}_{jk}$ allows us to write:
\begin{equation}
  \label{eq:defn_RicTilChi}
  \tilde{R}{}^\chi{}_{ij}
  =
  \frac{1}{2\chi}\left[
    \tilde{\D}{}_i[\tilde{\D}{}_j[\chi]]
    +\tilde{\gamma}{}_{ij}\tilde{\D}{}^l[\tilde{\D}{}_l[\chi]]
    -\frac{1}{2\chi}\tilde{\D}{}_i[\chi]\tilde{\D}{}_j[\chi]
  \right]
  -\frac{3}{4\chi^2} \tilde{\D}{}^l[\chi]\tilde{\D}{}_l[\chi] \tilde{\gamma}{}_{ij},
\end{equation}
and:
\begin{equation}
  \label{eq:defn_RicTil}
  \tilde{R}{}_{ij}
  =
  -\frac{1}{2}\tilde{\gamma}{}^{lm}
  \pd{}_l[\pd{}_m[\tilde{\gamma}{}_{ij}]]
  +\tilde{\gamma}{}_{k(i}
  \pd{}_{j)}[\tilde{\Gamma}{}^k]
  +\defG{}^k\tilde{\Gamma}{}_{(ij)k}
  +\tilde{\gamma}{}^{lm}
  (
    2\tilde{\Gamma}{}^k{}_{l(i}\tilde{\Gamma}{}_{j)km}
    +\tilde{\Gamma}{}^k{}_{im}\tilde{\Gamma}{}_{klj}
  ).
\end{equation}
Evolved variables must satisfy the \textit{dynamical constraints} which in terms of transformed variables $(\tilde{\mathcal{H}},\,\tilde{\mathcal{M}}{}_i,\,\Theta,\,\check{Z}{}^i)$:
\begin{equation}
  \label{eq:constr_ham}
  \tilde{\mathcal{H}} := R - \tilde{A}{}_{ij} \tilde{A}{}^{ij}
  +\frac{2}{3}\big(
  \hat{K}+2\Theta
  \big)^2 = 0,
\end{equation}
\begin{equation}
  \tilde{\mathcal{M}}{}_j
  :=
  \tilde{\mathrm{D}}_i[\tilde{A}{}^{i}{}_j]
  - \frac{3}{2}\tilde{A}{}^i{}_j\partial{}_i[\ln(\chi)]
  -\frac{2}{3}\pd{}_j[\hat{K}+2\Theta] =0,
\end{equation}
\begin{align}
  \Theta=&0, &
  \check{Z}{}^i=& \tilde{\Gamma}{}^i - \defG{}^i =0.
\end{align}
The transformation of Eq.\eqref{eq:confxform} also implies the \textit{algebraic constraints} $\mathcal{C}_A:=\big(\ln(\tilde{\gamma}),\,\tilde{\gamma}{}^{ij} \tilde{A}{}_{ij}\big)=0$ which are enforced during a numerical evolution for consistency\footnote{In particular, coupling to the puncture gauge with enforcement of $\mathcal{C}_A=0$ results in a strongly hyperbolic and well-posed system \cite{Cao:2011fu,Bernuzzi:2010ty}. Consequently the fully-discrete evolution enforces this condition at each time sub-step.}.

The \z4c{} system must be further supplemented by gauge conditions where the lapse $\alpha$ and shift $\beta{}^i$ describe how the elements of the foliation piece together. In this work we make use of the moving puncture gauge which consists of the Bona-M\'asso lapse \cite{Bona:1994b} and the gamma-driver shift \cite{Alcubierre:2002kk}:
\begin{align}
  \label{eq:gaugeBGDRV}
  \partial_t[\alpha] &= -\mu{}_L \alpha^2 \hat{K} +
    \beta{}^i \partial_i[\alpha], &
  \partial_t[\beta{}^i] &= \mu{}_S \alpha^2 \tilde{\Gamma}{}^i
    -\eta \beta{}^i + \beta{}^j \partial{}_j[\beta{}^i],
\end{align}
where the $1+\log$ lapse variant is selected through $\mu{}_L=2/\alpha$ together with $\mu{}_S=1/\alpha{}^2$, and $\eta$ is a specifiable damping parameter. During a subset of numerical tests we also make use of the harmonic gauge condition which sets $\mu_L=1$ in the dynamical relation for $\alpha$ of Eq.\eqref{eq:gaugeBGDRV} and the shift evolution becomes:
\begin{equation}
  \label{eq:gaugeBharm}
  \partial_t[\beta{}^i]
  =
  \alpha^2\chi\left[
    \tilde{\Gamma}
    +\frac{1}{2} \tilde{\gamma}^{ij} \partial_j[\chi]
    -\tilde{\gamma}{}^{ij}\partial{}_j[\log(\alpha)]
  \right]
  + \beta{}^j \partial{}_j[\beta{}^i].
\end{equation}

Semi-discretization proceeds as in prior sections however a few remarks are in order. In the evolution equations fields to be sampled over a domain $\Omega$ are sampled at points assembled from a tensor product grid of $\Omega_{x^i}$. Points are of the form $p\in\{(x_{I_1},\,y_{I_2},\,z_{I_3})\,|\,x_{I_1}\in\Omega_x,\, y_{I_2}\in\Omega_y,\,z_{I_3}\in\Omega_z\}$ where $I_i$ are to be understood as grid indices for a given axis. Sampled fields thus carry suppressed grid indices e.g.~$\left.\chi\right|_p=\chi(x_{I_1},\,y_{I_2},\,z_{I_3})$ and similarly $\left.\tilde{\Gamma}^i\right|_p=\tilde{\Gamma}^i(x_{I_1},\,y_{I_2},\,x_{I_3})$. Derivatives are approximated according to:
\begin{equation}
  \left(
    \partial{}^{\mathrm{d}}_i[\tilde{\Gamma}{}^j]
  \right){}_{I_1 I_2 I_3}
  =
  \sum_K
  \left(
    \tilde{D}{}^{(\mathrm{d})}_{\hpb{}i}
  \right){}_{I_i K}
  \delta{}_{I_i K}
  \tilde{\Gamma}{}^j
  \left(
    x{}_{I_1},\,y{}_{I_2},\,z_{I_3}
  \right),
  \quad
  \left(
    x{}_{I_1},\,y{}_{I_2},\,z_{I_3}
  \right) \in \Omega_x \times \Omega_y \times \Omega_z.
\end{equation}
Some care is required with mixed partial derivatives such as $\partial{}_i[\partial{}_j[\cdot]]$ as they commute when applied to $C^2$ functions. Consequently we explicitly symmetrize the discrete approximants $\partial{}_i[\partial{}_j[\cdot]]\rightarrow\frac{1}{2}\left(\tilde{D}{}^{(1)}_{\hpb{}i}[\tilde{D}{}^{(1)}_{\hpb{}j}[\cdot]]+\tilde{D}{}^{(1)}_{\hpb{}j}[\tilde{D}{}^{(1)}_{\hpb{}i}[\cdot]]\right)$ for $i\neq j$ and explicitly replace $\partial_i[\partial_j[\cdot]]\rightarrow \tilde{D}{}^{(2)}_{\hpb{}j}[\cdot]$ for $i=j$. In the case of shift-advective terms $\beta{}^j \partial{}_j[\cdot]$ derivative approximants are treated as in Eq.\eqref{eq:biasderapproxadv} where the value of $c$ is replaced by the pointwise value of the relevant shift vector component sampled on the underlying grid. In passing to the fully-discrete setting the implementation within \GRA{} performs time-evolution using the $\4th$ order RK$4()4[2S]$ low-storage method of \cite{ketcheson2010rungekuttamethods}. To ensure numerical stability Kreiss-Oliger dissipation is incorporated according to Eq.\eqref{eq:dissdef} and is applied to each field component, in each spatial direction.

\subsection{Z4c: numerical tests - gauge wave evolution}
\label{ssec:z4c_tests}

Armed with the \z4c{} system (\S\ref{ssec:z4c}) our first goal is to ensure that coupling our header-only, templated, generalized finite-difference code to \GRAthena{} leads to successful evolution on simple test problems. For this we make use of suitably modified Apples with Apples (\AwA) test-beds \cite{Alcubierre:2003pc,Babiuc:2007vr,Daverio:2018tjf}. The intention here is to quantify solution quality on small scale idealized problems and to probe for any sources of potential instability that may have been introduced through modifying derivative approximants away from the well-known properties of standard finite-difference (FD) explored in \cite{Daszuta:2021ecf}. Furthermore it allows for evolution of \z4c{} while gradually bridging the complexity gap from linear propagation problems of \S\ref{ssec:wave_eqn} and \S\ref{ssec:shifted_wave} towards binary black hole merger discussed later. Due to the task-based infrastructure and sophisticated treatment of sub-domain communication as a first step, compact FD approximants computed by the code we have coupled to \GRAthena{} relies solely on the closures described in \S\ref{ssec:domdec_drp} and does \textit{not} feature the hybrid strategy procedure described in \S\ref{ssec:domdec_iter}.

The \AwA{} test-beds are specified for $\Sigma$ of $\mathbb{T}^3$ topology hence $\gamma{}_{ij}$ is considered as periodic in each spatial direction. The effective dynamics occur over one (or two) spatial dimensions depending on the details of the test. In these directions the grid is taken as $\mathcal{G}_\mathrm{VC}[\Omega_{x^i};\,a=-1/2,\,b=1/2]=\{-1/2+k/N_M\,|\, k\in\{0,\,\dots,\,N_M\}\}$ such that $\delta x = 1/N_M$. For verification of the full system the remaining direction(s) fix this spacing and take the sampling parameter as $4$. During domain-decomposition partitioning is performed as discussed previously over directions with effective dynamics where sampled sub-domains have $N_B=16$ and are extended by $N_g$ ghost points to facilitate communication and derivative stencil evaluation. Overall we set $N_M=\rho N_B$ with $\rho\in\mathbb{N}$ serving to adjust resolution during convergence tests as required. This choice is motivated by the formal spatial order $\mathcal{O}(\delta x^6)$ of the schemes we employ. For the tests presented here numerical evolution is performed over $t\in[0,\,1000]$ with CFL of $\mathcal{C}_1=1/10$ and we select constraint damping parameters $\kappa_1=2/100$ and $\kappa_2=0$ together with Kreiss-Oliger dissipation $\sigma=2/100$ unless otherwise stated.

The initial \AwA{} test we performed was that of robust stability. An initial spatial slice of Minkowski space-time and to each sampled grid point an independent uniform random value drawn from $(-A_\rho,\,A_\rho)$ with $A_\rho=10^{-10}/\rho^2$ is added. This choice of $A_\rho$ effectively linearizes the system. We utilized the moving puncture gauge of Eq.\eqref{eq:gaugeBGDRV} with initial conditions $\left.\alpha\right|_{t=0}=1$ and $\left.\beta{}^i\right|_{t=0}=0$. The shift-damping parameter is set as $\eta=2$. The dynamics are considered to be effectively one-dimensional. The quantity $\Vert \gamma{}_{ij}-\delta{}_{ij}\Vert_\infty$ together with constraints such as $\Vert \mathcal{H}\Vert_\infty$ were monitored over the course of a calculation. As in the case of finite-difference tests made in \cite{Daszuta:2021ecf} we found that when adopting a wide variety of combinations of compact stencils as selected in the toy problems of \S\ref{ssec:wave_eqn} and \S\ref{ssec:shifted_wave} leads to contraint quantities decaying towards a plateau in norm with values comparable to the FD case. This indicates that error associated with numerical evolution of the principal part of the \z4c{} system does not appear to induce spurious growth of unstable exponential modes when utilizing compact finite-difference (CFD) approximants. The second \AwA{} test is the linearized wave test. Effective one-dimensional dynamics are induced through $\gamma{}_{ij}\dot{=}\mathrm{diag}(1,\,1+H(t,\,x),\,1-H(t,\,x))$ where $H=A\sin(2\pi(x-t))$ together with $K{}_{yy}=-\frac{1}{2}\partial_t[H(t,\,x)]=-K_{zz}$ with remaining components zero. Gauge is chosen as in the robust stability test. An amplitude $A=10^{-8}$ forces non-linear terms to numerical round-off thus linearizing the \z4c{} system when numerical calculations are performed in double-precision arithmetic. While these choices lead to a numerical solution that is well described as a simple travelling (i.e.~advected) wave, the puncture gauge is not necessarily compatible with pure advection, and furthermore the initial data are constraint violating \cite{Cao:2011fu}. We thus focus instead on the gauge wave tests as they describe propagation of simple constraint satisfying data to the full non-linear \z4c{} system.

\begin{figure}[htbp]
  \subfloat[\label{subfig:z4c_awa_gw1_a}]{%
    \includegraphics[width=0.49\textwidth]{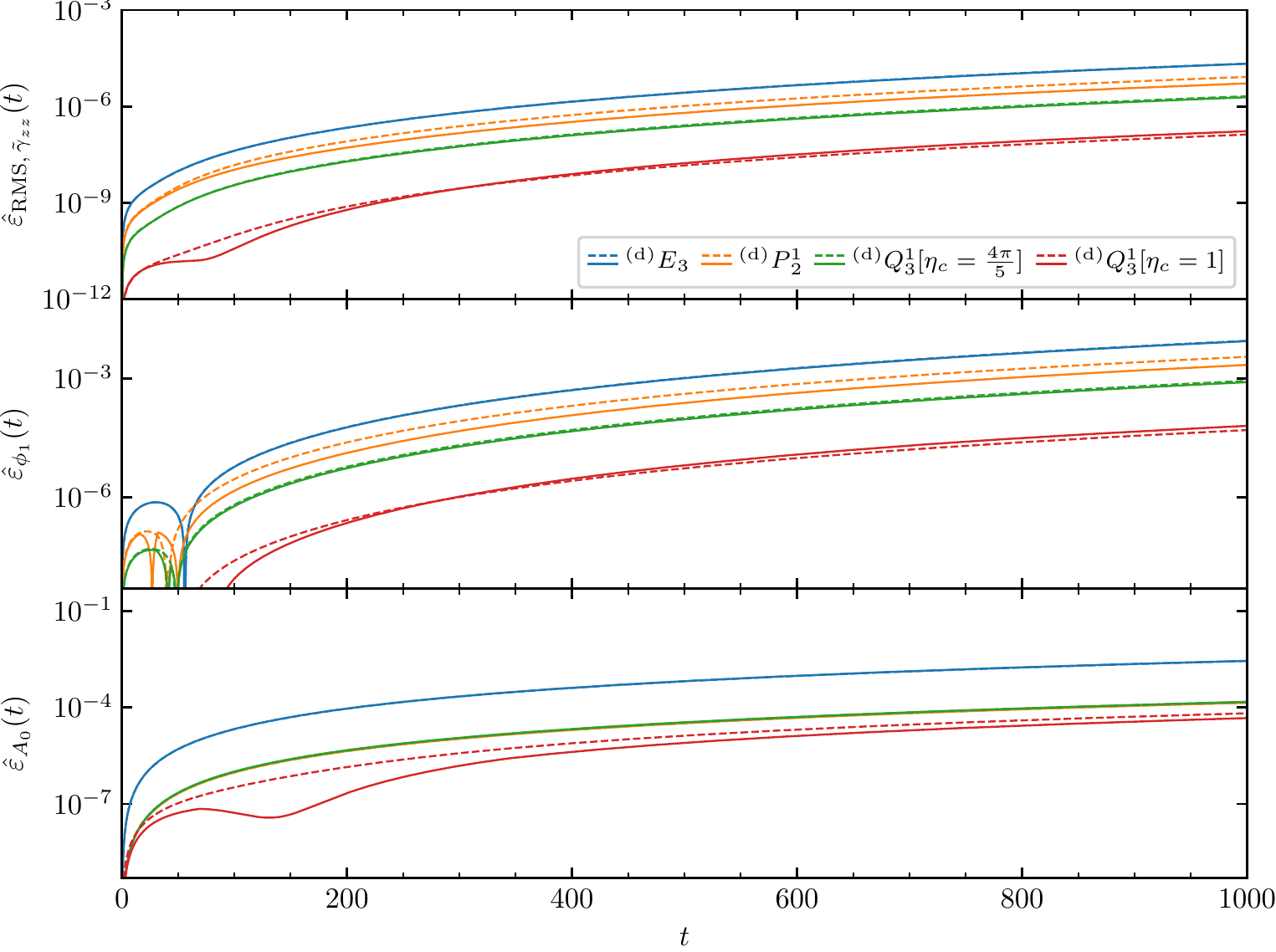}%
  }\hfill
  \subfloat[\label{subfig:z4c_awa_gw1_b}]{%
    \includegraphics[width=0.49\textwidth]{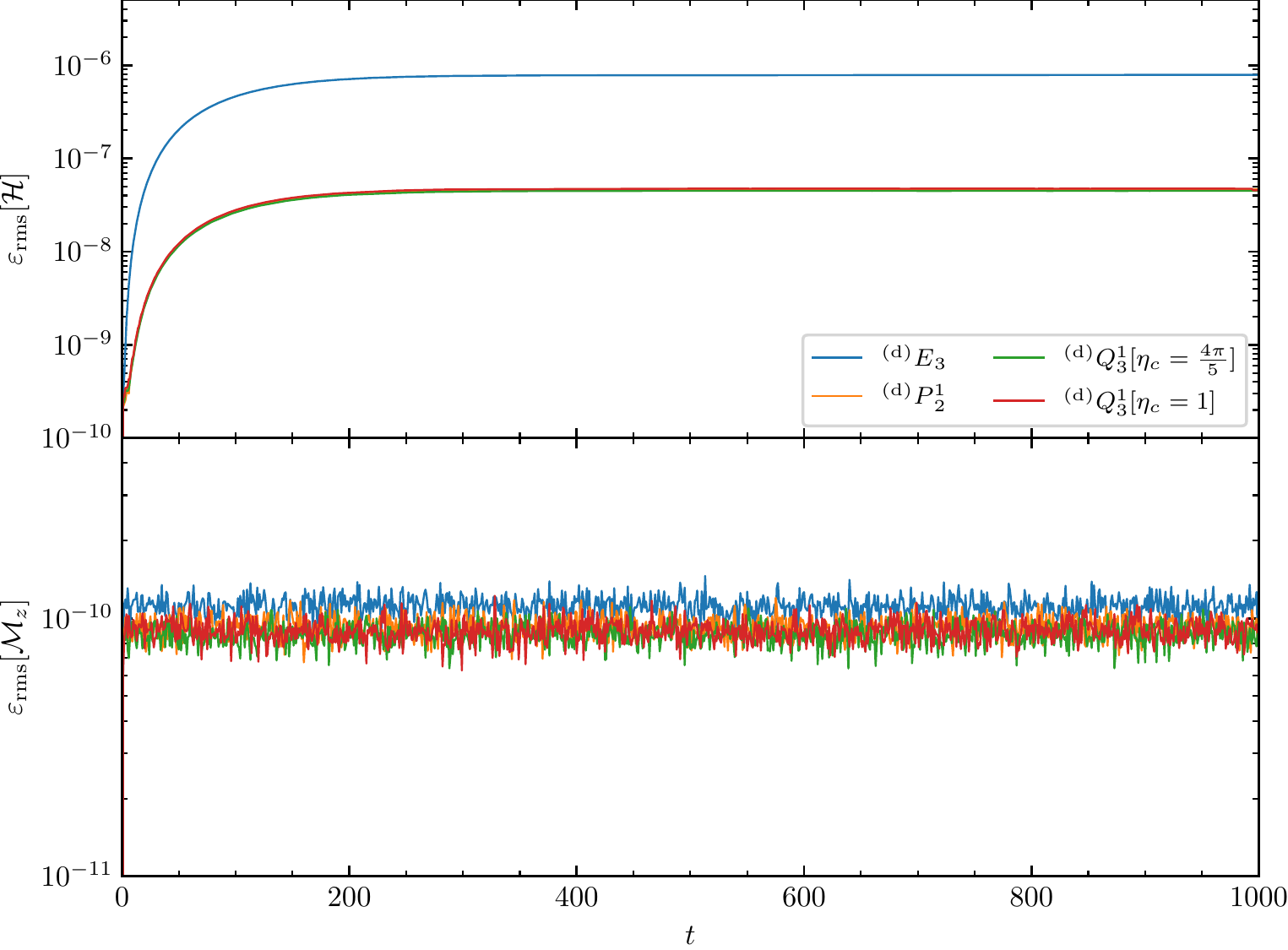}%
  }\hfill
  \caption{Error associated with aligned, unshifted gauge wave \AwA{} test with $(\rho_c,\,\rho_m,\,\rho_f)=(2,\,3,\,4)$ for a variety of derivative schemes of formal order of accuracy $\mathcal{O}(\delta x^6)$. In (a) we inspect the RMS error associated with the evolved $\tilde{\gamma}{}_{zz}$ (top), the phase error (middle), and the offset of the numerical profile relative to the amplitude (bottom). Error has been rescaled according to design order of the scheme with $Q_6$ of Eq.\eqref{eq:conv_fact} and utilizing the normalization of Eq.\eqref{eq:rescnormerr} with absolute error scale corresponding to evolution with $\rho_\star=\rho_f$. In (b) the RMS values of the Hamiltonian constraint $\mathcal{H}$ and $z$ component of the momentum constraint $\mathcal{M}_z$ are shown for $\rho_f$. See text for discussion.}
  \label{fig:z4c_awa_gw1}
\end{figure}

Consider the \AwA{} aligned, unshifted, gauge wave test in the form presented in \cite{Daverio:2018tjf} with components permuted for propagation along $x{}^1$:
\begin{align}
  \label{eq:gw1dsola}
  \alpha & = \sqrt{1 - H_s}, &
  \beta{}^i &= 0, &
  \chi   & = (1-H_s)^{-1/3}, &
  K      & = -\frac{H_c}{(1 - H_s)^{3/2}};
\end{align}
\begin{align}
  \label{eq:gw1dsolb}
  \tilde{\Gamma}{}^1 &\dot{=}
  -\frac{4}{3}\frac{H_c}{(1 - H_s)^{5/3}}, &
  \tilde{\Gamma}{}^2 &= \tilde{\Gamma}{}^3 \dot{=} 0;
\end{align}
\begin{align}
  \label{eq:gw1dsolc}
  \tilde{\gamma}{}_{ij} &\dot{=}
  \mathrm{diag}\left(
    (1-H_s)^{2/3},\,
    (1-H_s)^{-1/3},\,
    (1-H_s)^{-1/3}
  \right);
\end{align}
\begin{align}
  \label{eq:gw1dsold}
  \tilde{A}{}_{ij} & \dot{=}
  \frac{1}{3} H_c
  \mathrm{diag}\left(
    -2 (1 - H_s)^{-5/6},\,
    (1-H_s)^{-11/6},\,
    (1-H_s)^{-11/6}
  \right);
\end{align}
where:
\begin{align}
  \label{eq:defnHsHc}
  H_s & := A\sin(2\pi(x-t)), &
  H_c & := A\pi\cos(2\pi(x-t)).
\end{align}
Evaluating \Cref{eq:gw1dsola,eq:gw1dsolb,eq:gw1dsolc,eq:gw1dsold} at $t=0$ and setting $\Theta=0$ together with $\check{Z}^i=0$ yields a one-parameter family of initial data parametrized by amplitude $A$. We select $A=1/100$ as it is known that large values (e.g.~$A=1/2$) can lead to issues with stability in a variety of formulations and regardless of puncture or harmonic gauge choice \citep{Daverio:2018tjf,Cao:2011fu,boyle2007testingaccuracystability}. For compatiblity with the analytical gauge we make use of the harmonic prescription of Eq.\eqref{eq:gaugeBGDRV} for the lapse evolution with $\mu_L=1$ and Eq.\eqref{eq:gaugeBharm} for the shift. To assess solution quality we consider numerical evolution repeated at a triplet of resolutions $(\delta x_c,\, \delta x_m,\, \delta x_f)$ where $\delta x_c < \delta x_m < \delta x_f$. Convergence rates of the overall approximation of the corresponding field data $\mathcal{F}$ may be examined by comparing differences of solutions $\delta\mathcal{F}_{ab}:=\mathcal{F}_a-\mathcal{F}_b$ at distinct resolutions. For an approximation of order $n$ one finds based on Taylor expansion that $\delta\mathcal{F}_{mf} \simeq \delta\mathcal{F}_{cm} / Q_n$ where we have introduced the so-called convergence factor:
\begin{eqnarray}
  \label{eq:conv_fact}
  Q_n: = \frac{\delta x_c^n - \delta x_m^n}{\delta x_m^n - \delta x_f^n}.
\end{eqnarray}
As we know the space-time metric over the full foliation we may directly compare the RMS error of the numerical solution at any sampled $t\in[0,T]$. Additionally, we may inspect the associated phase error as suggested in \cite{Daverio:2018tjf}. To do this evolved, field data on sub-domains is reassembled on a single, discretized domain (e.g.~$\tilde{\gamma}{}_{zz}(t,\,x_I)$) with respect to which we define:
\begin{equation}
  \label{eq:Fkcoeffs}
  F_{K}(t):=\frac{1}{N_M}\sum_I \left(
    1-(\tilde{\gamma}{}_{zz}(t,\,x_I))^{-3}
  \right)\exp(-2\pi i K(x_I-t)),
\end{equation}
where we denote the phase of each complex coefficient $\phi_K(t)=\arg(F_K(t))$. Phase error can thus be quantified as $\varepsilon_{\phi_{\pm 1}}:=|\phi_{\pm1}(t)\mp \pi / 2|$. We also also compute the offset of the numerical profile relative to the amplitude through $\varepsilon_{A_0}:=|F_0(t)|/A$. To simultaneously assess convergence and absolute error define the normalized error:
\begin{equation}
  \label{eq:rescnormerr}
  \hat{\varepsilon}(t;\,T,\,
    \rho_a,\,\rho_b;\,
    \rho_c,\,\rho_d;\,
    \rho_\star
  ):=
  \left|
  \frac{\delta\varepsilon_{\rho_a\rho_b}(t)}{\max_t\delta\varepsilon_{\rho_c\rho_d}(t)}
  \left. \varepsilon(t)\right|_{\rho=\rho_\star}
  \right|.
\end{equation}
In the convergent regime the resolution triplet induced by $(\rho_c,\,\rho_m,\,\rho_f)$ satisfies $\hat{\varepsilon}(t;\,T,\,\rho_m,\,\rho_f;\,\rho_m,\,\rho_f;\,\rho_\star)\simeq \hat{\varepsilon}(t;\,T,\,\rho_c,\,\rho_m;\,\rho_m,\,\rho_f,\,\rho_\star)/Q_n$ with absolute scale given by $\left.\varepsilon\right|_{\rho=\rho_\star}(T)$. Results of a calculation involving $(\rho_c,\,\rho_m,\,\rho_f)=(2,\,3,\,4)$ are shown in Fig.\ref{fig:z4c_awa_gw1}.
In the case of the aligned, unshifted, gauge wave test we see that shift-advective terms are analytically zero. Thus we restrict discussion to replacement of standard FD approximants with centered CFD. As can be seen in Fig.\ref{subfig:z4c_awa_gw1_a} we find clean $\6th$ order convergence for the derivative schemes investigated. Furthermore we see that all tested CFD schemes outperform the FD scheme. In particular, for the phase error at $T=1000$ we find that ${}^{(\mathrm{d})}P{}^1_2$ reduces error by a factor of $\simeq 4$ when compared with ${}^{(\mathrm{d})}E_3$ whereas the spectrally-tuned ${}^{(\mathrm{d})}Q{}^1_3[\eta_c=\frac{4\pi}{5}]$ reduces error by a factor of $\simeq 10.8$. The best improvement in phase error is found for the approximant ${}^{(\mathrm{d})}Q{}^1_3[\eta_c=1]$ which when compared to FD results in a reduction in error of a factor of $\simeq 137.5$. Considering instead $\tilde{\gamma}{}_{xx}$ leads to qualitatively similar conclusions. Constraints are also well satisfied and better preserved when using CFD schemes -- indeed for the RMS of $\mathcal{H}$ we find a reduction of a factor of $\simeq 18$ when compared with FD.

\begin{figure}[htbp]
  \subfloat[\label{subfig:z4c_awa_gw1s_a}]{%
    \includegraphics[width=0.49\textwidth]{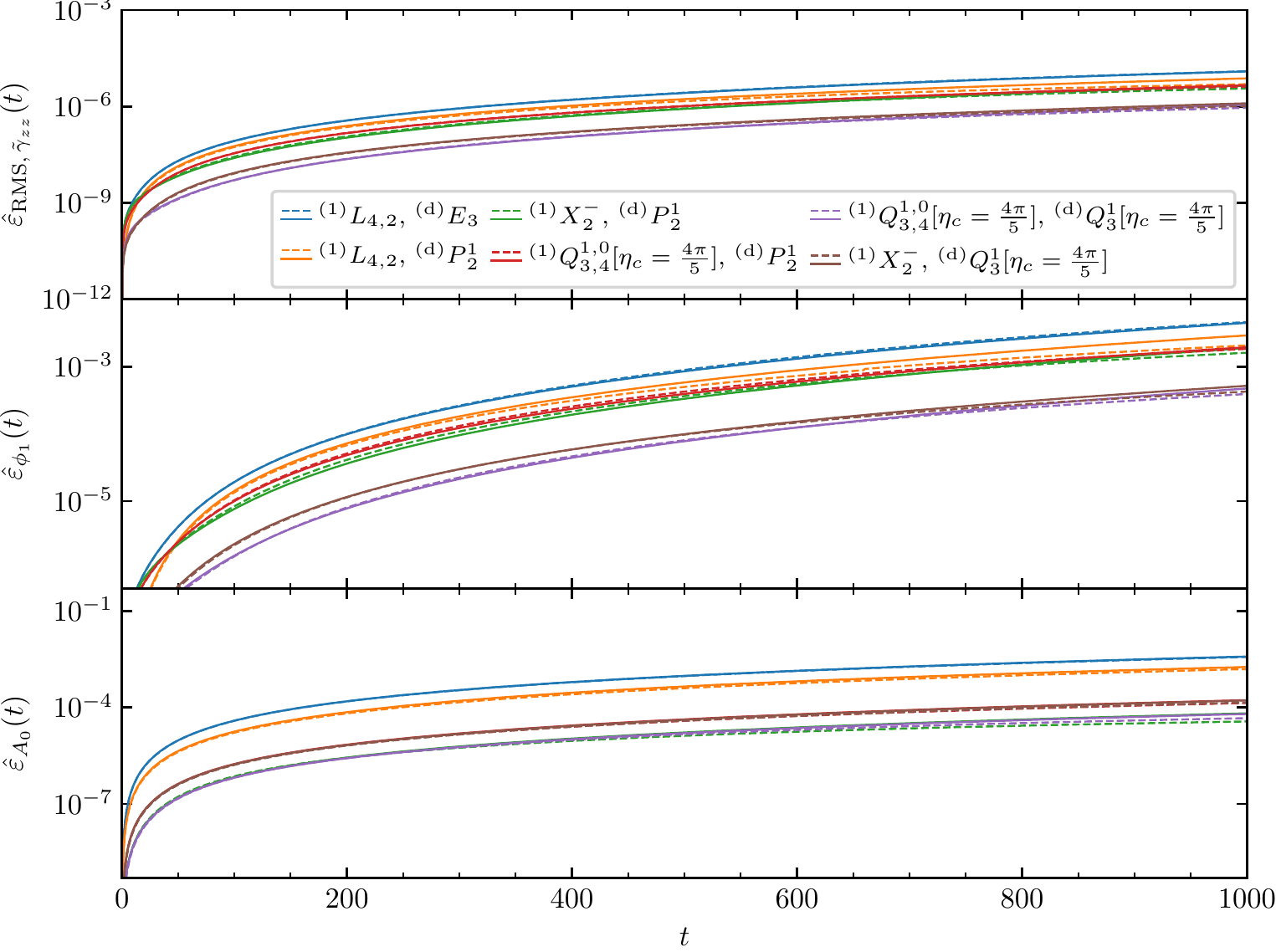}%
  }\hfill
  \subfloat[\label{subfig:z4c_awa_gw1s_b}]{%
    \includegraphics[width=0.49\textwidth]{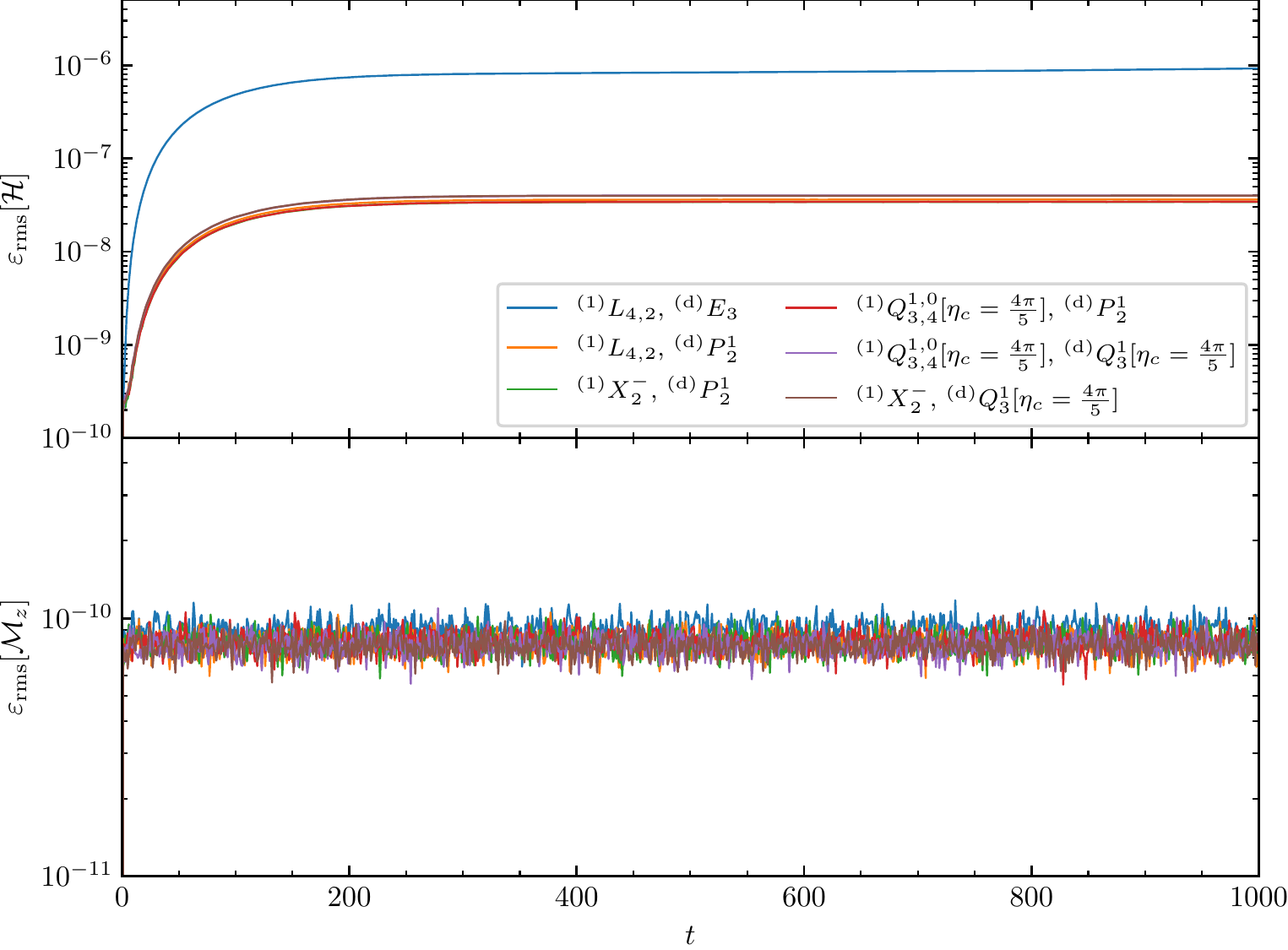}%
  }\hfill
  \caption{Error associated with aligned, shifted gauge wave \AwA{} test where for standard FD $(\rho_c,\,\rho_m,\,\rho_f)=(2,\,3,\,4)$ whereas for CFD $(\rho_c,\,\rho_m,\,\rho_f)=(1,\,2,\,4)$. In (a) we inspect the RMS error associated with the evolved $\tilde{\gamma}{}_{zz}$ (top), the phase error (middle), and the offset of the numerical profile relative to the amplitude (bottom). Error has been rescaled according to design order of the scheme with $Q_6$ of Eq.\eqref{eq:conv_fact} and utilizing the normalization of Eq.\eqref{eq:rescnormerr} with absolute error scale corresponding to evolution with $\rho_\star=\rho_f$. In (b) the RMS values of the Hamiltonian constraint $\mathcal{H}$ and $z$ component of the momentum constraint $\mathcal{M}_z$ are shown for $\rho_f$. See text for discussion.}\label{fig:z4c_awa_gw1s}
\end{figure}

In order to test shift-advective terms we consider the aligned, shifted gauge wave test where \cite{Daverio:2018tjf}:
\begin{align}
  \alpha & = (1 + H_s)^{-1/2}, &
  \chi   & = (1+H_s)^{-1/3}, &
  K      & = -\frac{H_c}{(1 + H_s)^{3/2}};
\end{align}
\begin{align}
  \beta{}^1 &= -\frac{H_s}{1+H_s},
  & \beta{}^2 = \beta{}^3 = 0;
\end{align}
\begin{align}
  \tilde{\Gamma}{}^1 &\dot{=}
  \frac{4}{3}\frac{H_c}{(1 + H_s)^{5/3}}, &
  \tilde{\Gamma}{}^2 &= \tilde{\Gamma}{}^3 \dot{=} 0;
\end{align}
\begin{align}
  \tilde{\gamma}{}_{ij} &\dot{=}
  \mathrm{diag}\left(
    (1-H_s)^{2/3},\,
    (1-H_s)^{-1/3},\,
    (1-H_s)^{-1/3}
  \right);
\end{align}
\begin{align}
  \tilde{A}{}_{ij} & \dot{=}
  \frac{1}{3} H_c
  \mathrm{diag}\left(
    -2 (1 + H_s)^{-5/6},\,
    (1+H_s)^{-11/6},\,
    (1+H_s)^{-11/6}
  \right);
\end{align}
with $H_s$ and $H_c$ are defined as in Eq.\eqref{eq:defnHsHc}. The method of setup and quantities analyzed are as in the unshifted case and we once again select $A=1/100$. During this test for standard FD resolution in induced through the triplet $(\rho_c,\,\rho_m,\,\rho_f)=(2,\,3,\,4)$ whereas for CFD we set $(\rho_c,\,\rho_m,\,\rho_f)=(1,\,2,\,4)$. As $\rho_f$ is common we may compare again with the normalization prescription of Eq.\eqref{eq:rescnormerr}. We depict the result of numerical evolution in Fig.\ref{fig:z4c_awa_gw1s}.
As shift-advective terms are now non-zero we additionally make use of upwinded stencils. In Fig.\ref{subfig:z4c_awa_gw1s_a} we again find clean $\6th$ order convergence for the derivative schemes investigated (cf.~Fig.\ref{subfig:z4c_awa_gw1_a}). In a similar vein as the unshifted test errors are reduced when making use of CFD schemes. Comparing phase error at $T=1000$ between schemes shows that replacing only the centered derivatives with ${}^{(\mathrm{d})}P{}^1_2$ reduces error by a factor of $\simeq1.5$ when compared with the fully explicit $\{{}^{(1)}L_{4,2},\,{}^{(\mathrm{d})}E_3\}$. If $\{{}^{(1)}X{}^-_2,\,{}^{(\mathrm{d})}P{}^1_2\}$ is utilized then instead find a reduction in phase error of a factor of $\simeq 2.4$ when compared with FD. The best improvement in phase error for schemes tested here is found for the approximants $\{{}^{(1)}X{}^-_2,\,{}^{(\mathrm{d})}Q{}^1_3[\eta_c=\frac{4\pi}{5}]\}$ which when compared to FD results in a reduction in error of a factor of $\simeq 8.6$.

\subsection{Z4c: numerical tests - binary black hole evolution}
\label{ssec:z4c_tests_2}

We close our numerical tests with a preliminary investigation of binary black hole (BBH) evolution utilizing CFD schemes. This test departs from those presented earlier in this work as the underlying computational domain is no longer periodic and non-trivial boundary conditions (BC) must be applied on evolved field components. In particular, the \z4c{} dynamical equations supplemented by gauge conditions populate $\{\chi,\,\tilde{\gamma}{}_{ij},\,\alpha,\,\beta{}^i\}$ on $\partial\Omega$ whereas Sommerfeld BC are applied to the field components $\{\hat{K},\,\tilde{\Gamma},\,\Theta,\,\tilde{A}{}_{ij}\}$. In addition, due to the range of spatial scales, we make use of adaptive mesh refinement (AMR) for computational efficiency. Suppose $\Omega_I$ is an element of a domain-decomposition of an $\Omega$ of interest. Within \GRAthena{} one can prescribe a conditional (i.e.~a target resolution over a region of $\Omega$ within a given distance of some feature described by the evolved fields) which controls the AMR. The sub-domain $\Omega_I$ is then recursively (de)refined to satisfy the conditional under the further restriction the nearest-neighbour sub-domains can differ in resolution by at most a $2:1$ ratio.  Extensive details on the treatment of BC and AMR made in \GRAthena{} which we utilize for this problem are described in \cite{Daszuta:2021ecf}.

To model the BBH evolution itself initial data compatible with the constraints must first be provided. To this end we consider the initial geometry as modelled by Brill-Lindquist wormhole topology describing $N$ black holes with $N+1$ disconnected, asymptotically flat ends. Each disconnected end is diffeomorphic to $\mathbb{R}^3$ minus a compact ball \cite{dain2002asymptoticallyflatregular}. An end is compactified and identified with a point $\mathbf{x}_p$ on $\mathbb{R}^3$. The coordinate singularity that occurs at a given $\mathbf{x}_p$ is a so-called puncture which describes the location of a black hole. This allows the constraints to be solved based on \cite{Ansorg:2004ds} thus providing initial data. Gauge conditions are initialized based on a ``precollapsed'' lapse and zero-shift \cite{Campanelli:2005dd}. The damping parameter in Eq.\eqref{eq:gaugeBGDRV} is now taken as $\eta=2/M$ which is fixed in terms of the ADM mass $M$ \cite{Arnowitt:1962hi} of the underlying system. The AMR criterion is based on a mock ``box-in-box'' oct-tree structure which adapts resolution based on tracking puncture centers $\mathbf{x}_p(t)$ during the course of a simulation. Given an overall $\Omega=[-x_M,\,x_M]^3$ that has been domain-decomposed and refined the resolution at both punctures is controlled by the maximum number of refinement levels $N_L$ as $\delta x_p = 2 x_M / (N_M 2^{N_L-1})$. Unless otherwise stated we make use of $N_B=16$ for the number of samples along each direction taken on a sub-domain.

As we would like to inspect convergence for a variety of derivative approximants we investigate an equal-mass initial configuration leading to a short evolution. The BBH system has two non-spinning punctures, initially centered on-axis at $\left.\mathbf{x}_{p^\pm}(t)\right|_{t=0}=(\pm 3.257,\,0,\,0) M$ with initial momenta $\left.\mathbf{p}_{p^\pm}(t)\right|_{t=0}=(0,\,\mp0.133,\,0)M$, and with bare-masses $m_{p^\pm}=0.483M$. This configuration results in $\sim2.5$ orbits before merger at evolution time $T\sim 170M$. For the overall grid extent we select $x_M=1536\,M$ such that $\partial \Omega$ is causally disconnected from the interior strong-field dynamics during the course of the initial inspiral through merger. As a diagnostic for assessing numerical simulation quality we consider extracted gravitational wave (GW) content associated with the strong-field dynamics. This is done by first assembling the four-dimensional Weyl tensor from the evolved \z4c{} variables. Subsequent projection over a suitable null tetrad \cite{Brugmann:2008zz,Daszuta:2021ecf} yields the complex, out-going Weyl scalar $\Psi_4$. A mode-decomposition with respect to spherical harmonics of spin-weight $s=-2$ at extraction radius $R$ based on numerical quadrature over geodesic spheres \cite{Daszuta:2021ecf} furnishes us with radiated GW content in the $(l,\,m)$ mode from $\psi{}_{lm}$.

\begin{figure}[htbp]
  \subfloat[\label{subfig:z4c_calib_a}]{%
    \includegraphics[width=0.49\textwidth]{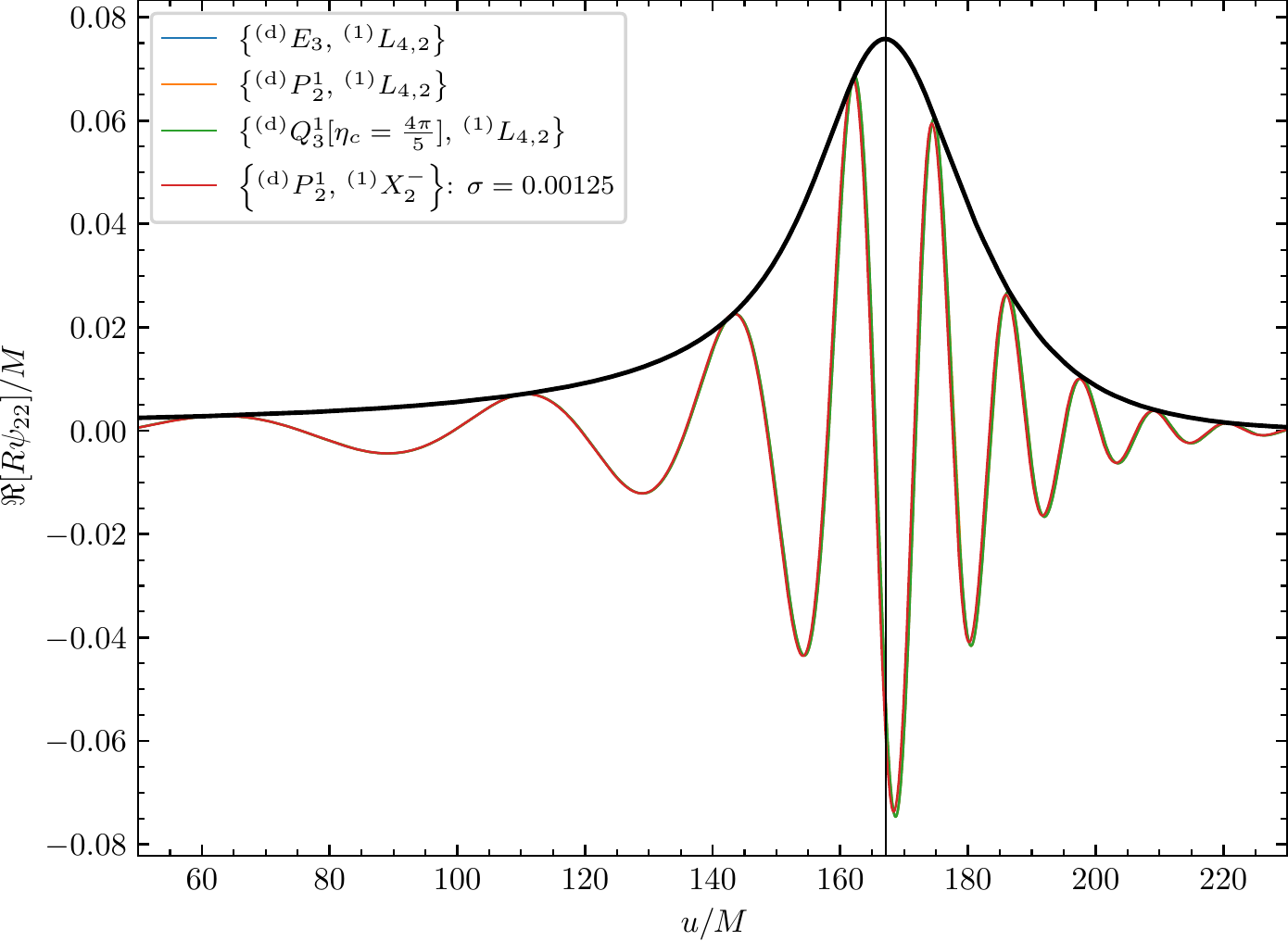}%
  }\hfill
  \subfloat[\label{subfig:z4c_calib_b}]{%
    \includegraphics[width=0.49\textwidth]{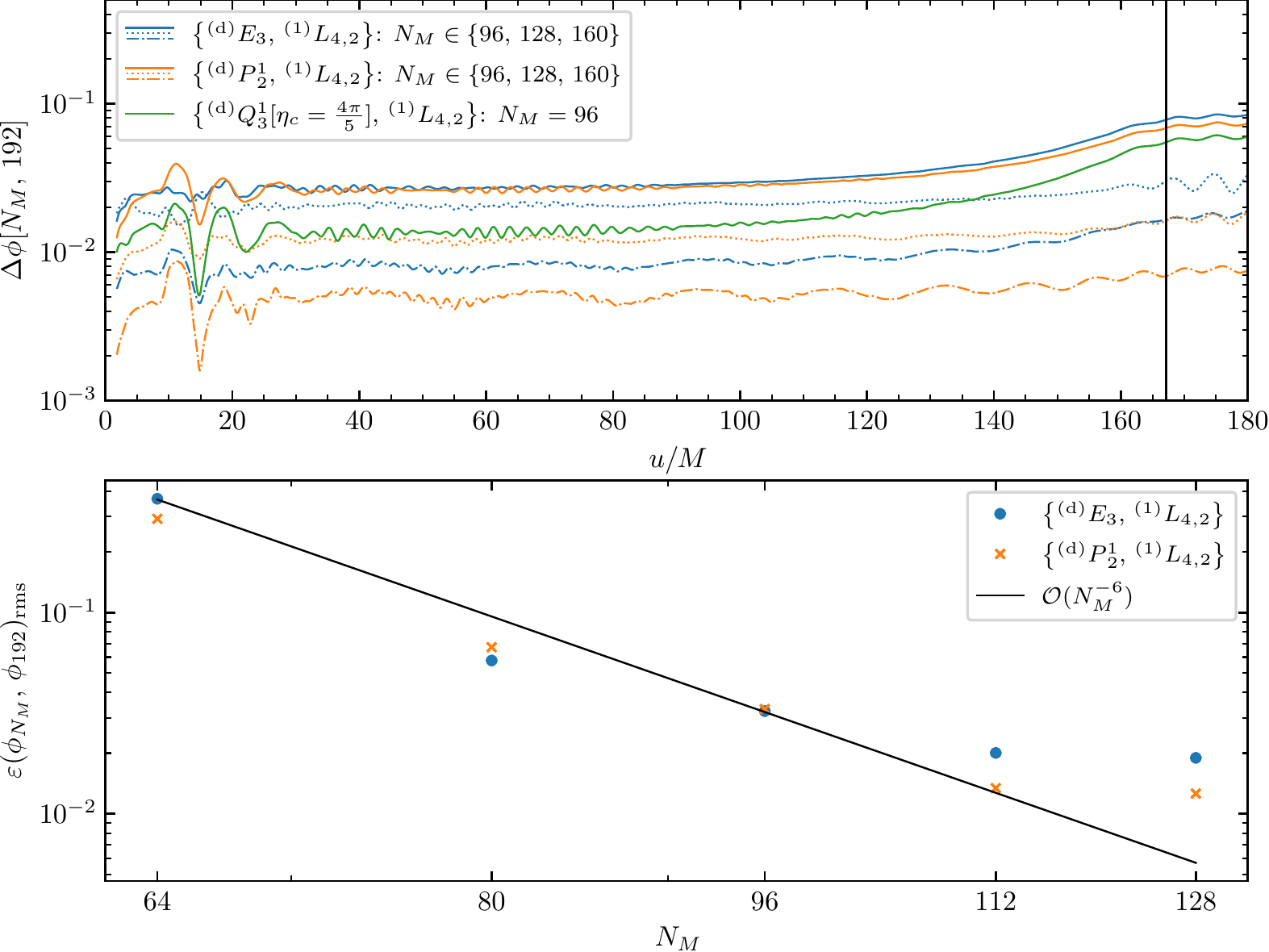}%
  }\hfill
  \caption{The real part of the dominant $(2,\,2)$ mode of GW content extracted at $R=70\,M$ and associated phase error. In (a) we depict the real part of $R\psi{}_{22}$ for a variety of schemes with $N_M=96$. Good agreement is found for this choice of resolution. In each case we select CFL $\mathcal{C}_1=1/2$ together with KO-dissipation $\sigma=0.02$ apart from $\left\{{}^{(\mathrm{d})}P{}^1_{2},\,{}^{(1)}X_2^-\right\}$ where $\mathcal{C}_1=1/10$ and $\sigma=0.00125$ are selected. Envelope curve indicates the amplitude $|\psi{}_{22}|$ as computed with $N_M=192$ based on simulations involving explicit FD schemes. Vertical black line indicates time of merger which is time of peak amplitude of the $(2,\,2)$ mode. In (b, upper) the difference in phase of $\psi{}_{lm}$ computed with the scheme $\left\{{}^{(\mathrm{d})}P{}^1_{2},\,{}^{(1)}L_{4,2}\right\}$ at $N_M=192$ and a selection of $N_M$ is shown for differing schemes at fixed dissipation $\sigma=0.02$ and $\mathcal{C}_1=1/2$. In (b, lower) the RMS error in phase up to merger is shown. Note: as is conventional we make use of retarded time $u:=T-r^*$ where $r^*:=r+2M\log|R/(2M)-1|$. See text for further discussion.}\label{fig:z4c_calib}
\end{figure}

We compute the dominant $(2,\,2)$ mode for simulations involving a variety of derivative approximants and choices of $N_M$ and show the result in Fig.\ref{fig:z4c_calib}.
We find that stable BBH evolution is possible utilizing the CFD schemes discussed in this work with resulting phase error of extracted $\psi{}_{22}$ compatible with the $\6th$ order of accuracy design of the underlying spatial derivative approximant schemes. In Fig.\ref{subfig:z4c_calib_b} we observe that replacing centered FD with CFD leads to a reduction in the associated $\psi{}_{22}$ phase error (when comparing fixed $N_M$) of a factor $\sim2$. A variety of effects influence this factor. As shown in \S\ref{ssec:z4c_tests} it may be important to adequately treat shift-advective terms for maximum improvement. Another delicate matter is transferring field data between between sub-domains at differing levels of refinement. In \GRAthena{} this is achieved through use of prolongation and restriction operations based on centered polynomial interpolation at formal order of accuracy matched to the underlying (C)FD scheme. Without additional care this may potentially degrade properties of the modified wavenumber discussed in \S\ref{ssec:scmp}. Additionally as observed during grid convergence tests (see e.g.~Fig.\ref{subfig:der_01_decomp_a}) when CFD stencils are utilized in the context of domain-decomposition error tends to accumulate at sub-domain boundaries. Transferring data between differing levels of refinement also occurs in this region and it is not entirely evident as to whether polynomial interpolation will amplify or diminish this source of error. Nonetheless we have described how error in the vicinity of $\partial \Omega_I$ can be mitigated through usage of a hybrid-communication strategy in \S\ref{ssec:domdec_iter}. We have not yet implemented this strategy as this would require involved modification to core \GRAthena{} functionality. We aim to address this in future.

\section{Summary and conclusion}
\label{sec:conc}

In this work we have shown that the unified compact finite difference (CFD) framework of \cite{Deshpande:2019uf} may be extended in numerical generation of new schemes which may be biased or centered, of arbitrary extent, and involve not only function data but also prescribed function derivative data. Upon fixing formal order of accuracy Taylor matching yields a linear system specifying a stencil. When the aforementioned system is underdetermined we may minimize a functional characterizing spectral error. This allows us to further extend the approach to construct implicit extensions to the Hermite methods described in \cite{Fornberg:2020ch}. Facilitating simpler construction of such schemes allows for rapid experimentation on practical problems and consequently we also have open-sourced our notebook \cite{notebook_repo}.

Large-scale problems crucially depend on exploiting parallelism for efficiency and consequently in order to treat solution of the implicit problem specifying a given CFD scheme we investigated modification of a dispersion-relation-preserving (DRP) method due to \cite{chen2021novelparallelcomputing}. This featured treating domain-decomposition with tailored closures for decoupling the implicit, linear system specifying a derivative approximant to individual, decoupled sub-domains. We applied the method to generalize the biased, first degree $\4th$ order CCU$(4,5)$ scheme \cite{chen2021novelparallelcomputing} to CCU$(6,7)$ and CCU$(8,9)$ which have formal order of accuracy of $\6th$ and $\8th$ order respectively. We also show directly that the DRP method may be exactly employed for biased second degree schemes, and furthermore, applied approximately in decoupling centered CFD methods. Grid convergence tests involving smooth functions revealed that edge artifacts can be induced under domain-decomposition with this strategy. We therefore proposed a hybrid-communication strategy that iterates upon results from decoupled sub-domains and allows for mitigation of error. Efficacy was verified through grid convergence testing on smooth functions. As a further test geared towards wave-propagation problems the numerical spectra of the semi-discretized two-dimensional advection equation, and shifted wave equation were directly inspected. This was achieved through an embedded description of sub-domain communication, iterated closure, and differentiation as block-partitioned system. Indeed domain-decomposition and DRP leads to deformed spectra which we demonstrate can be mitigated with our proposed hybrid-communication approach.

We numerically verified anticipated stability properties for the aforementioned toy-problems through solution of semi-discretized and fully-discretized formulations based on exponential integration and explicit Runge-Kutta methods respectively. Spatial discretization was fixed to have formal order of accuracy $\mathcal{O}(\delta x^6)$. In the case of two-dimensional advection we found that when compared to standard finite-difference (FD) the well-known Pad\'{e} scheme ${}^{(1)}P{}^1_2$ allowed for a reduction in maximum error by a factor of $\simeq 13.6$ whereas our new spectrally-tuned scheme ${}^{(1)}Q{}^1_3[\eta_c=1]$ attained a factor of $\simeq 65.9$ both of which remained robust under domain-decomposition. Comparable improvements were observed for the shifted wave equation.

As a first novel application we considered evolution of the \z4c{} formulation of numerical relativity. To this end we wrote a stand-alone, header-only, templated, generalized finite difference library which we coupled to the \GRAthena{} \cite{Daszuta:2021ecf} code. This is a dramatic increase in complexity of the underlying system being evolved as not only is the system now quasi-linear, with significantly more involved stability criteria, the number of independent field components is larger. Indeed this precluded a numerical investigation of spectra associated with semi-discretization. As a first step we coupled our code to \GRAthena{} and investigated introducing CFD with the (approximate) DRP prescription for closure under domain-decomposition. In the context of the \AwA{} aligned, unshifted, gauge wave evolution test (without shift-advective terms) this allowed for reducing the phase error of propagated metric components when compared with standard FD by a factor of $\simeq 4$ for ${}^{(\mathrm{d})}P{}^1_2$ and $\simeq 137.5$ for ${}^{(\mathrm{d})}Q{}^1_3[\eta_c=1]$. For the case of the shifted variant of this test we found that it is important to replace not only centered FD terms with an appropriate centered CFD prescription but to also treat lop-sided FD terms with biased CFD. The best improvement here we found for the combinations test was $\{{}^{(1)}X{}^-_2,\,{}^{(\mathrm{d})}Q{}^1_3[\eta_c=\frac{4\pi}{5}]\}$ which reduced phase error by a factor of $\simeq 8.6$ when compared with standard FD.

We also considered a preliminary application to binary-black-hole evolution. For this class of problem the discrete treatment of the underlying domain in \GRAthena{} features adaptive mesh refinement (AMR) for computational efficiency. During investigation of a $\sim2.5$ orbit, equal mass, non-spinning BBH simulation we found that utilizing CFD leads to stable evolution. By inspecting the phase of the dominant mode of the extracted gravitational waves we verified convergence compatible with a $\6th$ order trend. A factor $\sim2$ reduction in phase error at fixed resolution when compared with standard FD was also observed. This reduced efficiency in error reduction can potentially be attributed to a combination of: insufficiently aggressive CFL, approximate DRP and lack of closure iteration based on hybrid-communication, and the overly simple treatment of level-to-level transfer of field data between sub-domains at differing levels of refinement. The influence and precise tuning of dissipation together with potential replacement by compact filters \cite{kim2013quasidisjointpentadiagonalmatrix} may also significantly affect error reduction efficiency. We leave a thorough investigation to future work.


\begin{acknowledgments}
  The author thanks David Radice and Sebastiano Bernuzzi for discussions and constructive input during several stages of preparing this work. Special thanks to Martin Bernreuther for assistance with high performance computing support. The author is indebted to Beppe Starnazza.
  The author acknowledges support by the EU H2020 under ERC Starting Grant, no.~BinGraSp-714626. Simulations with {\GRA} were performed on the ARA cluster at Friedrich Schiller University Jena, SuperMUC-NG at the Leibniz-Rechenzentrum (LRZ, \url{www.lrz.de}) Munich, and HPE Apollo Hawk at the High Performance Computing Center Stuttgart (HLRS, \url{www.hlrs.de}). The ARA cluster is funded in part by DFG grants INST~275/334-1~FUGG and INST~275/363-1~FUGG, and ERC starting Grant, grant agreement no.~BinGraSp-714626. The author acknowledges the Gauss Centre for Supercomputing e.V. (\url{www.gauss-centre.eu}) for funding this project by providing computing time on the GCS Supercomputer SuperMUC-NG at LRZ (allocation \texttt{pn68wi}). The author acknowledges HLRS for funding this project by providing access to the supercomputer HPE Apollo Hawk under the grant number \texttt{INTRHYGUE/44215}.
    \end{acknowledgments}

\onecolumngrid
\appendix

\section{Collection of derivative approximants}
\label{app:sch_tuned_coeff}
For convenience we collect a variety of derivative approximants as specified through the $\boldsymbol{\alpha}$ coefficients that enter Eq.\eqref{eq:mdAnsatz} that have been constructed utilizing the method outlined in \S\ref{sec:method}. We tabulate expansions of the associated relative error in the normalized wavenumber $\varepsilon_{\tilde{\eta}}:=\tilde{\eta}^{d_r}/\eta^{d_r} - 1$ where $\tilde{\eta}$ is defined in Eq.\eqref{eq:fdpwerrdef}.
\begin{table}[htbp]
  \centering
  \begin{tabular}{c|l|l}
    \hline
    Scheme &
    $\boldsymbol{\alpha}^{(0)}$ &
    $\varepsilon_{\tilde{\eta}}$ \\
    \hline
    \hline
    ${}^{(1)}E{}_{2}\left[\mathcal{O}(\delta x^4)\right]$ &
    $\frac{1}{12} \left(1,\,-8,\,0,\,8,\,-1 \right)$ &
    $-\frac{1}{30}\eta^4 + \frac{1}{252}\eta^6 - \mathcal{O}(\eta^{8})$ \\
    ${}^{(2)}E{}_{2}\left[\mathcal{O}(\delta x^4)\right]$ &
    $\frac{1}{12} \left(-1,\,16,\,-30,\,16,\,-1 \right)$ &
    $-\frac{1}{90}\eta^4 + \frac{1}{1008}\eta^6 - \mathcal{O}(\eta^{8})$ \\
    ${}^{(1)}L{}_{3,1}\left[\mathcal{O}(\delta x^4)\right]$ &
    $\frac{1}{12} \left(-1,\,6,\,-18,\,10,\,3 \right)$ &
    $\frac{1}{20}\eta^4 - \frac{i}{24}\eta^5 - \mathcal{O}(\eta^{6})$ \\
    \hline
    ${}^{(1)}E{}_{3}\left[\mathcal{O}(\delta x^6)\right]$ &
    $\frac{1}{60} \left(-1,\,9,\,-45,\,0,\,45,\,-9,\,1 \right)$ &
    $-\frac{1}{140}\eta^6 + \frac{1}{720}\eta^8 - \mathcal{O}(\eta^{10})$ \\
    ${}^{(2)}E{}_{3}\left[\mathcal{O}(\delta x^6)\right]$ &
    $\frac{1}{2}\left(\frac{1}{45},\,-\frac{3}{10},\,3,\,-\frac{49}{9},\,3,\,-\frac{3}{10},\,\frac{1}{45} \right)$ &
    $-\frac{1}{560}\eta^6 + \frac{1}{3600}\eta^8 + \mathcal{O}(\eta^{10})$ \\
    ${}^{(1)}L{}_{4,2}\left[\mathcal{O}(\delta x^6)\right]$ &
    $\frac{1}{60} \left(1,\,-8,\,30,\,-80,\,35,\,24,\,-2 \right)$ &
    $\frac{1}{105}\eta^6 - \frac{i}{120}\eta^7 - \mathcal{O}(\eta^8)$ \\
    \hline
    ${}^{(1)}E{}_{4}\left[\mathcal{O}(\delta x^8)\right]$ &
    $\left(-\frac{1}{560},\,\frac{8}{315},\,-\frac{1}{5},\,\frac{8}{5},\,-\frac{205}{72},\,\frac{8}{5},\,-\frac{1}{5},\,\frac{8}{315},-\frac{1}{560} \right)$ &
    $-\frac{1}{630}\eta^8 + \frac{1}{2310}\eta^{10} - \mathcal{O}(\eta^{12})$ \\
    ${}^{(2)}E{}_{4}\left[\mathcal{O}(\delta x^8)\right]$ &
    $\frac{1}{10}\left(-\frac{1}{56},\,\frac{16}{63},\,-2,\,16,\,-\frac{1025}{36},\,16,\,-2,\,\frac{16}{63},\,-\frac{1}{56} \right)$ &
    $-\frac{1}{3150}\eta^8 + \frac{1}{13860}\eta^{10} - \mathcal{O}(\eta^{12})$ \\
    ${}^{(1)}L{}_{5,3}\left[\mathcal{O}(\delta x^8)\right]$ &
    $\frac{1}{12} \left(-\frac{3}{70},\,\frac{3}{7},\,-2,\,6,\,-15,\,\frac{27}{5},\,6,\,-\frac{6}{7},\,\frac{1}{14} \right)$ &
    $\frac{1}{504}\eta^8 - \frac{i}{560}\eta^9 - \mathcal{O}(\eta^{10})$ \\
    \hline
    ${}^{(1)}E{}_{5}\left[\mathcal{O}(\delta x^{10})\right]$ &
    $\frac{1}{21}\left(-\frac{1}{60},\,\frac{5}{24},\,-\frac{5}{4},\,5,\,-\frac{35}{2},\,0,\,\frac{35}{2},\,-5,\,\frac{5}{4},\,-\frac{5}{24},\,\frac{1}{60} \right)$ &
    $-\frac{1}{2772}\eta^{10} + \frac{5}{39312}\eta^{12} - \mathcal{O}(\eta^{14})$ \\
    ${}^{(2)}E{}_{5}\left[\mathcal{O}(\delta x^{10})\right]$ &
    $\frac{1}{21}\left(\frac{1}{150},\,-\frac{5}{48},\,\frac{5}{6},\,-5,\,35,\,-\frac{36883}{600},\,35,\,-5,\,\frac{5}{6},\,-\frac{5}{48},\,\frac{1}{150}\right)$ &
    $-\frac{1}{16632}\eta^{10} + \frac{5}{275184}\eta^{12} - \mathcal{O}(\eta^{14})$ \\
    ${}^{(1)}L{}_{6,4}\left[\mathcal{O}(\delta x^{10})\right]$ &
    $\frac{1}{21} \left(\frac{1}{60},\,-\frac{1}{5},\,\frac{9}{8},\,-4,\,\frac{21}{2},\,-\frac{126}{5},\,\frac{77}{10},\,12,\,-\frac{9}{4},\,\frac{1}{3},\,-\frac{1}{40} \right)$ &
    $\frac{1}{2310}\eta^{10} - \frac{i}{2520}\eta^{11} - \mathcal{O}(\eta^{12})$ \\
    \hline
  \end{tabular}
 \caption{Miscellaneous explicit derivative approximant schemes and associated relative error in the normalized wavenumber $\varepsilon_{\tilde{\eta}}$. The formal order of accuracy is indicated in parenthesis together with the scheme label. The lop-sided schemes ${}^{(1)}L{}_{N+1,N-1}$ have $\Im{}[\varepsilon_{\tilde{\eta}}]\neq 0$. The opposite lop-siding to ${}^{(1)}L{}_{N+1,N-1}$, that is, ${}^{(1)}L{}_{N-1,N+1}$ has coefficients $\alpha{}^{(0)}_{\hpb m}=-\alpha{}^{(0)}_{-m}$, and $\varepsilon_{\tilde{\eta}}\left[{}^{(1)}L{}_{N-1,N+1}\right]=\varepsilon_{\tilde{\eta}}\left[{}^{(1)}L{}_{N+1,N-1}\right]^*$.}
 \label{tab:misc_explicit}
\end{table}
\begin{table}[htbp]
  \centering
  \begin{tabular}{c|l|l}
    \hline
    Scheme &
    $\boldsymbol{\alpha}^{(0)}$ &
    $\varepsilon_{\tilde{\eta}}$ \\
    \hline
    \hline
    ${}^{(4)}E{}_{2}$ &
    $\left(1,\,-4,\,6,\,-4,\,1 \right)$ &
    $-\frac{1}{6}\eta^2 + \frac{1}{80}\eta^4 - \mathcal{O}(\eta^{6})$ \\
    ${}^{(6)}E{}_{3}$ &
    $\left(1,\,-6,\,15,\,-20,\,15,\,-6,\,1 \right)$ &
    $-\frac{1}{4}\eta^2 + \frac{7}{240}\eta^4 - \mathcal{O}(\eta^{6})$ \\
    ${}^{(8)}E{}_{4}$ &
    $\left(1,\,-8,\,28,\,-56,\,70,\,-56,\,28,\,-8,\,1 \right)$ &
    $-\frac{1}{3}\eta^2 + \frac{19}{360}\eta^4 - \mathcal{O}(\eta^{6})$ \\
    ${}^{(10)}E{}_{5}$ &
    $\left(1,\,-10,\,45,\,-120,\,210,\,-252,\,210,\,-120,\,45,\,-10,\,1 \right)$ &
    $-\frac{5}{12}\eta^2 + \frac{1}{12}\eta^4 - \mathcal{O}(\eta^{6})$ \\
    ${}^{(12)}E{}_{6}$ &
    $\left(1,\,-12,\,66,\,-220,\,495,\,-792,\,924,\,-792,\,495,\,-220,\,66,\,-12,\,1 \right)$ &
    $-\frac{1}{2}\eta^2 + \frac{29}{240}\eta^4 - \mathcal{O}(\eta^{6})$ \\
    \hline
  \end{tabular}
 \caption{Higher degree, centered explicit derivative approximant schemes with formal order of accuracy $\mathcal{O}(\delta x^2)$ and associated relative error in the normalized wavenumber $\varepsilon_{\tilde{\eta}}$. Used to e.g. tune $\tilde{\eta}$ in \S\ref{ssec:domdec_drp} by modifying the dissipation relation.}
 \label{tab:diss_scheme}
\end{table}
\begin{table}[htbp]
  \centering
  \begin{tabular}{c|ll|l}
    \hline
    Scheme &
    $\boldsymbol{\alpha}^{(1)}$ &
    $\boldsymbol{\alpha}^{(0)}$ &
    $\varepsilon_{\tilde{\eta}}$ \\
    \hline
    \hline
    ${}^{(1)}P{}^{1}_{1}\left[\mathcal{O}(\delta x^4)\right]$ &
    $\left(\frac{1}{4},\, 1,\, \frac{1}{4}\right)$ &
    $\frac{1}{4}\left(-3,\,0,\,3  \right)$ &
    $-\frac{1}{180}\eta^4 -\frac{1}{1512}\eta^6 -\mathcal{O}(\eta^8)$ \\
    ${}^{(1)}P{}^{1}_{2}\left[\mathcal{O}(\delta x^6)\right]$ &
    $\left(\frac{1}{3},\, 1,\, \frac{1}{3}\right)$ &
    $\frac{1}{36}\left(-1,\,-28,\,0,\,28,\,1  \right)$ &
    $-\frac{1}{2100}\eta^6 -\frac{1}{18000}\eta^8 -\mathcal{O}(\eta^{10})$ \\
    ${}^{(1)}P{}^{1}_{3}\left[\mathcal{O}(\delta x^8)\right]$ &
    $\left(\frac{3}{8},\, 1,\, \frac{3}{8}\right)$ &
    $\frac{1}{480}\left(1,\,-24,\,-375,\,0,\,375,\,24,\,-1  \right)$ &
    $-\frac{1}{17640}\eta^8 -\frac{1}{226380}\eta^{10} -\mathcal{O}(\eta^{12})$ \\
    \hline
 \end{tabular}
 \begin{tabular}{c|l|l}
  \hline
  Closure &
  $\boldsymbol{\alpha}^{(0)}$ &
  $\varepsilon_{\tilde{\eta}}$ \\
  \hline
  \hline
  ${}^{(1)}\overline{P}{}_{3}\left[\mathcal{O}(\delta x^4)\right]$ &
  $\frac{1}{72}\left(-1,\,10,\,-53,\,0,\,53,\,-10,\,1  \right)$ &
  $-\frac{1}{180}\eta^4 -\frac{1}{189}\eta^6 +\mathcal{O}(\eta^8)$ \\
  ${}^{(1)}\overline{P}{}_{4}\left[\mathcal{O}(\delta x^6)\right]$ &
  $\frac{1}{300}\left(1,\,-11,\,59,\,-239,\,0,\,239,\,-59,\,11,\,-1  \right)$ &
  $-\frac{1}{2100}\eta^6 -\frac{1}{720}\eta^8 +\mathcal{O}(\eta^{10})$ \\
  ${}^{(1)}\overline{P}{}_{5}\left[\mathcal{O}(\delta x^8)\right]$ &
  $\frac{1}{11760}\left(-9,\,114,\,-691,\,2784,\,-9786,\,0,\,9786,\,-2784,\,691,\,-114,\,9  \right)$ &
  $-\frac{1}{17640}\eta^8 -\frac{43}{129360}\eta^{10} +\mathcal{O}(\eta^{12})$ \\
  \hline
\end{tabular}
 \caption{First degree, centered Pad\'{e} derivative approximant schemes at a variety of approximation orders together with corresponding centered, explicit closures that have been tuned to match $\varepsilon_{\tilde{\eta}}$ at lowest order (see \S\ref{ssec:domdec_drp}).}
 \label{tab:pade_deg_one}
\end{table}
\begin{table}[htbp]
  \centering
  \begin{tabular}{c|ll|l}
    \hline
    Scheme &
    $\boldsymbol{\alpha}^{(2)}$ &
    $\boldsymbol{\alpha}^{(0)}$ &
    $\varepsilon_{\tilde{\eta}}$ \\
    \hline
    \hline
    ${}^{(2)}P{}^{1}_{1}\left[\mathcal{O}(\delta x^4)\right]$ &
    $\left(\frac{1}{10},\, 1,\, \frac{1}{10}\right)$ &
    $\frac{1}{5}\left(6,\,-12,\,6  \right)$ &
    $-\frac{1}{240}\eta^4 -\frac{1}{6048}\eta^6 +\mathcal{O}(\eta^8)$ \\
    ${}^{(2)}P{}^{1}_{2}\left[\mathcal{O}(\delta x^6)\right]$ &
    $\left(\frac{2}{11},\, 1,\, \frac{2}{11}\right)$ &
    $\frac{1}{44}\left(3,\,48,\,-102,\,48,\,3  \right)$ &
    $-\frac{23}{75600}\eta^6 -\frac{1}{54000}\eta^8 -\mathcal{O}(\eta^{10})$ \\
    ${}^{(2)}P{}^{1}_{3}\left[\mathcal{O}(\delta x^8)\right]$ &
    $\left(\frac{9}{38},\, 1,\, \frac{9}{38}\right)$ &
    $\frac{1}{152}\left(-\frac{23}{45},\, \frac{102}{5},\,147,\,-\frac{3004}{9},\,147,\,\frac{102}{5},\,-\frac{23}{45}  \right)$ &
    $-\frac{43}{1411200}\eta^8 -\frac{227}{173859840}\eta^{10} -\mathcal{O}(\eta^{12})$ \\
    \hline
 \end{tabular}
 \begin{tabular}{c|l|l}
  \hline
  Closure &
  $\boldsymbol{\alpha}^{(0)}$ &
  $\varepsilon_{\tilde{\eta}}$ \\
  \hline
  \hline
  ${}^{(2)}\overline{P}{}_{3}\left[\mathcal{O}(\delta x^4)\right]$ &
  $\frac{1}{144}\left(1,\,-18,\,207,\,-380,\,207,\,-18,\,1  \right)$ &
  $-\frac{1}{240}\eta^4 -\frac{1}{1344}\eta^6 +\mathcal{O}(\eta^8)$ \\
  ${}^{(2)}\overline{P}{}_{4}\left[\mathcal{O}(\delta x^6)\right]$ &
  $\frac{1}{675}\left(-1,\,\frac{31}{2},\,-\frac{517}{4},\,\frac{2137}{2},\,-\frac{3815}{2},\,\frac{2137}{2},\,-\frac{517}{4},\,\frac{31}{2},\,-1  \right)$ &
  $-\frac{23}{75600}\eta^6 -\frac{7}{32400}\eta^8 +\mathcal{O}(\eta^{10})$ \\
  ${}^{(2)}\overline{P}{}_{5}\left[\mathcal{O}(\delta x^8)\right]$ &
  $\frac{1}{128}\left(\frac{9}{245},\,-\frac{146}{245},\,\frac{10813}{2205},\,-\frac{7352}{245},\,\frac{7438}{35},\,-\frac{117716}{315},\,\frac{7438}{35},\,-\frac{7352}{245},\,\frac{10813}{2205},\,-\frac{146}{245},\,\frac{9}{245}  \right)$ &
  $-\frac{43}{1411200}\eta^8 -\frac{589}{12418560}\eta^{10} +\mathcal{O}(\eta^{12})$ \\
  \hline
\end{tabular}
 \caption{Second degree, centered Pad\'{e} derivative approximant schemes at a variety of approximation orders together with corresponding centered, explicit closures that have been tuned to match $\varepsilon_{\tilde{\eta}}$ at lowest order (see \S\ref{ssec:domdec_drp}).}
 \label{tab:pade_deg_two}
\end{table}
\begin{table}[htbp]
  \centering
  \begin{tabular}{c|l|l|l||l|l|l||l|l|l}
    \hline
      &
      ${}^{(1)}\overline{W}{}^{-}_{1}$ &
      ${}^{(1)}W{}^{-}_{1}$ &
      ${}^{(1)}\widetilde{W}{}^{-}_{1}$ &
      ${}^{(1)}\overline{W}{}^{-}_{2}$ &
      ${}^{(1)}W{}^{-}_{2}$ &
      ${}^{(1)}\widetilde{W}{}^{-}_{2}$ &
      ${}^{(1)}\overline{W}{}^{-}_{3}$ &
      ${}^{(1)}W{}^{-}_{3}$ &
      ${}^{(1)}\widetilde{W}{}^{-}_{3}$ \\
    \hline
    \hline
    $\alpha{}^{(1)}_{-1}$ &
     &
    $\frac{181}{300}$ &
    $\frac{181}{300}$ &
     &
    $\frac{829}{1200}$ &
    $\frac{829}{1200}$ &
     &
    $\frac{301}{400}$ &
    $\frac{301}{400}$\\
    $\alpha{}^{(1)}_{\hpb 0}$ &
    1 &
    1 &
    1 &
    1 &
    1 &
    1 &
    1 &
    1 &
    1 \\
    $\alpha{}^{(1)}_{\hpb 1}$ &
    $\frac{181}{300}$&
    &
    &
    $\frac{829}{1200}$&
    &
    &
    $\frac{301}{400}$&
    &
    \\
    \hline
    $\alpha{}^{(0)}_{-5}$ &
    &
    &
    &
    &
    &
    &
    $-\frac{121632291}{616413414400}$ &
    &
    $\frac{171}{1254400}$\\
    $\alpha{}^{(0)}_{-4}$ &
    &
    &
    &
    $\frac{638561861}{555773535000}$ &
    &
    $-\frac{71}{135000}$&
    $\frac{433223073}{154103353600}$ &
    $-\frac{1}{16800}$&
    $-\frac{4273}{1881600}$\\
    $\alpha{}^{(0)}_{-3}$ &
    $-\frac{24759259}{3331598400}$&
    &
    $\frac{19}{14400}$&
    $-\frac{773306021}{55577353500}$ &
    $\frac{29}{12000}$&
    $\frac{5707}{540000}$&
    $-\frac{7264078741}{369848048640}$ &
    $\frac{181}{50400}$&
    $\frac{81997}{3763200}$\\
    $\alpha{}^{(0)}_{-2}$ &
    $\frac{7699921}{104112450}$&
    $-\frac{31}{600}$&
    $-\frac{181}{2400}$&
    $\frac{24069573467}{277886767500}$ &
    $-\frac{877}{12000}$&
    $-\frac{19043}{135000}$&
    $\frac{88945079957}{924620121600}$ &
    $-\frac{13}{160}$&
    $-\frac{30201}{156800}$\\
    $\alpha{}^{(0)}_{-1}$ &
    $-\frac{504568931}{1110532800}$&
    $-\frac{2561}{1800}$&
    $-\frac{3311}{2880}$&
    $-\frac{1015672928093}{2223094140000}$ &
    $-\frac{24271}{14400}$&
    $-\frac{1213417}{1080000}$&
    $-\frac{2891351427}{6289932800}$ &
    $-\frac{15189}{8000}$&
    $-\frac{497279}{448000}$\\
    $\alpha{}^{(0)}_{\hpb 0}$ &
    $-\frac{1352100071}{1665799200}$&
    $\frac{69}{50}$&
    $\frac{3163}{3600}$&
    $-\frac{480625731203}{555773535000}$ &
    $\frac{151}{80}$&
    $\frac{5969}{6750}$&
    $-\frac{77442062}{85995175}$ &
    $\frac{1651}{720}$&
    $\frac{40601}{44800}$\\
    $\alpha{}^{(0)}_{\hpb 1}$ &
    $\frac{754089743}{666319680}$&
    $\frac{67}{600}$&
    $\frac{1847}{4800}$&
    $\frac{4974276451727}{4446188280000}$ &
    $-\frac{307}{2400}$&
    $\frac{234379}{540000}$&
    $\frac{244089838091}{220147648000}$ &
    $-\frac{269}{800}$&
    $\frac{40533}{89600}$\\
    $\alpha{}^{(0)}_{\hpb 2}$ &
    $\frac{15762473}{277633200}$&
    $-\frac{31}{1800}$&
    $-\frac{59}{1440}$&
    $\frac{1867414306}{13894338375}$ &
    $-\frac{167}{36000}$&
    $-\frac{9803}{135000}$&
    $\frac{5864828469}{30820670720}$ &
    $\frac{49}{2400}$&
    $-\frac{42779}{470400}$\\
    $\alpha{}^{(0)}_{\hpb 3}$ &
    $\frac{48956249}{3331598400}$&
    &
    $\frac{19}{14400}$&
    $-\frac{7783562141}{2223094140000}$ &
    $\frac{29}{24000}$&
    $\frac{10109}{1080000}$&
    $-\frac{35183198149}{1849240243200}$ &
    $-\frac{17}{20160}$&
    $\frac{65309}{3763200}$\\
    $\alpha{}^{(0)}_{\hpb 4}$ &
    $-\frac{682551}{185088800}$&
    &
    &
    $-\frac{159391966}{69471691875}$ &
    &
    $-\frac{71}{135000}$&
    $\frac{376907057}{462310060800}$ &
    $-\frac{1}{28000}$&
    $-\frac{7047}{3136000}$\\
    $\alpha{}^{(0)}_{\hpb 5}$ &
    &
    &
    &
    $\frac{1913697589}{4446188280000}$ &
    &
    &
    $\frac{161328177}{616413414400}$ &
    &
    $\frac{171}{1254400}$\\
    $\alpha{}^{(0)}_{\hpb 6}$ &
    &
    &
    &
    &
    &
    &
    $-\frac{10105731}{220147648000}$ &
    &
    \\
    \hline
    $\alpha{}^{(2)}_{\hpb 0}$ &
    &
    $\frac{19}{100}$ &
    &
    &
    $\frac{71}{200}$ &
    &
    &
    $\frac{19}{40}$ &
    \\
    \hline
  \end{tabular}\\
\begin{tabular}{c|l|l|l}
  \hline
  & ${}^{(1)}W{}^{-}_{1}$ & ${}^{(1)}W{}^{-}_{2}$ & ${}^{(1)}W{}^{-}_{3}$ \\
  \hline
  \hline
  $\sigma$ &
  $38/100$ &
  $71/100$ &
  $95/100$ \\
  $\varepsilon_{\tilde{\eta}}$ &
  $-\frac{19}{9620}\eta^4-\frac{11313 i}{3701776}\eta^5 +\mathcal{O}(\eta^6)$ &
  $-\frac{71}{213045}\eta^6-\frac{2308441 i }{7410313800}\eta^7 +\mathcal{O}(\eta^8)$ &
  $-\frac{19}{353304}\eta^8-\frac{235017 i}{7705167680}\eta^9 +\mathcal{O}(\eta^{10})$ \\
  \hline
\end{tabular}
  \caption{First degree derivative approximant schemes ${}^{(1)}W{}^{-}_{M}$ (i.e CCU$(M,M+1)$ with embedded term $f{}^{(2)}_{\hpb k}$ approximated through ${}^{(1)}P{}^1_M$ -- see \S\ref{ssec:domdec_drp}) of formal order of accuracy $\mathcal{O}(\delta x^{2M+2})$ together with left and right closures that have been tuned to match $\Im[\varepsilon_{\tilde{\eta}}]$ of the interior scheme at lowest order. The dispersion relation is approximately preserved at the lowest two orders under the left ${}^{(1)}\overline{W}{}^{-}_{M}$ and right ${}^{(1)}\widetilde{W}{}^{-}_{M}$ closures. Note that the oppositely biased schemes and corresponding closures are given by $\alpha{}^{(0)}_{\hpb m}[{}^{(1)}W{}^{+}_{M}] = - \alpha{}^{(0)}_{-m}[{}^{(1)}W{}^{-}_{M}]$ and $\alpha{}^{(1)}_{\hpb m}[{}^{(1)}W{}^{+}_{M}] = \alpha{}^{(1)}_{-m}[{}^{(1)}W{}^{-}_{M}]$.}
 \label{tab:ccuw_deg_one_col}
\end{table}
\begin{table}[htbp]
  \centering
  \begin{tabular}{c|l|l||l|l||l|l}
    \hline
      &
      ${}^{(1)}\overline{X}{}^{-}_{1}$ &
      ${}^{(1)}X{}^{-}_{1}$ &
      ${}^{(1)}\overline{X}{}^{-}_{2}$ &
      ${}^{(1)}X{}^{-}_{2}$ &
      ${}^{(1)}\overline{X}{}^{-}_{3}$ &
      ${}^{(1)}X{}^{-}_{3}$ \\
    \hline
    \hline
    $\alpha{}^{(1)}_{-1}$ &
    &
    $\frac{121}{200}$ &
    &
    $\frac{407}{600}$ &
    &
    $\frac{727}{1000}$ \\
    $\alpha{}^{(1)}_{\hpb 0}$ &
    1 &
    1 &
    1 &
    1 &
    1 &
    1 \\
    $\alpha{}^{(1)}_{\hpb 1}$ &
    $\frac{121}{200}$&
    &
    $\frac{407}{600}$&
    &
    $\frac{727}{1000}$&
    \\
    \hline
    $\alpha{}^{(0)}_{-5}$ &
      &
      &
      &
      &
    $-\frac{2248363}{11184483750}$ &
    $\frac{73}{315000}$ \\
    $\alpha{}^{(0)}_{-4}$ &
      &
      &
    $\frac{194871521}{170360232000}$ &
    $-\frac{43}{56000}$ &
    $\frac{32109283}{10676098125}$ &
    $-\frac{1549}{504000}$ \\
    $\alpha{}^{(0)}_{-3}$ &
    $-\frac{1488913}{206082000}$ &
    $\frac{37}{18000}$ &
    $-\frac{242883881}{17036023200}$ &
    $\frac{1523}{126000}$ &
    $-\frac{182778443}{8540878500}$ &
    $\frac{1019}{42000}$ \\
    $\alpha{}^{(0)}_{-2}$ &
    $\frac{628021}{8586750}$ &
    $-\frac{321}{4000}$ &
    $\frac{3813668231}{42590058000}$ &
    $-\frac{857}{6000}$ &
    $\frac{4735581811}{45551352000}$ &
    $-\frac{1607}{8400}$ \\
    $\alpha{}^{(0)}_{-1}$ &
    $-\frac{10390539}{22898000}$ &
    $-\frac{57}{50}$ &
    $-\frac{8537761009}{18252882000}$ &
    $-\frac{39697}{36000}$ &
    $-\frac{5482052561}{11387838000}$ &
    $-\frac{64709}{60000}$ \\
    $\alpha{}^{(0)}_{\hpb 0}$ &
    $-\frac{66763501}{82432800}$ &
    $\frac{3119}{3600}$ &
    $-\frac{3401756057}{4056196000}$ &
    $\frac{245}{288}$ &
    $-\frac{6941712887}{8134170000}$ &
    $\frac{17007}{20000}$ \\
    $\alpha{}^{(0)}_{\hpb 1}$ &
    $\frac{7728137}{6869400}$ &
    $\frac{79}{200}$ &
    $\frac{40137795113}{36505764000}$ &
    $\frac{2723}{6000}$ &
    $\frac{5852841859}{5422780000}$ &
    $\frac{2911}{6000}$ \\
    $\alpha{}^{(0)}_{\hpb 2}$ &
    $\frac{1473587}{22898000}$ &
    $-\frac{181}{4000}$ &
    $\frac{507715327}{3650576400}$ &
    $-\frac{1459}{18000}$ &
    $\frac{56883929}{295788000}$ &
    $-\frac{1493}{14000}$ \\
    $\alpha{}^{(0)}_{\hpb 3}$ &
    $\frac{1189063}{103041000}$ &
    $\frac{37}{18000}$ &
    $-\frac{338742841}{42590058000}$ &
    $\frac{2899}{252000}$ &
    $-\frac{97022683}{3795946000}$ &
    $\frac{379}{16800}$ \\
    $\alpha{}^{(0)}_{\hpb 4}$ &
    $-\frac{429671}{137388000}$ &
     &
    $-\frac{458269213}{511080696000}$ &
    $-\frac{43}{56000}$ &
    $\frac{42946231}{11387838000}$ &
    $-\frac{8297}{2520000}$ \\
    $\alpha{}^{(0)}_{\hpb 5}$ &
     &
     &
    $\frac{16295059}{63885087000}$ &
     &
    $-\frac{144978233}{341635140000}$ &
    $\frac{73}{315000}$ \\
    $\alpha{}^{(0)}_{\hpb 6}$ &
     &
     &
     &
     &
    $\frac{167862119}{7515973080000}$ &
     \\
    \hline
 \end{tabular}\\
\begin{tabular}{c|l|l|l}
  \hline
  & ${}^{(1)}X{}^{-}_{1}$ & ${}^{(1)}X{}^{-}_{2}$ & ${}^{(1)}X{}^{-}_{3}$ \\
  \hline
  \hline
  $\sigma$ &
  $37/100$ &
  $86/100$ &
  $146/100$ \\
  $\varepsilon_{\tilde{\eta}}$ &
  $-\frac{37 }{19260}\eta^4 -\frac{3551 i }{1373880}\eta^5 +\mathcal{O}(\eta^6)$ &
  $-\frac{43 }{105735}\eta^6-\frac{40037 i }{212950290}\eta^7 +\mathcal{O}(\eta^8)$ &
  $-\frac{73 }{870408}\eta^8+\frac{230897 i}{15031946160}\eta^9 -\mathcal{O}(\eta^{10})$ \\
  \hline
\end{tabular}
  \caption{First degree derivative approximant schemes ${}^{(1)}X{}^{-}_{M}$ (i.e CCU$(M,M+1)$ with embedded term $f{}^{(2)}_{\hpb k}$ approximated through ${}^{(2)}E_{M+2}$ -- see \S\ref{ssec:domdec_drp}) of formal order of accuracy $\mathcal{O}(\delta x^{2M+2})$ together with closures that have been tuned to match $\Im[\varepsilon_{\tilde{\eta}}]$ at lowest order. The dispersion relation is exactly preserved under the closures ${}^{(1)}\overline{X}{}^{-}_{M}$. Note that the oppositely biased schemes and corresponding closures are given by $\alpha{}^{(0)}_{\hpb m}[{}^{(1)}X{}^{+}_{M}] = - \alpha{}^{(0)}_{-m}[{}^{(1)}X{}^{-}_{M}]$ and $\alpha{}^{(1)}_{\hpb m}[{}^{(1)}X{}^{+}_{M}] = \alpha{}^{(1)}_{-m}[{}^{(1)}X{}^{-}_{M}]$.}
 \label{tab:ccux_deg_one_col}
\end{table}
\begin{table}[htbp]
  \centering
  \begin{tabular}{c|r|r||r||r|r||r}
    \hline
      &
      ${}^{(1)}\overline{Q}{}_{4}[\eta_c=1]$ &
      ${}^{(1)}Q{}^1_{3}[\eta_c=1]$ &
      ${}^{(1)}Q{}^{1,0}_{3,4}[\eta_c=1]$ &
      ${}^{(1)}\overline{Q}{}_{4}\left[\eta_c=\frac{4\pi}{5}\right]$ &
      ${}^{(1)}Q{}^1_{3}\left[\eta_c=\frac{4\pi}{5}\right]$ &
      ${}^{(1)}Q{}^{1,0}_{3,4}\left[\eta_c=\frac{4\pi}{5}\right]$ \\
    \hline
    \hline
    $\alpha{}^{(1)}_{-1}$ &
    &                    
    $0.37987923$ &       
    $0.61258918$ &       
                 &       
    $0.41825851$ &       
    $0.71391655$         
    \\
    $\alpha{}^{(1)}_{\hpb 0}$ &
    1 & 
    1 & 
    1 & 
    1 & 
    1 & 
    1   
    \\
    $\alpha{}^{(1)}_{\hpb 1}$ &
                                   & 
    $0.37987923$                   & 
                                   & 
                                   & 
    $0.41825851$                   & 
    \\
    \hline
    $\alpha{}^{(0)}_{-4}$ &
    $0.0035978349$                 & 
                                   & 
                                   & 
    $0.0037957585$                 & 
                                   & 
    \\
    $\alpha{}^{(0)}_{-3}$ &
    $-0.038253676$                    & 
    $0.0023272948$                    & 
    $0.0054439068$                    & 
    $-0.039441218$                    & 
    $0.0042462587$                    & 
    $0.011541239$                       
    \\
    $\alpha{}^{(0)}_{-2}$ &
    $0.20036969$                      & 
    $-0.052602255$                    & 
    $-0.10687221$                     & 
    $0.20314062$                      & 
    $-0.073071204$                    & 
    $-0.16644143$                       
    \\
    $\alpha{}^{(0)}_{-1}$ &
    $-0.80036969$                     & 
    $-0.78165660$                     & 
    $-1.0718341$                      & 
    $-0.80314062$                     & 
    $-0.78485488$                     & 
    $-1.0738269$                        
    \\
    $\alpha{}^{(0)}_{\hpb 0}$ &
    $0$                              & 
    $0$                              & 
    $0.75760288$                     & 
    $0$                              & 
    $0$                              & 
    $0.79751467$                       
    \\
    $\alpha{}^{(0)}_{\hpb 1}$ &
    $0.80036969$                      & 
    $0.78165660$                      & 
    $0.50288811$                      & 
    $0.80314062$                      & 
    $0.78485488$                      & 
    $0.54741579$                        
    \\
    $\alpha{}^{(0)}_{\hpb 2}$ &
    $-0.20036969$                  & 
    $0.052602255$                  & 
    $-0.10383106$                  & 
    $-0.20314062$                  & 
    $0.073071204$                  & 
    $-0.14743556$                    
    \\
    $\alpha{}^{(0)}_{\hpb 3}$ &
    $0.038253676$                    & 
    $-0.0023272948$                  & 
    $0.018293386$                    & 
    $0.039441218$                    & 
    $-0.0042462587$                  & 
    $0.035642870$                      
    \\
    $\alpha{}^{(0)}_{\hpb 4}$ &
    $-0.0035978349$                   & 
                                      & 
    $-0.0016909342$                   & 
    $-0.0037957585$                   & 
                                      & 
    $-0.0044106875$                     
    \\
    \hline
 \end{tabular}\\
\begin{tabular}{c|l|l||l|l}
  \hline
  & ${}^{(1)}Q{}^{1}_{3}[\eta_c=1]$ & ${}^{(1)}Q{}^{1,0}_{3,4}[\eta_c=1]$ & ${}^{(1)}Q{}^{1}_{3}\left[\eta_c=\frac{4\pi}{5}\right]$ & ${}^{(1)}Q{}^{1,0}_{3,4}\left[\eta_c=\frac{4\pi}{5}\right]$ \\
  \hline
  \hline
  $\varepsilon_{\tilde{\eta}}$ &
  $5.3\times 10^{-5} \eta^6 - \mathcal{O}(\eta^{8})$ &
  $2.4\times 10^{-4} \eta^6 + \mathcal{O}(\eta^7)$ &
  $4.5\times 10^{-4} \eta^6 - \mathcal{O}(\eta^{8})$ &
  $2.4\times 10^{-3} \eta^6 + \mathcal{O}(\eta^7)$
  \\
  \hline
\end{tabular}
  \caption{Numerically tuned first degree derivative approximant schemes with formal order of accuracy $\mathcal{O}(\delta x^6)$. Closures ${}^{(1)}\overline{Q}{}_{4}$ for ${}^{(1)}Q{}^{1}_{3}$ are tuned to match $\varepsilon_{\tilde{\eta}}$ at lowest order. Note that biased schemes ${}^{(1)}Q{}^{1,0}_{3,4}$ may be used to construct ${}^{(1)}Q{}^{0,1}_{4,3}$ as $\alpha{}^{(0)}_{\hpb m}[{}^{(1)}Q{}^{0,1}_{4,3}] = - \alpha{}^{(0)}_{-m}[{}^{(1)}Q{}^{1,0}_{3,4}]$ and $\alpha{}^{(1)}_{\hpb m}[{}^{(1)}Q{}^{0,1}_{4,3}] = \alpha{}^{(1)}_{-m}[{}^{(1)}Q{}^{1,0}_{3,4}]$. As described in \S\ref{ssec:domdec_drp} the biased schemes here also satisfy: $\varepsilon_{\tilde{\eta}}[{}^{(1)}Q{}^{0,1}_{4,3}]=\varepsilon_{\tilde{\eta}}[{}^{(1)}Q{}^{1,0}_{3,4}]^*$.}
 \label{tab:num_deg_one}
\end{table}
\begin{table}[htbp]
  \centering
  \begin{tabular}{c|r|r||r||r|r||r}
    \hline
      &
      ${}^{(2)}\overline{Q}{}_{4}[\eta_c=1]$ &
      ${}^{(2)}Q{}^1_{3}[\eta_c=1]$ &
      ${}^{(2)}Q{}^{0,1}_{4,3}[\eta_c=1]$ &
      ${}^{(2)}\overline{Q}{}_{4}\left[\eta_c=\frac{4\pi}{5}\right]$ &
      ${}^{(2)}Q{}^1_{3}\left[\eta_c=\frac{4\pi}{5}\right]$ &
      ${}^{(2)}Q{}^{0,1}_{4,3}\left[\eta_c=\frac{4\pi}{5}\right]$ \\
    \hline
    \hline
    $\alpha{}^{(2)}_{-1}$ &
    &                    
    $0.24246603$ &       
    &                    
                 &       
    $0.28533501$ &       
    \\
    $\alpha{}^{(2)}_{\hpb 0}$ &
    1 & 
    1 & 
    1 & 
    1 & 
    1 & 
    1   
    \\
    $\alpha{}^{(2)}_{\hpb 1}$ &
                                   & 
    $0.24246603$                   & 
    $0.20617275$                   & 
                                   & 
    $0.28533501$                   & 
    $0.33257004$                     
    \\
    \hline
    $\alpha{}^{(0)}_{-4}$ &
    $-0.0018142695$                & 
                                   & 
    $-0.0022908083$                & 
    $-0.0020184946$                & 
                                   & 
    $-0.0036952226$                  
    \\
    $\alpha{}^{(0)}_{-3}$ &
    $0.025625267 $                    & 
    $-0.0037062571$                   & 
    $0.029437578$                     & 
    $0.027259068$                     & 
    $-0.0063260285$                   & 
    $0.040672892$                       
    \\
    $\alpha{}^{(0)}_{-2}$ &
    $-0.20079955$                     & 
    $0.14095923$                      & 
    $-0.21185182$                     & 
    $-0.20651785$                     & 
    $0.19240201$                      & 
    $-0.24977101$                       
    \\
    $\alpha{}^{(0)}_{-1}$ &
    $1.6015991$                       & 
    $0.95445144$                      & 
    $1.5973594$                       & 
    $1.6130357$                       & 
    $0.85799622$                      & 
    $1.6570470$                         
    \\
    $\alpha{}^{(0)}_{\hpb 0}$ &
    $-2.8492211$                     & 
    $-2.1834088$                     & 
    $-2.5733197$                     & 
    $-2.8635168$                     & 
    $-2.0881444$                     & 
    $-2.4820328$                       
    \\
    $\alpha{}^{(0)}_{\hpb 1}$ &
    $1.6015991$                       & 
    $0.95445144$                      & 
    $1.0670372$                       & 
    $1.6130357$                       & 
    $0.85799622$                      & 
    $0.80160292$                        
    \\
    $\alpha{}^{(0)}_{\hpb 2}$ &
    $-0.20079955$                  & 
    $0.14095923$                   & 
    $0.095116489$                  & 
    $-0.20651785$                  & 
    $0.19240201$                   & 
    $0.24538882$                     
    \\
    $\alpha{}^{(0)}_{\hpb 3}$ &
    $0.025625267$                    & 
    $-0.0037062571$                  & 
    $-0.0014883346$                  & 
    $0.027259068$                    & 
    $-0.0063260285$                  & 
    $-0.0092126135$                    
    \\
    $\alpha{}^{(0)}_{\hpb 4}$ &
    $-0.0018142695$                   & 
                                      & 
                                      & 
    $-0.0020184946$                   & 
                                      & 
    \\
    \hline
 \end{tabular}\\
\begin{tabular}{c|l|l||l|l}
  \hline
  & ${}^{(2)}Q{}^{1}_{3}[\eta_c=1]$ & ${}^{(2)}Q{}^{1,0}_{3,4}[\eta_c=1]$ & ${}^{(2)}Q{}^{1}_{3}\left[\eta_c=\frac{4\pi}{5}\right]$ & ${}^{(2)}Q{}^{1,0}_{3,4}\left[\eta_c=\frac{4\pi}{5}\right]$ \\
  \hline
  \hline
  $\varepsilon_{\tilde{\eta}}$ &
  $2.9\times 10^{-5} \eta^6 - \mathcal{O}(\eta^{8})$ &
  $1.1\times 10^{-4} \eta^6 - \mathcal{O}(\eta^7)$ &
  $2.3\times 10^{-4} \eta^6 - \mathcal{O}(\eta^{8})$ &
  $9.9\times 10^{-4} \eta^6 - \mathcal{O}(\eta^7)$
  \\
  \hline
\end{tabular}
  \caption{Numerically tuned second degree derivative approximant schemes with formal order of accuracy $\mathcal{O}(\delta x^6)$. Closures ${}^{(2)}\overline{Q}{}_{4}$ for ${}^{(2)}Q{}^{1}_{3}$ are tuned to match $\varepsilon_{\tilde{\eta}}$ at lowest order. Note that biased schemes ${}^{(2)}Q{}^{0,1}_{4,3}$ may be used to construct ${}^{(1)}Q{}^{1,0}_{3,4}$ as $\alpha{}^{(d)}_{\hpb m}[{}^{(2)}Q{}^{1,0}_{3,4}] = \alpha{}^{(d)}_{-m}[{}^{(2)}Q{}^{0,1}_{4,3}]$ where $d\in\{0,\,2\}$. The biased schemes here satisfy: $\varepsilon_{\tilde{\eta}}[{}^{(2)}Q{}^{0,1}_{4,3}]=\varepsilon_{\tilde{\eta}}[{}^{(2)}Q{}^{1,0}_{3,4}]^*$.}
 \label{tab:num_deg_two}
\end{table}

\end{document}